\documentclass[12pt]{article}
\usepackage[utf8]{inputenc}
\usepackage[english]{babel}
\usepackage{amsmath,amsfonts,pifont}

\usepackage{graphicx,booktabs,array,multicol,multirow,rotating}

\usepackage{url,xcolor}

\usepackage{comment,xr}
\usepackage{multirow}
\usepackage{threeparttable}
\usepackage{rotating,enumitem}
\usepackage{adjustbox}

\usepackage{natbib}
\usepackage[colorlinks=true,linkcolor=red,citecolor=blue]{hyperref}

\newcommand{\blind}{1}

\newcommand{\bZ}{\mathbf{Z}}
\newcommand{\bX}{\mathbf{X}}

\newcommand{\bx}{\mathbf{x}}

\newcommand{\bz}{\mathbf{z}}

\newcommand{\bU}{\mathbf{U}}

\newcommand{\bW}{\mathbf{W}}

\newcommand{\bR}{\mathbf{R}}

\newcommand{\bw}{\mathbf{w}}
\newcommand{\bo}{\mathbf{0}}

\newcommand{\bbeta}{\boldsymbol{\beta}}

\newcommand{\bSigma}{\boldsymbol{\Sigma}}
\newcommand{\balpha}{\boldsymbol{\alpha}}

\newcommand{\btheta}{\boldsymbol{\theta}}

\newcommand{\bmu}{\boldsymbol{\mu}}

\allowdisplaybreaks
\begin{document}

\def\spacingset#1{\renewcommand{\baselinestretch}%
{#1}\small\normalsize} \spacingset{1}


\if1\blind
{
  \title{\bf Regression calibration for logistic regression with correlated measurement error and multiple misclassification in covariates}

  \author{
    Zhiqiang Cao$^{1}$,
    Man Yu Wong$^{2,*}$
    \vspace{0.2cm} 
    \\
    $^{1}$School of Artificial Intelligence, Shenzhen Technology University, Guangdong, China \\
    $^{2}$Department of Mathematics, The Hong Kong University of Science and Technology, \\Hong Kong, China \\
    $^{*}$\emph{email}: mamywong@ust.hk
  }
  \maketitle
} \fi

\if0\blind
{
  \bigskip
  \bigskip
  \bigskip
  \begin{center}
    {\LARGE\bf Regression calibration for logistic regression with correlated measurement error and multiple misclassification in covariates}
  \end{center}
  \medskip
} \fi

\begin{abstract}
Epidemiological studies frequently face two major challenges of data quality on covariates: continuous predictors subject to measurement error and categorical predictors subject to misclassification. Existing correction methods for logistic regression typically address either measurement error in continuous covariates or misclassification for a single categorical variable. Few approaches simultaneously handle multiple misclassified categorical exposures alongside correlated measurement error in several continuous predictors. To fill this methodological gap, we extend the widely used regression calibration for logistic regression to accommodate multiple continuous covariates with correlated additive measurement error and two or more misclassified categorical covariates. We construct a unified modelling framework combining additive measurement error models and misclassification transition matrices, and derive the regression calibration estimator (RCE) via Kronecker product algebra. The proposed method is computationally
tractable and can in principle be generalized to more than two misclassified categorical variables. We conduct extensive simulation studies across various misclassification rates, measurement error magnitudes, sample sizes, and true effect sizes to compare finite‑sample performance of RCE with the naive estimator that ignores error. We further apply RCE to the EPIC‑InterAct case‑cohort study to correct for correlated measurement error in protein and energy intakes, together with misclassification in two categorical variables: two-level physical activity status and three‑level categorical alcohol consumption among male participants. Real data analysis indicates that ignoring both error sources yields severely biased coefficients and misleading inference for main and interaction effects in nutritional epidemiology. The proposed RCE provides a practical correction tool, though it treats validation‑study nuisance parameters as known and ignores their estimation uncertainty.
\end{abstract}


\newpage
\spacingset{1.8} 

\section{Introduction}\label{sec1}
Logistic regression is one of the most widely used statistical tools in medical, epidemiological, social, and behavioral sciences for modeling the relationship between a binary outcome and a set of covariates. However, covariates are not measured perfectly in many practical applications \citep{Young2018}. Continuous covariates are often subject to measurement error and categorical covariates are frequently misclassified \citep{Keogh2020,Shaw2020}. These data quality issues are particularly prevalent in nutritional epidemiology \citep{Shaw2018}, as dietary intakes assessed by food frequency questionnaires (FFQ) are prone to measurement error due to recall bias and lifestyle factors, such as physical activity or dietary habits, are often misclassified due to self-reporting inaccuracies \citep{Sobiecki2023,Li2026}.

It is well established that ignoring measurement error or misclassification in covariates can lead to severely biased parameter estimates, loss of statistical power, and invalid inference \citep{Carroll2006}. In logistic regression, measurement errors in continuous covariates typically attenuate their estimated regression coefficients toward zero, whereas misclassification in categorical covariates can bias estimates in either direction, depending on the misclassification mechanism \citep{Keogh2020,Fraser2001}. 
When both types of errors are present in large nutritional epidemiology studies simultaneously, i.e., correlated measurement errors in multiple continuous covariates such as protein, fat, and energy, and misclassified status of physical activity, the problem becomes considerably more complex and challenging \citep{Cao2023,Cao2025}.     
This problem merits careful methodological investigation, as epidemiological regression models frequently contain multiple covariates: some continuous covariates are measured with error, and some categorical covariates suffer from misclassification. Analyses that fail to handle both error types jointly may produce unreliable regression estimates.

For the problem of measurement error and misclassification in regression models, most existing references focus on addressing one type of error \citep{Keogh2020,Shaw2020}, and substantive methodologies have been developed to solve measurement error and misclassification in logistic regression separately. For example, methods for handling measurement error in continuous covariates such as regression calibration (RC) \citep{Rosner1989}, simulation extrapolation (SIMEX) \citep{Cook1994}, conditional score approaches \citep{Stefanski1987}, and various score or likelihood-based methods \citep{Novick2002,Schneeweiss2009,Zucker2013,Cao2021} have been proposed. For handling misclassification in categorical covariates, common correction methods include maximum likelihood approaches \citep{Spiegelman2000,Yi2015}, Bayesian methods \cite{Falley2018,MacLehose2018}, and misclassification SIMEX \citep{Kuchenhoff2006}.

However, far less attention has been devoted to the simultaneous treatment of both measurement error and misclassification in covariates. Notable among the sparse contributions, \citet{Spiegelman2000} developed estimation and inference procedures for logistic regression with  misclassification and measurement error in covariates under main study/validation study designs. \citet{Yi2015} systematically studied functional and structural methods for generalized linear models with mixed measurement error and misclassification in covariates under the scenario that an external validation study is available. But, both of their validation studies included true and error-prone measurements for covariates with measurement error or misclassification. Thus, estimates of all unknown parameters in logistic or GLM model can be obtained by maximizing a full data likelihood on contributions from both main and validation studies. In addition, \citet{Cao2023} proposed an approximate likelihood estimation (ALE) method for logistic regression
on multivariate continuous covariates
with correlated error and one misclassified categorical variable simultaneously based on the  EPIC-InterAct Study \citep{InterAct2011}. It is noted that their validation study only contains a single measure from FFQ and a single measure from a 24-hour recall (24HR) on each error-prone covariate, which has no true covariates and is different from the validation study scenario in \citet{Spiegelman2000} and \citet{Yi2015}
Furthermore, \citet{Cao2025} applied the ALE method to the Weibull regression model with both measurement error and misclassification in covariates.

Despite these advances, existing methods are largely limited to situations involving at most one misclassified categorical variable. However, in many real-world applications, multiple categorical covariates may be subject to misclassification. For example, physical activity level (Cambridge index) in the EPIC-InterAct Study \citep{InterAct2011} is self-reported and thus prone to misclassification. Also, 
in other epidemiological cohorts \citep{Nishinakagawa2023,Hu2022} variables, e.g., education level, occupation and dietary habits, may be misclassified. 
It is noted that published studies suggested a U-shaped relationship between alcohol and type 2 diabetes \citep{Carlsson2005,Takahashi2024}, so if we classify alcohol consumption into three levels, that is, low alcohol consumption, moderate alcohol consumption and high alcohol consumption, then it is likely that the resulting categorical alcohol consumption levels will subject to misclassification since alcohol consumption is collected by self-reporting. When two or more categorical variables are misclassified, their misclassification errors may be correlated, e.g., individuals who over-report physical activity may also under-report smoking. In this case, the correction methods developed for a single misclassified variable are no longer directly applicable.

To the best of our knowledge, no existing regression calibration framework for logistic regression permits multiple categorical covariates to be simultaneously misclassified and also accounts for correlated measurement error among continuous predictors. This gap is significant because: (1) multiple categorical covariates (e.g., physical activity, discretized alcohol consumption, dietary patterns) are often measured with error (i.e. misclassification) in practice; (2) ignoring the presence of multiple misclassified variables can lead to substantial bias in estimated regression coefficients and incorrect inferences about interaction effects.

The objective of this paper is to fill this gap by extending the regression calibration (RC) method developed in \citet{Cao2023} to accommodate two misclassified categorical variables in addition to multiple continuous covariates with correlated measurement errors. In particular, this paper makes four main contributions. First, we develop a unified modeling framework that incorporates an additive measurement error model for continuous covariates and different misclassification models for two categorical variables. Second, we derive the RC estimation for a logistic regression model with correlated measurement errors in continuous covariates and misclassification in two or more
categorical variables. Third, we carry out extensive simulation studies to evaluate the finite-sample performance of the proposed method under various cases of measurement error magnitude, misclassification rate, and sample size. Finally, we apply the proposed method to analyze data from the EPIC-InterAct Study, where multiple nutrients are measured with correlated errors, and both physical activity and discretized alcohol consumption are subject to misclassification.

The remainder of this paper is organized as follows. In Section 2, we introduce the logistic regression model, the misclassification model for two categorical variables, and the additive measurement error model for continuous covariates. Section 3 demonstrates how to conduct the RC estimation in logistic regression model when two or more categorical variables are misclassified as well as multiple continuous covariates are measured with correlated additive error. We conduct extensive simulation studies to examine the finite-sample performance of RC in Section 4, and show its advantage in bias reduction and other metrics over naive estimation (i.e., the method ignores both measurement error and misclassification). 
In Section 5, we apply the proposed method to the EPIC-InterAct Study data to illustrate its application value. Section 6 concludes with discussion and future directions.

\section{Notation and Models for logistic regression}\label{sec:notation}
\subsection{Logistic regression model}\label{subsec:logistic}
Throughout the article, we use the following notations as illustration. Let $Y$ be a response variable; $\bX=(X_1,\cdots,X_p)^{\top}$ be a $p$-dimensional vector of true values for error-prone continuous covariates; and ${\bZ}=(Z_1,\cdots,Z_q)^{\top}$ be a $q$-dimensional vector of error-free covariates, in which each covariate in $\bZ$ can be either continuous or discrete. In addition, ${\bX}$ and ${\bZ}$ are assumed to be independent. In this article, we consider a logistic regression of $Y$ on two categorical variables $V$ and $A$ together with 
${\bX}$ and ${\bZ}$, where $V$ has two levels (e.g., 0 and 1) and $A$ has three levels (e.g., 1, 2 and 3). Combination of $V$ and $A$ creates a new variable $G$ such that $G=3m+j$ when $V=m$ and $A=j$ for $m=0,1$ and $j=1,2,3$. Assume the form of the logistic model is:
\begin{eqnarray}\label{logistic}
	{\rm log}\bigg({P\over 1-P}\bigg)=\sum_{l=1}^6(\beta_{0l}+\beta_l^{\top}\bX)I(G=l)+\balpha^T {\bZ}=
	\begin{cases}
		\beta_{01}+\bbeta_1^{\top} {\bX}+\balpha^{\top} {\bZ} ~~~~~G=1, \\
		\beta_{02}+\bbeta_2^{\top} {\bX}+\balpha^{\top} {\bZ} ~~~~~G=2, \\
		\beta_{03}+\bbeta_3^{\top} {\bX}+\balpha^{\top} {\bZ} ~~~~~G=3, \\
		\beta_{04}+\bbeta_4^{\top} {\bX}+\balpha^{\top} {\bZ} ~~~~~G=4, \\
		\beta_{05}+\bbeta_5^{\top} {\bX}+\balpha^{\top} {\bZ} ~~~~~G=5, \\
		\beta_{06}+\bbeta_6^{\top} {\bX}+\balpha^{\top} {\bZ} ~~~~~G=6, \\
	\end{cases},
\end{eqnarray}
where $P=Pr(Y=1|{\bX},{\bZ}, V, A)$, $I(\cdot)$ is an indicator function, $\beta_{0l}$, $\bbeta_l=(\beta_{1l},\dots,\beta_{pl})^{\top}$ for $l=1,\dots,6$
and $\balpha=(\alpha_1,\dots,\alpha_q)^{\top}$ are unknown parameters. If we choose 0 and 1 as reference groups for $V$ and $A$, then the model \eqref{logistic} above is equivalent to a model (i.e., model \eqref{logistic4}) with categorical variables in terms of indicator (or dummy) variables, covariates and their interactions, that is,
\begin{eqnarray}\label{logistic4}
	{\rm log}\bigg({P\over 1-P}\bigg)&=&\gamma_{010}+\gamma_{110}V+\gamma_{020}I(A=2)+	\gamma_{030}I(A=3)+\gamma_{120}V*I(A=2)+  \nonumber \\
	&&\gamma_{130}V*I(A=3)+\sum_{k=1}^p\left[\gamma_{01k}X_k+\gamma_{11k}V*X_k +\gamma_{02k}I(A=2)*X_k+ \right.  \\ 
	&&\left. \gamma_{03k}I(A=3)*X_k+\gamma_{12k}V*I(A=2)*X_k+\gamma_{13k}V*I(A=3)*X_k\right],\nonumber
\end{eqnarray}	
where $\gamma_{mjk}$ is the regression coefficient of $V=m$, $A=j$ and $X_k$ for $k=1,2,\dots,p$. Then, $\gamma_{010}=\beta_{01}$, $\gamma_{020}=\beta_{02}-\beta_{01}$, $\gamma_{030}=\beta_{03}-\beta_{01}$, $\gamma_{110}=\beta_{04}-\beta_{01}$,
$\gamma_{120}=\beta_{05}-\beta_{02}-(\beta_{04}-\beta_{01})$, $\gamma_{130}=\beta_{06}-\beta_{03}-(\beta_{04}-\beta_{01})$, $\gamma_{01k} = \beta_{k1}$, 
$\gamma_{02k}=\beta_{k2}-\beta_{k1}$, $\gamma_{03k}=\beta_{k3}-\beta_{k1}$, $\gamma_{11k}=\beta_{k4}-\beta_{k1}$,
$\gamma_{12k}=\beta_{k5}-\beta_{k2}-(\beta_{k4}-\beta_{k1})$,
$\gamma_{13k}=\beta_{k6}-\beta_{k3}-(\beta_{k4}-\beta_{k1})$. Therefore, based on model (\ref{logistic}), we can easily obtain interaction effects between $\bX$, $V$ and $A$. Next, we focus on model  (\ref{logistic}) for parameter estimation.  

Denote ${n}^*_l$ as number of observations for $G$ with value $l$ (defined  as $G^{(l)}$) for $l=1,\dots,6$. 
Based on model (\ref{logistic}) and sample data $(y_i, {\bx}_{i},{\bz}_i,v_i,a_i)=(y_i, {\bx}_{i},{\bz}_i,G^{(l)}_i)$, for $i=1,2,\dots,{n}^*_l$ and $l=1,\dots, 6$, the likelihood function of $\btheta=(\beta_{01},\beta_{02},\beta_{03},\beta_{04},\beta_{05},\beta_{06},\bbeta_1^{\top},\bbeta_2^{\top},\bbeta_3^{\top},
\bbeta_4^{\top},\bbeta_5^{\top},\bbeta_6^{\top},\balpha^{\top})^{\top}$ can be written as
\begin{eqnarray*}\label{like_log}
	L(\theta|{\bx},{\bz},G)&=&\prod_{l=1}^6 \prod_{i=1}^{n^*_l}\bigr[Pr(y_i=1|{\bx}_i,{\bz}_i,G^{(l)}_i)^{y_i}Pr(y_i=0|{\bx}_i,{\bz}_i,G^{(l)}_i)^{1-y_i}\bigr]\\
	& =&\prod_{l=1}^6\prod_{i=1}^{n^*_l}\bigg[{e^{(\beta_{0l}+\bbeta_l^{\top} {\bx}_i+\balpha^{\top} {\bz}_i)y_i} \over 1+e^{\beta_{0l}+\bbeta_l^{\top} {\bx}_i+\balpha^{\top} {\bz}_i}}\bigg].
\end{eqnarray*}
Maximizing the above function $L(\btheta|{\bx},{\bz},G)$ gives the MLE of $\btheta$, denoted as $\widehat{\btheta}$. We should use numerical method such as Newton-Raphson iteration to obtain $\widehat{\btheta}$ since logistic regression has no closed-form expression for MLE. Under standard regularity conditions, $\widehat{\btheta}$ is consistent and asymptotically normality.\cite{Shao1999}

\subsection{Misclassification and measurement error models}\label{subsec:misclassification_measurement_error}
The basic preparation for any measurement error correction method is to lay down on the assumptions of the relationship between unobserved variables and their observable counterparts carefully. In short, the assumptions listed below and the subsequent comments form the basis for the correction of measurement error and misclassification in Section \ref{sec:estimation_method}.

Throughout this article, we make the following assumptions on misclassification and measurement error models:
\begin{enumerate}
	\item[(A1)] For misclassification and additive measurement error models, we make a standard assumption that both types of errors are non-differential \citep{Carroll2006,Cao2023,Cao2025,Spiegelman2000,Yi2015}. 
	\item[(A2)] For the misclassification model of a binary variable $V$, let $V_o$ be the corresponding observed version of true $V$. Define observed and true proportions as $P^v_{obs}=Pr(V_o=1)$ and $P^v_t=Pr(V=1)$, respectively. Misclassification probabilities of $V$ equal to $m$ are defined as 
	\begin{eqnarray*}
		\pi^v_m=Pr(V_o=m'|V=m),~~m \neq m',~m, ~m' \in \{0,1\}.	
	\end{eqnarray*}
	We assume misclassification probabilities for different values of $V$ are independent of each other, i.e., $\pi^v_1 \perp \pi^v_0$. Based on this assumption, we have $P^v_{obs}=\pi^v_0(1-P^v_t)+(1-\pi^v_1)P^v_t$.
	
	Similarly, for a three-level categorical variable $A$, let $A_o$ be its corresponding observed value. Define observed proportion as $P^a_{obs,j}=Pr(A_o=j)$, true proportion as $P^a_{T,j}=Pr(A=j)$ and independent misclassification probabilities as $\phi_{jj'}=Pr(A_o=j'|A=j)$ for $j,j'=1,2,3$. Then we have
	\begin{eqnarray*}
		P^a_{obs,1}&=&(1-\phi_{12}-\phi_{13})P^a_{T,1}+\phi_{21}P^a_{T,2}+\phi_{31}P^a_{T,3},\\
		P^a_{obs,2}&=&\phi_{12}P^a_{T,1}+(1-\phi_{21}-\phi_{23})P^a_{T,2}+\phi_{32}P^a_{T,3},\\
		P^a_{obs,3}&=&\phi_{13}P^a_{T,1}+\phi_{23}P^a_{T,2}+(1-\phi_{31}-\phi_{32})P^a_{T,3}.
	\end{eqnarray*}
	\item[(A3)] Due to measurement errors in covariates, we do not observe true values of $\bX=(X_1,\dots,X_p)^{\top}$. Instead, their respective surrogates ${\bW}=(W_1,\cdots,W_p)^{\top}$ are available. We assume $\bX$ and $\bW$ are related in an additive error model, i.e.,  
	\begin{eqnarray}\label{add_error_m}	
		W_k=a_k+b_kX_k+U_k,\quad k=1,\cdots,p,
	\end{eqnarray}
	where $a_k$ and $b_k$ are unknown parameters.
	\item[(A4)] $\bX$ is assumed to follow multivariate normal distribution, i.e., ${\bX} \sim N(\bmu_x,\bSigma_{xx})$, where $\bmu_x=(\mu_{x_1},\cdots,\mu_{x_p})^{\top}$,
	the $k$th diagonal element of $\bSigma_{xx}$ is denoted as $\sigma^2_{x_k}$ for $k=1,\cdots,p$, and the $(i,j)$th element of $\bSigma_{xx}$ is $\rho_{x_i x_j}\sigma_{x_i}\sigma_{x_j}$ for $i \neq j,~i,j=1,\cdots,p$, with $\rho_{x_i x_j}$ as the Pearson correlation coefficient of $X_i$ and $X_j$.
	\item[(A5)] $\bU=(U_1,\dots,U_p)^{\top}$ is assumed to be independent of $Y$, $\bX$, $\bZ$, $V$ and $A$.
	\item[(A6)] $\bU$ is assumed to be multivariate normally distributed, i.e., ${\bU} \sim N({\bo},\bSigma_{uu})$, where the $k$th diagonal element of $\bSigma_{uu}$ is denoted as $\sigma^2_{u_k}$ for $k=1,\dots,p$, and the $(i,j)$th element of $\bSigma_{uu}$ is $\rho_{u_i u_j}\sigma_{u_i}\sigma_{u_j}$ for $i \neq j, i,j=1,\dots,p$, with $\rho_{u_i u_j}$ as the Pearson correlation coefficient of $U_i$ and $U_j$. 
	\item[(A7)] The conditional distribution of $\bX$ given $\bW$ and $\bZ$ does not depend on $\bZ$.
	\item[(A8)] Unknown parameters in misclassification and measurement error models can be estimated from a validation study or are known from an external source.
\end{enumerate}

The misclassification models in (A2) is commonly assumed when the categorical variable is misclassified \citep{Cao2023,Zuck2008,Sevilimedu2023}. By contrast, the additive error model (\ref{add_error_m}) in (A3) is also widely used for the FFQs and nutrient biomarkers in epidemiological studies \citep{Keogh2013}. The normality assumption of $\bX$ in (A4) provides an indispensable requirement for the construction of the covariate measurement error correction method based on the conditional distribution of $\bX$ given $\bW$. Assumption (A5) is required to make the non-differential error. As usual, we have no a priori knowledge of measurement error $\bU$. It is common to make the normality assumption (A6) on the error term \citep{Carroll2006}. In addition, the distribution of $\bX$ given $\bW$ and $\bZ$ is required for the construction of our proposed method of estimation. However, it cannot be estimated if not all error-free covariates are available in the dataset of the validation study, e.g., the real data example in Section \ref{sec:data_analysis}. Therefore, we adopt the assumption (A7) used by \citet{Augustin1999} for the correction of measurement error in parametric duration models as one of our assumptions. This assumption makes sure that the conditional distribution of $\bX$ given $\bW$ and $\bZ$ is equal to the conditional distribution of $\bX$ given $\bW$, i.e., $(\bX|\bW,\bZ)\sim (\bX|\bW)$, which is crucial for the method developed in Section \ref{sec:estimation_method}. Also, the measurement error model (\ref{add_error_m}) is reasonable under the assumption (A7). For assumption (A8), these parameters are estimated from validation data in practice, and their estimation uncertainty is not propagated in our subsequent estimation procedure.

Based on the additive measurement error model (\ref{add_error_m}) and assumptions of (A4)-(A7), we have
\begin{eqnarray}\label{con_xw}
	(\bX|\bW,\bZ)\sim (\bX| \bW) \sim N(\bmu_{x|w},\bSigma_{x|w}),
\end{eqnarray}
where $\bmu_{x|w}=\bmu_x+\bSigma_{xw}\bSigma^{-1}_{ww}(\bW-\bmu_{w})$, $\bmu_w$ is the mean vector of ${\bW}$, $\bSigma_{xw}={\rm Cov}(\bX,\bW)$, $\bSigma_{ww}={\rm Var}(\bW)$ and $\bSigma_{x|w}=\bSigma_{xx}-\bSigma_{xw}\bSigma^{-1}_{ww}\bSigma^{\top}_{xw}$.

For estimating regression coefficients in the logistic regression \eqref{logistic}, we first estimate unknown parameters in misclassification and measurement error models from a validation study and then treat them as true parameter values. The uncertainty in the estimation of these unknown parameters is not taken into account since it is difficult to obtain. Detailed discussion on methods for estimating unknown parameters in misclassification and measurement error models using the dataset from the EPIC-InterAct Study is given in Section 5.

\section{Estimation Methods}\label{sec:estimation_method}
\subsection{Regression calibration for one misclassified categorical variable}\label{subsec:one_categorical_variable}
Regression calibration (RC) is one of the most widely used measurement error correction methods in application \citep{Keogh2020}. The idea of RC is simple and intuitive, that is, when a true covariate $X$ is unobservable, the RC estimator of the regression coefficient is obtained by replacing the true covariate by its expectation given observed covariate $W$, that is, using $E(X|W)$ to replace $X$ in the model. When a logistic regression model only has measurement error in continuous covariates, \citet{Cao2021} showed that RC estimation is a special case of the approximate maximum likelihood estimation (AMLE). When a logistic model has both error-prone covariates and one two-level misclassified categorical variable, i.e., $A$ is not included in model \eqref{logistic}, \citet{Cao2023} showed that RC estimation can be obtained by replacing $\tilde{\bx}_i$ by $E(\tilde{\bx}_i|\tilde{\bw}_i)=(\bR^V)^{\top} \tilde{\bw}_i$, for $i=1,2,\dots n$, or replaced $\tilde{\bX}$ in model (\ref{logistic}) by $\tilde{\bW}\bR^V$, where 
\begin{eqnarray}\label{matrix_RV}
	\bR^V=\left\{\begin{pmatrix}
		1 & \bmu_x^{\top}-\bmu_w^{\top}\bSigma^{-1}_{ww}\bSigma_{wx} \\
		{\bo} & \bSigma^{-1}_{ww}\bSigma_{wx} \\
	\end{pmatrix}\right\} \otimes \left(\triangle^V\right)^{-1}\triangle^V_e, 
\end{eqnarray} 
with $\otimes$ as Kronecker product and $\bSigma_{wx}=\bSigma^{\top}_{xw}$, 
\begin{eqnarray*}
	\triangle^V=\begin{pmatrix}
		Pr(V_o=0) & 0 \\
		0 & Pr(V_o=1) 
	\end{pmatrix} 
	=\begin{pmatrix}
		(1-\pi^v_0)(1-P^v_t)+\pi^v_1 P^v_t & 0 \\
		0 &\pi^v_0(1-P^v_t)+(1-\pi^v_1) P^v_t \\
	\end{pmatrix},
\end{eqnarray*}
and 
\begin{eqnarray*}
	\triangle^V_e&=&\begin{pmatrix}
		Pr(V_o=0,V=0) & Pr(V_o=0,V=1) \\
		Pr(V_o=1,V=0) & Pr(V_o=1,V=1) 
	\end{pmatrix}
	=\begin{pmatrix}
		(1-\pi^v_0)(1-P^v_t) & \pi^v_1 P^v_t \\
		\pi^v_0(1-P^v_t) &(1-\pi^v_1) P^v_t 
	\end{pmatrix}.
\end{eqnarray*}
Specifically, $\tilde{\bX}_{n\times 2(p+1)}=(\tilde{\bx}_1,\dots,\tilde{\bx}_n)^{\top}$, $\tilde{\bW}_{n\times 2(p+1)}=(\tilde{\bw}_1,\dots,\tilde{\bw}_n)^{\top}$ as realized values for true and observed designed matrices in model (\ref{logistic}), respectively, where $\tilde{\bx}_i=(I(v_i=0),I(v_i=1),x_{i1}I(v_i=0),x_{i1}I(v_i=1),\cdots,x_{ip}I(v_i=0),x_{ip}I(v_i=1))^{\top}$
and $\tilde{\bw}_i=(I(v_{oi}=0),I(v_{oi}=1),w_{i1}I(v_{oi}=0),w_{i1}I(v_{oi}=1),
\cdots,w_{ip}I(v_{oi}=0),w_{ip}I(v_{oi}=1))^{\top}$, for $i=1, \cdots,n$. Furthermore, if we define
$\wp_V=(1-P^v_t,P^v_t)$, $\Lambda_V=\begin{pmatrix}
	1-P^v_t & 0\\0 & P^v_t
\end{pmatrix}$ and $\Gamma_V=\begin{pmatrix}
	1-\pi_o^v & \pi_o^v \\ \pi_1^v & 1-\pi_1^v
\end{pmatrix}$, then it is easy to obtain that the $i$th diagonal element of $\triangle^V$, which is equal to the $i$th element of $\wp_V\times \Gamma_V$ for $i=1,2$, and $\triangle^V_e=\Gamma_V^{\top}\times \Lambda_V$. 

If a logistic model only has error-prone covariates and one three-level categorical variable with misclassification, i.e., model \eqref{logistic} includes $A$ but not $V$. For constructing the corresponding $R^A$ matrix like $R^V$ in \eqref{matrix_RV} and then obtaining RC estimation, we define 
$\wp_A=(P^a_{T,1},P^a_{T,2},P^a_{T,3})$,
\begin{eqnarray*}
	\Lambda_A=\begin{pmatrix}
		P^a_{T,1}&0&0\\
		0 &P^a_{T,2} &0\\
		0 &0 &P^a_{T,3}
	\end{pmatrix} \quad {\rm and} \quad \Gamma_A=\begin{pmatrix}
		1-\phi_{12}-\phi_{13} & \phi_{12} & \phi_{13}\\
		\phi_{21} & 1-\phi_{21}-\phi_{23} & \phi_{23} \\
		\phi_{31} &\phi_{32}  &1-\phi_{31}-\phi_{32}
	\end{pmatrix}.
\end{eqnarray*}
Then $\triangle^A$ is a $3 \times 3$ diagonal matrix, in which the $i$th diagonal element is equal to the $i$th element of $\wp_A\times \Gamma_A$ for $i=1,2,3$, $\triangle^A_e=\Gamma_A^{\top}\times \Lambda_A$, the resulting $R^A$ matrix can be constructed as 
\begin{eqnarray}\label{matrix_RA}
	\bR^A=\left\{\begin{pmatrix}
		1 & \bmu_x^{\top}-\bmu_w^{\top}\bSigma^{-1}_{ww}\bSigma_{wx} \\
		{\bo} & \bSigma^{-1}_{ww}\bSigma_{wx} \\
	\end{pmatrix}\right\} \otimes \left(\triangle^A\right)^{-1}\triangle^A_e. 
\end{eqnarray} 
Therefore, RC estimation can be obtained by replacing $\tilde{\bx}_i$ by $E(\tilde{\bx}_i|\tilde{\bw}_i)=(\bR^A)^{\top} \tilde{\bw}_i$, for $i=1,2,\dots n$, or replace $\tilde{\bX}$ in model (\ref{logistic}) by $\tilde{\bW}\bR^A$. 

\subsection{Regression calibration for more than one misclassified categorical variable}\label{subsec:two_categorical_variable}
In addition to measurement error in covariates, a logistic regression model on two categorical variables with misclassification, like model \eqref{logistic}, can be treated as a model on one misclassified categorical variable with 6 ($=2*3$ for 2 levels of $V$ and 3 levels of $A$) levels. Similar to those definitions in (A2) of Section \ref{subsec:misclassification_measurement_error}, we define true and observed probabilities for each level of the combination $G$ as $P^g_{T,l}$ and $P^g_{obs,l}$ for $l=1,\dots,6$, respectively. Besides, a critical assumption underlying this construction is that misclassification for $V$ and misclassification for $A$ operate independently. That is, the misclassification probability of one variable does not depend on the true or observed status of the other categorical covariate. 
Based on this assumption, we can construct a table $G$ vs $G_o$, like tables of misclassifications of $V$ and $A$, and show that
$\Gamma_G=\Gamma_V \otimes \Gamma_A$.

Specifically, define $\wp_G=(P^g_{T,1},P^g_{T,2},P^g_{T,3},P^g_{T,4},P^g_{T,5},P^g_{T,6})$ and $\Lambda_G$ as a diagonal matrix with $l$th diagonal element $P^g_{T,l}$. Then, based on results in Section \ref{subsec:one_categorical_variable}, we have:
\begin{enumerate}
	\item $\wp_G=\wp_V \otimes \wp_A$ and $\Lambda_G=\Lambda_V\otimes \Lambda_A$.
	\item The $l$th diagonal element of $\triangle^G$ is equal to the  $l$th diagonal element of $\wp_G\times \Gamma_G$ for $l=1,\dots,6$, and all off-diagonal elements of $\triangle^G$ are equal to zero, where $\wp_G\times \Gamma_G=(\wp_V \otimes \wp_A)(\Gamma_V \otimes \Gamma_A)=\wp_V\Gamma_V\otimes \wp_A\Gamma_A=\triangle^V\otimes \triangle^A$.
	\item $\triangle_e^G=\Gamma_G^{\top}\times \Lambda_G=(\Gamma_V \otimes \Gamma_A)^{\top}(\Lambda_V\otimes \Lambda_A)=\Gamma_V^{\top}\Lambda_V\otimes  \Gamma_A^{\top}\Lambda_A= \triangle_e^V \otimes \triangle_e^A$.
\end{enumerate}
The corresponding matrix $\bR^G$ is, then,
\begin{eqnarray}\label{matrix_RG}
	\bR^G=\left\{\begin{pmatrix}
		1 & \bmu_x^{\top}-\bmu_w^{\top}\bSigma^{-1}_{ww}\bSigma_{wx} \\
		{\bo} & \bSigma^{-1}_{ww}\bSigma_{wx} \\
	\end{pmatrix}\right\} \otimes \left(\triangle^G\right)^{-1}\triangle^G_e,
\end{eqnarray} 
where $\left(\triangle^G\right)^{-1}\triangle^G_e=\left(\triangle^V \otimes \triangle^A\right)^{-1}\left( \triangle_e^V \otimes \triangle_e^A\right)=\left(\left(\triangle^V\right)^{-1}\triangle^V_e\right)\otimes \left(\left(\triangle^A\right)^{-1}\triangle^A_e\right)$.
Thus, RC estimation can be obtained by replacing $\tilde{\bx}_i$ by $E(\tilde{\bx}_i|\tilde{\bw}_i)=(\bR^G)^{\top} \tilde{\bw}_i$, for $i=1,2,\dots n$, or replace $\tilde{\bX}$ in model (\ref{logistic}) by $\tilde{\bW}\bR^G$. 

If there are three misclassified categorical variables, e.g., we have another categorical variable $C$ with 2 levels in addition to $V$ and $A$, we then create a new categorical variable $G$ with 12 $(=2*3*2)$ levels and  $\left(\triangle^G\right)^{-1}\triangle^G_e=\left(\left(\triangle^V\right)^{-1}\triangle^V_e\right)\otimes \left(\left(\triangle^A\right)^{-1}\triangle^A_e\right) \otimes \left(\left(\triangle^C\right)^{-1}\triangle^C_e\right)$, where $\triangle^C$ and $\triangle^C_e$ have the same form as  $\triangle^V$ and $\triangle^V_e$ in \eqref{matrix_RV}, respectively.
In principle, our strategy for constructing $R^G$ can be extended to settings with three and more than three misclassified categorical variables.

\section{Simulation Studies}
\subsection{Simulation setup}
In this Section, we conduct simulation studies to examine the finite-sample performance of RC estimation (RCE) described in Section \ref{sec:estimation_method}
as well as naive estimation (Naive), i.e., ignoring measurement error and misclassification.  The proposed method  RCE is developed for settings involving one three-level categorical variable and one two-level categorical variable. For simplicity, the simulation study considers two binary (two-level) categorical variables. Nonetheless, the method is fully applicable to pairs of two-level categorical variables as well as pairs of three-level categorical variables, and this simulation design does not alter the core nature of the proposed approach. Therefore, we consider the following true logistic regression, that is, 
\begin{eqnarray}\label{simu_model}
	{\rm log}\bigg({P\over 1-P}\bigg)&=&
	\begin{cases}
		\beta_{01}+\beta_{11}X_1+\beta_{21}X_2, \quad G=1,\\
		\beta_{02}+\beta_{12}X_1+\beta_{22}X_2, \quad G=2, \\
		\beta_{03}+\beta_{13}X_1+\beta_{23}X_2, \quad G=3,\\
		\beta_{04}+\beta_{14}X_1+\beta_{24}X_2, \quad G=4,
	\end{cases} 
\end{eqnarray}
where $P=Pr(Y=1\mid X_1,X_2,V,A)$, $\bX=(X_1,X_2)^{\top}$ are true covariates, $G=2*i+j+1$ when $V=i$ and $A=j$ for $i,j=0,1$. Both $V$ and $A$ are generated from Bernoulli distributions with probability $P^v_{t}$ and $P^a_{t}$ for $V=1$ and $A=1$, respectively. For the additive error model, we assume observed covariates are related to true covariates as $W_j=0.4X_j+U_j$ for $j=1,2$, which is similar to the relationship between true observed nutritions (e.g., protein and fat) by using data from male population in EPIC-InterAct Study \citep{Cao2022}.
Also, we let ${\bX}=(X_1,X_2)^{\top}$ and ${\bU}=(U_1,U_2)^{\top}$
be generated from multivariate normal distribution $N(0,\bSigma_{xx})$ and $N(0,\bSigma_{uu})$, respectively, where
\begin{eqnarray*}
	\bSigma_{xx}=\begin{pmatrix}
		1 & 0 \\
		0 & 1 \\
	\end{pmatrix}, \quad 
	\bSigma_{uu}=\begin{pmatrix}
		0.8 & \rho_{uu}\sqrt{0.8*0.85} \\
		\rho_{uu}\sqrt{0.8*0.85} & 0.85 \\
	\end{pmatrix}, 
\end{eqnarray*}
and $\rho_{uu}=0.75$. 
Values in $\bSigma_{xx}$ and $\bSigma_{uu}$ are the same as those used in simulation setup of \citet{Cao2023}, which means correlation between two additive errors is relatively high (0.75), and ratios of estimated error variance to estimated true variance for $X_1$ and $X_2$ roughly equal to 80\% and 85\%, respectively.

Model \eqref{simu_model} has two categorical variables, both of them are assumed to be misclassified.
For the first categorical variable $V$, we consider the same three scenarios of the misclassification model as \citet{Cao2023}. That is, (1) $P^v_t=0.15$ ($P^v_o=0.22$) and $\pi^v_0=\pi^v_1=0.1$, which represents a relatively weak misclassification effect; (2) $P^v_t=0.13$ ($P^v_o=0.22$), $\pi^v_0=0.14$ and $\pi^v_1=0.23$, which values are similar to estimated values of $P^v_t$, $\pi^v_0$ and $\pi^v_1$ for male based on the estimation strategy described in Section 5.3 of \citet{Cao2023}; (3) $P^v_t=0.4$ ($P^v_o=0.44$) and $\pi^v_0=\pi^v_1=0.2$, 
which has larger $P^v_t$ for evaluating the effect of the proportion of $V_o=1$ on the performance of RCE. For misclassified category variable $A$, we also consider three scenarios of misclassification rates: (1) $P^a_{t}=0.5$,  $\pi^a_0=\pi^a_1=0$; (2) $P^a_{t}=0.5$,  $\pi^a_0=\pi^a_1=0.1$; (3) $P^a_{t}=0.5$,  $\pi^a_0=\pi^a_1=0.2$. Therefore, we totally have 9 cases of $P^v_t$, $\pi^v_0$, $\pi^v_1$, $P^a_{t}$, $\pi^a_0$ and $\pi^a_1$, that is.
\begin{enumerate}
	\item[(1)]  $P^v_t=0.15$, $\pi^v_0=\pi^v_1=0.1$ and $P^a_{t}=0.5$,  $\pi^a_0=\pi^a_1=0$;
	\item[(2)]  $P^v_t=0.15$, $\pi^v_0=\pi^v_1=0.1$ and $P^a_{t}=0.5$,  $\pi^a_0=\pi^a_1=0.1$;
	\item[(3)]  $P^v_t=0.15$, $\pi^v_0=\pi^v_1=0.1$ and $P^a_{t}=0.5$,  $\pi^a_0=\pi^a_1=0.2$;
	\item[(4)]  $P^v_t=0.13$, $\pi^v_0=0.14$, $\pi^v_1=0.23$ and $P^a_{t}=0.5$,  $\pi^a_0=\pi^a_1=0$;
	\item[(5)]  $P^v_t=0.13$, $\pi^v_0=0.14$, $\pi^v_1=0.23$ and $P^a_{t}=0.5$,  $\pi^a_0=\pi^a_1=0.1$;
	\item[(6)]  $P^v_t=0.13$, $\pi^v_0=0.14$, $\pi^v_1=0.23$ and $P^a_{t}=0.5$,  $\pi^a_0=\pi^a_1=0.2$;
	\item[(7)]  $P^v_t=0.4$, $\pi^v_0=\pi^v_1=0.2$ and $P^a_{t}=0.5$,  $\pi^a_0=\pi^a_1=0$;
	\item[(8)]  $P^v_t=0.4$, $\pi^v_0=\pi^v_1=0.2$ and $P^a_{t}=0.5$,  $\pi^a_0=\pi^a_1=0.1$;
	\item[(9)]  $P^v_t=0.4$, $\pi^v_0=\pi^v_1=0.2$ and $P^a_{t}=0.5$,  $\pi^a_0=\pi^a_1=0.2$.
\end{enumerate}

Furthermore, since estimated regression coefficients of
nutrients from EPIC-InterAct study are usually small \citep{Agogo2014,Harding2008},
we then primarily consider two choices of
$\btheta=(\beta_{01},\beta_{11},\beta_{21},\beta_{02},\beta_{12},\beta_{22},\beta_{03},\beta_{13},\beta_{23},\beta_{04},\beta_{14},\beta_{24})^{\top}$, that is,
$\btheta=(0, 0.3, 0.3, -0.1, 0.25, 0.25, 0.1, 0.24, 0.24, 0, 0.2, 0.2)^{\top}$ and $\btheta=(0, 0.25, 0.25, -0.1, 0.2, 0.2, 0.1$,\\$0.3, 0.3, 0, 0.24, 0.24)^{\top}$. The former means both effects of $V$ and $A$ conditional on $X_1$ and $X_2$ make ${\rm logit}(P)$ smaller, while the later indicates the effect of $V$ conditional on $X_1$ and $X_2$ makes ${\rm logit}(P)$ larger but the effect of $A$ conditional on $X_1$ and $X_2$ makes ${\rm logit}(P)$ smaller. In addition, we consider larger regression coefficient,
say, $\btheta=(0, 0.5, 0.5, -0.1, 0.42, 0.42, 0.1$,\\$0.4, 0.4, 0, 0.33, 0.33)^{\top}$,  to evaluate the performance of different estimation methods. These simulation configurations are motivated by effect sizes, misclassification rates and error variances observed in the EPIC‑InterAct real-data application. Besides, when applying our proposed method RCE in simulations, we directly use the true values of parameters in measurement error and misclassification models.

For each parameter configuration, we generate 2000 simulations
with sample sizes of 2000 and 6000, respectively. The latter sample size is close to the number of males in the main study of the EPIC-InterACT Study after cleaning \citep{Cao2022}. 
For assessing performances of Naive and RCE, we also include the true estimator (TRUE), which is obtained based on true values of $\bX$, $V$ and $A$, as benchmark. 
We report relative bias percentage $\text{RB}(\%) = 100 \times (\bar{\hat{\beta}}-\beta_{\text{true}})/\beta_{\text{true}})$ (where \(\bar{\hat{\beta}}\) is the Monte‑Carlo mean of parameter estimates across replicates), Monte-Carlo standard 
deviation (SD),  Monte-Carlo standard error (SE) (i.e., average of standard error reported from standard software for logistic regression over 2000 replications), and the empirical coverage proportion of nominal 95\% confidence intervals (CP). Simulation results for interested regression coefficients $\beta_{11}$, $\beta_{21}$, $\beta_{12}$, $\beta_{22}$, $\beta_{13}$, $\beta_{23}$, $\beta_{14}$ and $\beta_{24}$  under $\btheta=(0, 0.3, 0.3, -0.1, 0.25, 0.25, 0.1, 0.24, 0.24, 0, 0.2, 0.2)^{\top}$, different sample sizes and  misclassification rate cases are summarized in Tables 1-4. Results for the remaining two $\btheta$ scenarios are given in Tables S1–S8 of the Web Appendix. 

\subsection{Simulation Results}
\begin{table}[h]
	\caption{\footnotesize Simulation results for regression coefficients of $X_1$ and $X_2$ at $G=1$ and $G=2$, i.e., $\beta_{11}$, $\beta_{21}$, $\beta_{12}$ and $\beta_{22}$ when $\btheta=(0, 0.3, 0.3, -0.1, 0.25, 0.25, 0.1, 0.24, 0.24, 0, 0.2, 0.2)^{\top}$ under $n=2000$ and different cases. RB (\%) = relative bias (percent), SD = Monte-Carlo standard 
		deviation, SE = Monte-Carlo standard error, CP = empirical 95\% coverage probability.}
	\label{simu:tab1}
	\begin{center}
		\begin{adjustbox}{width=1\textwidth}
			\begin{tabular}{llrrrrrrrrrrrrrrrrrrr}
				\hline
				&& \multicolumn{9}{c}{$G=1$} & &\multicolumn{9}{c}{$G=2$} \\
				\cline{3-11}\cline{13-21}
				& &\multicolumn{4}{c}{$\hat{\beta}_{11}$}& 
				&\multicolumn{4}{c}{ $\hat{\beta}_{21}$} & &\multicolumn{4}{c}{ $\hat{\beta}_{12}$}
				& &\multicolumn{4}{c}{ $\hat{\beta}_{22}$}
				\\
				\cline{3-6}\cline{8-11}\cline{13-16}\cline{18-21} Case & Method & RB(\%) & SD & SE & CP & &RB(\%) & SD & SE & CP & & RB(\%) & SD& SE & CP & &RB(\%) & SD & SE & CP\\
				\hline
				1   & Naive & -74.256 & 0.092 & 0.095 & 0.346 && -77.079 & 0.095 & 0.092 & 0.298 && -74.406 & 0.094 & 0.095 & 0.489 && -77.320 & 0.094 & 0.092 & 0.450 \\ 
				& TRUE &-0.181 & 0.070 & 0.072 & 0.963 && 0.782 & 0.071 & 0.072 & 0.951 && 0.969 & 0.071 & 0.071 & 0.953 && 0.839 & 0.072 & 0.071 & 0.945 \\
				& RCE & -2.464 & 0.177 & 0.184 & 0.957 && -1.995 & 0.192 & 0.188 & 0.942 && -3.209 & 0.181 & 0.183 & 0.961 && -2.828 & 0.190 & 0.188 & 0.947 \\ 
				\\
				2   & Naive & -74.163 & 0.092 & 0.095 & 0.347 && -77.550 & 0.088 & 0.092 & 0.282 && -73.418 & 0.095 & 0.095 & 0.511 && -77.017 & 0.095 & 0.092 & 0.450 \\ 
				& TRUE & 1.391 & 0.072 & 0.072 & 0.951 && 0.623 & 0.073 & 0.072 & 0.948 && -0.219 & 0.071 & 0.071 & 0.947 && -0.070 & 0.074 & 0.071 & 0.934 \\
				& RCE & -1.162 & 0.207 & 0.208 & 0.955 && -1.290 & 0.210 & 0.213 & 0.955 && -2.380 & 0.210 & 0.207 & 0.953 && -2.582 & 0.221 & 0.213 & 0.943 \\ 
				\\
				3 & Naive &  -74.567 & 0.092 & 0.095 & 0.329 && -77.959 & 0.087 & 0.092 & 0.270 && -72.931 & 0.096 & 0.095 & 0.515 && -76.508 & 0.096 & 0.092 & 0.461 \\ 
				& TRUE & 1.391 & 0.072 & 0.072 & 0.951 && 0.623 & 0.073 & 0.072 & 0.948 && -0.219 & 0.071 & 0.071 & 0.947 && -0.070 & 0.074 & 0.071 & 0.934 \\ 
				& RCE & -0.802 & 0.255 & 0.252 & 0.943 && -1.025 & 0.255 & 0.259 & 0.953 && -2.769 & 0.262 & 0.252 & 0.939 && -2.830 & 0.272 & 0.258 & 0.940 \\ 
				\\
				4 & Naive &  -74.473 & 0.095 & 0.095 & 0.356 && -77.636 & 0.092 & 0.092 & 0.279 && -74.879 & 0.094 & 0.095 & 0.498 && -76.750 & 0.092 & 0.093 & 0.454 \\ 
				& TRUE & 0.514 & 0.071 & 0.071 & 0.948 && 1.082 & 0.071 & 0.071 & 0.950 && 0.581 & 0.071 & 0.070 & 0.951 && 0.200 & 0.069 & 0.070 & 0.959 \\
				& RCE & -3.672 & 0.205 & 0.197 & 0.939 && -3.455 & 0.207 & 0.202 & 0.944 && -3.903 & 0.197 & 0.197 & 0.951 && -2.531 & 0.203 & 0.202 & 0.954 \\ 
				\\
				5 & Naive & -74.555 & 0.098 & 0.095 & 0.353 && -78.197 & 0.093 & 0.093 & 0.276 && -73.604 & 0.098 & 0.095 & 0.506 && -76.009 & 0.093 & 0.092 & 0.460 \\
				& TRUE & 1.108 & 0.071 & 0.071 & 0.960 && 0.248 & 0.072 & 0.071 & 0.954 && -0.311 & 0.072 & 0.070 & 0.946 && 0.287 & 0.070 & 0.070 & 0.955 \\ 
				& RCE &-3.326 & 0.229 & 0.223 & 0.941 && -3.963 & 0.231 & 0.229 & 0.946 && -1.013 & 0.227 & 0.222 & 0.950 && 0.196 & 0.228 & 0.228 & 0.949 \\
				\\
				6 & Naive &-75.594 & 0.098 & 0.095 & 0.339 && -78.198 & 0.093 & 0.093 & 0.277 && -72.434 & 0.098 & 0.095 & 0.527 && -76.008 & 0.092 & 0.092 & 0.460 \\ 
				& TRUE &1.108 & 0.071 & 0.071 & 0.960 && 0.248 & 0.072 & 0.071 & 0.954 && -0.311 & 0.072 & 0.070 & 0.946 && 0.287 & 0.070 & 0.070 & 0.955 \\
				& RCE & -5.347 & 0.279 & 0.271 & 0.944 && -4.671 & 0.279 & 0.277 & 0.952 && 1.156 & 0.278 & 0.270 & 0.946 && 0.884 & 0.275 & 0.277 & 0.955 \\ 
				\\
				7 & Naive &-75.301 & 0.115 & 0.112 & 0.468 && -77.537 & 0.109 & 0.109 & 0.436 && -73.746 & 0.113 & 0.112 & 0.627 && -78.412 & 0.111 & 0.109 & 0.559 \\ 
				& TRUE & 1.142 & 0.084 & 0.086 & 0.949 && 0.644 & 0.086 & 0.086 & 0.953 && 0.144 & 0.087 & 0.085 & 0.944 && 1.076 & 0.084 & 0.085 & 0.956 \\ 
				& RCE & -3.798 & 0.266 & 0.268 & 0.960 && -2.523 & 0.268 & 0.274 & 0.951 && -0.976 & 0.270 & 0.267 & 0.948 && -2.668 & 0.278 & 0.274 & 0.947 \\ 
				\\
				8 & Naive & -75.205 & 0.111 & 0.112 & 0.475 && -77.812 & 0.110 & 0.109 & 0.426 && -74.021 & 0.114 & 0.112 & 0.622 && -77.805 & 0.110 & 0.109 & 0.550 \\ 
				& TRUE &  1.787 & 0.087 & 0.086 & 0.952 && 0.997 & 0.087 & 0.086 & 0.954 && 0.833 & 0.086 & 0.085 & 0.951 && 1.525 & 0.086 & 0.085 & 0.949 \\
				& RCE & -2.613 & 0.305 & 0.303 & 0.946 && -1.417 & 0.317 & 0.311 & 0.952 && -2.173 & 0.307 & 0.302 & 0.949 && -3.410 & 0.316 & 0.310 & 0.954 \\
				\\
				9 & Naive & -76.005 & 0.111 & 0.112 & 0.472 && -78.084 & 0.110 & 0.109 & 0.420 && -73.151 & 0.114 & 0.112 & 0.616 && -77.500 & 0.111 & 0.109 & 0.571 \\ 
				& TRUE &  1.787 & 0.087 & 0.086 & 0.952 && 0.997 & 0.087 & 0.086 & 0.954 && 0.833 & 0.086 & 0.085 & 0.951 && 1.525 & 0.086 & 0.085 & 0.949 \\ 
				& RCE & -4.181 & 0.367 & 0.368 & 0.951 && -1.521 & 0.383 & 0.377 & 0.950 && -0.636 & 0.371 & 0.367 & 0.953 && -3.488 & 0.386 & 0.377 & 0.943 \\ 
				\hline 
			\end{tabular}
		\end{adjustbox}
	\end{center}
\end{table}

\begin{table}[h]
	\caption{\footnotesize Simulation results for regression coefficients of $X_1$ and $X_2$ at $G=3$ and $G=4$, i.e., $\beta_{13}$, $\beta_{23}$, $\beta_{14}$ and $\beta_{24}$ when $\btheta=(0, 0.3, 0.3, -0.1, 0.25, 0.25, 0.1, 0.24, 0.24, 0, 0.2, 0.2)^{\top}$ under $n=2000$ and different cases. RB (\%) = relative bias (percent), SD = Monte-Carlo standard 
		deviation, SE = Monte-Carlo standard error, CP = empirical 95\% coverage probability.}
	\label{simu:tab2}
	\begin{center}
		\begin{adjustbox}{width=1\textwidth}
			\begin{tabular}{llrrrrrrrrrrrrrrrrrrr}
				\hline
				&& \multicolumn{9}{c}{$G=3$} & &\multicolumn{9}{c}{$G=4$} \\
				\cline{3-11}\cline{13-21}
				& &\multicolumn{4}{c}{$\hat{\beta}_{13}$}& 
				&\multicolumn{4}{c}{ $\hat{\beta}_{23}$} & &\multicolumn{4}{c}{ $\hat{\beta}_{14}$}
				& &\multicolumn{4}{c}{ $\hat{\beta}_{24}$}
				\\
				\cline{3-6}\cline{8-11}\cline{13-16}\cline{18-21} Case & Method & RB(\%) & SD & SE & CP & &RB(\%) & SD & SE & CP & & RB(\%) & SD& SE & CP & &RB(\%) & SD & SE & CP\\
				\hline
				1  & Naive & -70.796 & 0.186 & 0.181 & 0.834 & & -74.493 & 0.180 & 0.177 & 0.826 & & -70.532 & 0.181 & 0.181 & 0.881 & & -75.206 & 0.175 & 0.176 & 0.863 \\
				& TRUE & 4.697 & 0.180 & 0.173 & 0.948 & & 4.729 & 0.178 & 0.173 & 0.952 & & 6.364 & 0.180 & 0.172 & 0.951 & & 3.067 & 0.174 & 0.172 & 0.954 \\
				& RCE & 1.724 & 0.593 & 0.571 & 0.941 & & 1.408 & 0.607 & 0.586 & 0.951 & & 1.547 & 0.582 & 0.571 & 0.954 & & -0.205 & 0.589 & 0.585 & 0.954 \\
				\\
				2  & Naive & -72.262 & 0.188 & 0.181 & 0.832 & & -74.092 & 0.176 & 0.177 & 0.828 & & -70.720 & 0.184 & 0.181 & 0.877 & & -74.816 & 0.183 & 0.176 & 0.858 \\
				& TRUE & 1.706 & 0.176 & 0.173 & 0.949 & & 2.299 & 0.176 & 0.173 & 0.953 & & 4.858 & 0.181 & 0.172 & 0.945 & & 5.054 & 0.178 & 0.172 & 0.942 \\
				& RCE & -0.839 & 0.672 & 0.646 & 0.944 & & 2.393 & 0.667 & 0.663 & 0.957 & & -2.676 & 0.638 & 0.646 & 0.960 & & -3.567 & 0.669 & 0.662 & 0.954 \\
				\\
				3  & Naive & -73.563 & 0.189 & 0.181 & 0.828 & & -75.259 & 0.177 & 0.177 & 0.819 & & -69.017 & 0.182 & 0.181 & 0.889 & & -73.764 & 0.181 & 0.176 & 0.866 \\
				& TRUE & 1.706 & 0.176 & 0.173 & 0.949 & & 2.299 & 0.176 & 0.173 & 0.953 & & 4.858 & 0.181 & 0.172 & 0.945 & & 5.054 & 0.178 & 0.172 & 0.942 \\
				& RCE & -9.116 & 0.813 & 0.784 & 0.948 & & -5.016 & 0.810 & 0.804 & 0.950 & & 6.895 & 0.769 & 0.784 & 0.956 & & 4.189 & 0.812 & 0.804 & 0.947 \\
				\\
				4  & Naive & -69.148 & 0.180 & 0.180 & 0.845 & & -74.628 & 0.176 & 0.176 & 0.830 & & -67.727 & 0.186 & 0.181 & 0.878 & & -71.044 & 0.176 & 0.176 & 0.877 \\
				& TRUE & 5.135 & 0.195 & 0.187 & 0.947 & & 0.568 & 0.193 & 0.187 & 0.946 & & 2.263 & 0.188 & 0.185 & 0.948 & & 11.814 & 0.189 & 0.186 & 0.959 \\
				& RCE & 4.588 & 0.828 & 0.820 & 0.952 & & 0.917 & 0.841 & 0.841 & 0.954 & & 24.787 & 0.852 & 0.823 & 0.952 & & 24.447 & 0.853 & 0.842 & 0.957 \\
				\\
				5  & Naive & -70.392 & 0.185 & 0.180 & 0.838 & & -73.456 & 0.182 & 0.176 & 0.808 & & -67.504 & 0.183 & 0.181 & 0.878 & & -71.352 & 0.178 & 0.176 & 0.875 \\
				& TRUE & 3.551 & 0.188 & 0.186 & 0.954 & & 6.421 & 0.194 & 0.187 & 0.948 & & 4.669 & 0.188 & 0.185 & 0.956 & & 4.252 & 0.196 & 0.186 & 0.949 \\
				& RCE & 4.287 & 0.951 & 0.929 & 0.948 & & 6.937 & 0.981 & 0.952 & 0.948 & & 17.133 & 0.960 & 0.931 & 0.945 & & 15.799 & 0.985 & 0.953 & 0.946 \\
				\\
				6  & Naive & -70.601 & 0.185 & 0.180 & 0.836 & & -73.620 & 0.181 & 0.176 & 0.817 & & -66.945 & 0.181 & 0.181 & 0.887 & & -71.197 & 0.177 & 0.176 & 0.876 \\
				& TRUE & 3.551 & 0.188 & 0.186 & 0.954 & & 6.421 & 0.194 & 0.187 & 0.948 & & 4.669 & 0.188 & 0.185 & 0.956 & & 4.252 & 0.196 & 0.186 & 0.949 \\
				& RCE & 8.867 & 1.149 & 1.128 & 0.943 & & 9.984 & 1.188 & 1.156 & 0.944 & & 13.528 & 1.160 & 1.130 & 0.944 & & 13.214 & 1.196 & 1.158 & 0.942 \\
				\\
				7  & Naive & -71.304 & 0.131 & 0.127 & 0.714 & & -75.792 & 0.130 & 0.124 & 0.676 & & -71.652 & 0.129 & 0.127 & 0.786 & & -74.790 & 0.125 & 0.123 & 0.763 \\
				& TRUE & 1.100 & 0.104 & 0.104 & 0.953 & & 1.856 & 0.105 & 0.104 & 0.950 & & 2.345 & 0.103 & 0.103 & 0.947 & & 1.064 & 0.102 & 0.103 & 0.960 \\
				& RCE & 1.065 & 0.364 & 0.361 & 0.953 & & -0.641 & 0.381 & 0.371 & 0.945 & & 0.729 & 0.368 & 0.361 & 0.947 & & 2.171 & 0.374 & 0.370 & 0.953 \\
				\\
				8  & Naive & -70.796 & 0.130 & 0.127 & 0.717 & & -76.049 & 0.127 & 0.124 & 0.673 & & -72.811 & 0.125 & 0.126 & 0.788 & & -74.185 & 0.126 & 0.123 & 0.773 \\
				& TRUE & 2.392 & 0.104 & 0.104 & 0.955 & & 1.151 & 0.106 & 0.104 & 0.949 & & 0.826 & 0.103 & 0.103 & 0.956 & & 1.734 & 0.104 & 0.103 & 0.949 \\
				& RCE & 4.990 & 0.422 & 0.409 & 0.940 & & 1.829 & 0.432 & 0.420 & 0.950 & & -5.269 & 0.404 & 0.408 & 0.954 & & -1.204 & 0.433 & 0.419 & 0.939 \\
				\\
				9  & Naive & -70.897 & 0.130 & 0.127 & 0.721 & & -76.978 & 0.127 & 0.124 & 0.673 & & -72.486 & 0.127 & 0.126 & 0.791 & & -73.275 & 0.127 & 0.123 & 0.773 \\
				& TRUE & 2.392 & 0.104 & 0.104 & 0.955 & & 1.151 & 0.106 & 0.104 & 0.949 & & 0.826 & 0.103 & 0.103 & 0.956 & & 1.734 & 0.104 & 0.103 & 0.949 \\
				& RCE & 7.390 & 0.511 & 0.497 & 0.938 & & 0.881 & 0.529 & 0.510 & 0.940 & & -7.750 & 0.491 & 0.496 & 0.958 & & -0.250 & 0.526 & 0.508 & 0.945 \\
				\hline 
			\end{tabular}
		\end{adjustbox}
	\end{center}
\end{table}

\begin{table}[h]
	\caption{\footnotesize Simulation results for regression coefficients of $X_1$ and $X_2$ at $G=1$ and $G=2$, i.e., $\beta_{11}$, $\beta_{21}$, $\beta_{12}$ and $\beta_{22}$ when $\btheta=(0, 0.3, 0.3, -0.1, 0.25, 0.25, 0.1, 0.24, 0.24, 0, 0.2, 0.2)^{\top}$ under $n=6000$ and different cases. RB (\%) = relative bias (percent), SD = Monte-Carlo standard 
		deviation, SE = Monte-Carlo standard error,  CP = empirical 95\% coverage probability.}
	\label{simu:tab3}
	\begin{center}
		\begin{adjustbox}{width=1\textwidth}
			\begin{tabular}{llrrrrrrrrrrrrrrrrrrr}
				\hline
				&& \multicolumn{9}{c}{$G=1$} & &\multicolumn{9}{c}{$G=2$} \\
				\cline{3-11}\cline{13-21}
				& &\multicolumn{4}{c}{$\hat{\beta}_{11}$}& 
				&\multicolumn{4}{c}{ $\hat{\beta}_{21}$} & &\multicolumn{4}{c}{ $\hat{\beta}_{12}$}
				& &\multicolumn{4}{c}{ $\hat{\beta}_{22}$}
				\\
				\cline{3-6}\cline{8-11}\cline{13-16}\cline{18-21} Case & Method & RB(\%) & SD & SE & CP & &RB(\%) & SD & SE & CP & & RB(\%) & SD& SE & CP & &RB(\%) & SD & SE & CP\\
				\hline
				1  & Naive & -75.067 & 0.055 & 0.054 & 0.018 & & -77.746 & 0.053 & 0.053 & 0.005 & & -75.128 & 0.053 & 0.055 & 0.072 & & -76.251 & 0.053 & 0.053 & 0.056 \\
				& TRUE & -0.199 & 0.042 & 0.041 & 0.946 & & 0.081 & 0.044 & 0.041 & 0.939 & & -0.404 & 0.041 & 0.041 & 0.952 & & 0.352 & 0.042 & 0.041 & 0.950 \\
				& RCE & -5.456 & 0.103 & 0.105 & 0.955 & & -4.977 & 0.104 & 0.108 & 0.957 & & -3.320 & 0.102 & 0.105 & 0.958 & & -1.171 & 0.107 & 0.108 & 0.955 \\
				\\
				2  & Naive & -74.740 & 0.054 & 0.054 & 0.016 & & -78.004 & 0.053 & 0.053 & 0.006 & & -73.475 & 0.053 & 0.054 & 0.075 & & -77.402 & 0.053 & 0.053 & 0.054 \\
				& TRUE & 0.583 & 0.040 & 0.041 & 0.956 & & 0.397 & 0.041 & 0.041 & 0.950 & & 0.211 & 0.041 & 0.041 & 0.953 & & 0.877 & 0.041 & 0.041 & 0.951 \\
				& RCE & -3.491 & 0.118 & 0.119 & 0.951 & & -3.468 & 0.122 & 0.122 & 0.954 & & -2.955 & 0.117 & 0.119 & 0.954 & & -3.644 & 0.124 & 0.122 & 0.947 \\
				\\
				3  & Naive & -75.145 & 0.054 & 0.054 & 0.012 & & -78.292 & 0.053 & 0.053 & 0.006 & & -73.024 & 0.054 & 0.054 & 0.082 & & -77.049 & 0.054 & 0.053 & 0.046 \\
				& TRUE & 0.583 & 0.040 & 0.041 & 0.956 & & 0.397 & 0.041 & 0.041 & 0.950 & & 0.211 & 0.041 & 0.041 & 0.953 & & 0.877 & 0.041 & 0.041 & 0.951 \\
				& RCE & -3.200 & 0.145 & 0.145 & 0.951 & & -3.023 & 0.148 & 0.149 & 0.955 & & -3.381 & 0.145 & 0.145 & 0.943 & & -4.219 & 0.152 & 0.149 & 0.949 \\
				\\
				4  & Naive & -74.289 & 0.054 & 0.055 & 0.015 & & -78.436 & 0.052 & 0.053 & 0.002 & & -73.655 & 0.056 & 0.055 & 0.090 & & -77.428 & 0.054 & 0.053 & 0.050 \\
				& TRUE & 0.424 & 0.041 & 0.041 & 0.948 & & -0.105 & 0.040 & 0.041 & 0.950 & & 0.670 & 0.040 & 0.040 & 0.953 & & -0.089 & 0.041 & 0.040 & 0.953 \\
				& RCE & -4.341 & 0.113 & 0.113 & 0.951 & & -5.409 & 0.114 & 0.116 & 0.950 & & -1.003 & 0.114 & 0.113 & 0.950 & & -1.438 & 0.117 & 0.116 & 0.955 \\
				\\
				5  & Naive & -74.289 & 0.054 & 0.055 & 0.015 & & -78.436 & 0.052 & 0.053 & 0.002 & & -73.655 & 0.056 & 0.055 & 0.090 & & -77.428 & 0.054 & 0.053 & 0.050 \\
				& TRUE & 0.424 & 0.041 & 0.041 & 0.948 & & -0.105 & 0.040 & 0.041 & 0.950 & & 0.670 & 0.040 & 0.040 & 0.953 & & -0.089 & 0.041 & 0.040 & 0.953 \\
				& RCE & -4.341 & 0.113 & 0.113 & 0.951 & & -5.409 & 0.114 & 0.116 & 0.950 & & -1.003 & 0.114 & 0.113 & 0.950 & & -1.438 & 0.117 & 0.116 & 0.955 \\
				\\
				6  & Naive & -75.492 & 0.056 & 0.055 & 0.016 & & -78.271 & 0.053 & 0.053 & 0.009 & & -73.104 & 0.055 & 0.055 & 0.093 & & -77.280 & 0.053 & 0.053 & 0.051 \\
				& TRUE & 0.481 & 0.041 & 0.041 & 0.946 & & 0.330 & 0.041 & 0.041 & 0.947 & & 0.111 & 0.042 & 0.040 & 0.944 & & 0.410 & 0.041 & 0.040 & 0.943 \\
				& RCE & -3.879 & 0.158 & 0.155 & 0.949 & & -3.367 & 0.159 & 0.159 & 0.953 & & -3.299 & 0.155 & 0.155 & 0.951 & & -4.420 & 0.157 & 0.159 & 0.954 \\
				\\
				7  & Naive & -75.427 & 0.066 & 0.064 & 0.066 & & -77.693 & 0.063 & 0.063 & 0.040 & & -75.217 & 0.064 & 0.064 & 0.170 & & -78.124 & 0.063 & 0.063 & 0.134 \\
				& TRUE & -0.202 & 0.048 & 0.049 & 0.956 & & 0.398 & 0.048 & 0.049 & 0.951 & & -0.116 & 0.049 & 0.049 & 0.945 & & -0.046 & 0.050 & 0.049 & 0.946 \\
				& RCE & -4.035 & 0.157 & 0.154 & 0.945 & & -2.914 & 0.158 & 0.157 & 0.951 & & -4.981 & 0.153 & 0.153 & 0.949 & & -4.570 & 0.158 & 0.157 & 0.943 \\
				\\
				8  & Naive & -76.023 & 0.065 & 0.064 & 0.057 & & -78.529 & 0.062 & 0.063 & 0.034 & & -74.236 & 0.064 & 0.064 & 0.184 & & -77.408 & 0.062 & 0.063 & 0.127 \\
				& TRUE & 0.552 & 0.048 & 0.049 & 0.958 & & 0.681 & 0.049 & 0.049 & 0.953 & & -0.147 & 0.048 & 0.049 & 0.950 & & 0.214 & 0.049 & 0.049 & 0.945 \\
				& RCE & -5.937 & 0.174 & 0.174 & 0.949 & & -4.612 & 0.175 & 0.178 & 0.954 & & -2.272 & 0.176 & 0.174 & 0.945 & & -2.072 & 0.177 & 0.178 & 0.946 \\
				\\
				9  & Naive & -76.496 & 0.066 & 0.064 & 0.054 & & -78.725 & 0.063 & 0.063 & 0.042 & & -73.674 & 0.065 & 0.064 & 0.180 & & -77.151 & 0.063 & 0.063 & 0.136 \\
				& TRUE & 0.552 & 0.048 & 0.049 & 0.958 & & 0.681 & 0.049 & 0.049 & 0.953 & & -0.147 & 0.048 & 0.049 & 0.950 & & 0.214 & 0.049 & 0.049 & 0.945 \\
				& RCE & -6.399 & 0.212 & 0.211 & 0.948 & & -4.569 & 0.213 & 0.216 & 0.957 & & -1.705 & 0.213 & 0.211 & 0.945 & & -2.094 & 0.215 & 0.216 & 0.952 \\
				\hline 
			\end{tabular}
		\end{adjustbox}
	\end{center}	
\end{table}

\begin{table}[h]
	\caption{\footnotesize Simulation results for regression coefficients of $X_1$ and $X_2$ at $G=3$ and $G=4$, i.e., $\beta_{13}$, $\beta_{23}$, $\beta_{14}$ and $\beta_{24}$ when $\btheta=(0, 0.3, 0.3, -0.1, 0.25, 0.25, 0.1, 0.24, 0.24, 0, 0.2, 0.2)^{\top}$ under $n=6000$ and different cases. RB (\%) = relative bias (percent), SD = Monte-Carlo standard 
		deviation, SE = Monte-Carlo standard error, CP = empirical 95\% coverage probability.}
	\label{simu:tab4}
	\begin{center}
		\begin{adjustbox}{width=1\textwidth}
			\begin{tabular}{llrrrrrrrrrrrrrrrrrrr}
				\hline
				&& \multicolumn{9}{c}{$G=3$} & &\multicolumn{9}{c}{$G=4$} \\
				\cline{3-11}\cline{13-21}
				& &\multicolumn{4}{c}{$\hat{\beta}_{13}$}& 
				&\multicolumn{4}{c}{ $\hat{\beta}_{23}$} & &\multicolumn{4}{c}{ $\hat{\beta}_{14}$}
				& &\multicolumn{4}{c}{ $\hat{\beta}_{24}$}
				\\
				\cline{3-6}\cline{8-11}\cline{13-16}\cline{18-21} Case & Method & RB(\%) & SD & SE & CP & &RB(\%) & SD & SE & CP & & RB(\%) & SD& SE & CP & &RB(\%) & SD & SE & CP\\
				\hline
				1  & Naive & -72.146 & 0.101 & 0.103 & 0.614 & & -74.457 & 0.100 & 0.100 & 0.573 & & -69.386 & 0.107 & 0.103 & 0.715 & & -76.223 & 0.103 & 0.100 & 0.664 \\
				& TRUE & 1.521 & 0.099 & 0.098 & 0.952 & & 2.808 & 0.097 & 0.098 & 0.958 & & 2.338 & 0.100 & 0.097 & 0.951 & & 2.518 & 0.099 & 0.097 & 0.957 \\
				& RCE & -1.110 & 0.327 & 0.325 & 0.946 & & 0.500 & 0.339 & 0.334 & 0.952 & & 3.554 & 0.327 & 0.324 & 0.952 & & -2.805 & 0.329 & 0.333 & 0.956 \\
				\\
				2  & Naive & -71.619 & 0.104 & 0.103 & 0.614 & & -75.311 & 0.100 & 0.101 & 0.562 & & -71.612 & 0.103 & 0.103 & 0.704 & & -72.382 & 0.101 & 0.100 & 0.683 \\
				& TRUE & 2.040 & 0.099 & 0.098 & 0.951 & & 0.282 & 0.098 & 0.098 & 0.950 & & 0.993 & 0.096 & 0.097 & 0.956 & & 1.629 & 0.095 & 0.097 & 0.959 \\
				& RCE & 0.089 & 0.356 & 0.368 & 0.959 & & -0.524 & 0.361 & 0.377 & 0.958 & & 0.836 & 0.369 & 0.367 & 0.952 & & 6.517 & 0.382 & 0.377 & 0.947 \\
				\\
				3  & Naive & -72.681 & 0.103 & 0.103 & 0.617 & & -75.207 & 0.101 & 0.101 & 0.564 & & -70.347 & 0.103 & 0.103 & 0.707 & & -72.486 & 0.102 & 0.100 & 0.678 \\
				& TRUE & 2.040 & 0.099 & 0.098 & 0.951 & & 0.282 & 0.098 & 0.098 & 0.950 & & 0.993 & 0.096 & 0.097 & 0.956 & & 1.629 & 0.095 & 0.097 & 0.959 \\
				& RCE & -1.898 & 0.434 & 0.447 & 0.957 & & -0.721 & 0.444 & 0.458 & 0.955 & & 3.296 & 0.444 & 0.446 & 0.950 & & 6.846 & 0.459 & 0.457 & 0.948 \\
				\\
				4  & Naive & -71.304 & 0.104 & 0.103 & 0.624 & & -73.001 & 0.098 & 0.100 & 0.590 & & -70.378 & 0.104 & 0.103 & 0.722 & & -74.191 & 0.101 & 0.100 & 0.673 \\
				& TRUE & 1.205 & 0.108 & 0.106 & 0.947 & & 2.904 & 0.107 & 0.106 & 0.950 & & 1.702 & 0.104 & 0.105 & 0.946 & & 0.671 & 0.106 & 0.104 & 0.948 \\
				& RCE & -0.285 & 0.479 & 0.468 & 0.944 & & 5.608 & 0.473 & 0.480 & 0.962 & & -4.517 & 0.456 & 0.467 & 0.959 & & -3.921 & 0.470 & 0.479 & 0.958 \\
				\\
				5  & Naive & -71.670 & 0.103 & 0.103 & 0.611 & & -73.655 & 0.100 & 0.100 & 0.589 & & -70.035 & 0.103 & 0.103 & 0.717 & & -73.428 & 0.099 & 0.100 & 0.685 \\
				& TRUE & 1.210 & 0.108 & 0.106 & 0.945 & & 1.625 & 0.107 & 0.106 & 0.945 & & 1.940 & 0.105 & 0.105 & 0.945 & & 2.972 & 0.101 & 0.104 & 0.963 \\
				& RCE & -2.172 & 0.528 & 0.529 & 0.949 & & 1.387 & 0.540 & 0.543 & 0.950 & & -0.308 & 0.521 & 0.529 & 0.955 & & 1.420 & 0.536 & 0.542 & 0.953 \\
				\\
				6  & Naive & -72.183 & 0.105 & 0.103 & 0.602 & & -73.641 & 0.100 & 0.100 & 0.588 & & -69.343 & 0.103 & 0.102 & 0.729 & & -73.516 & 0.098 & 0.100 & 0.704 \\
				& TRUE & 1.210 & 0.108 & 0.106 & 0.945 & & 1.625 & 0.107 & 0.106 & 0.945 & & 1.940 & 0.105 & 0.105 & 0.945 & & 2.972 & 0.101 & 0.104 & 0.963 \\
				& RCE & 1.068 & 0.659 & 0.643 & 0.940 & & 6.505 & 0.662 & 0.659 & 0.949 & & -3.913 & 0.644 & 0.642 & 0.949 & & -4.774 & 0.651 & 0.658 & 0.954 \\
				\\
				7  & Naive & -72.117 & 0.071 & 0.073 & 0.327 & & -75.989 & 0.071 & 0.071 & 0.277 & & -71.020 & 0.075 & 0.073 & 0.503 & & -75.229 & 0.070 & 0.071 & 0.442 \\
				& TRUE & -0.323 & 0.062 & 0.060 & 0.943 & & 0.366 & 0.059 & 0.060 & 0.955 & & 0.999 & 0.059 & 0.059 & 0.952 & & 0.779 & 0.059 & 0.059 & 0.951 \\
				& RCE & -1.923 & 0.207 & 0.207 & 0.955 & & -2.868 & 0.216 & 0.212 & 0.949 & & 3.756 & 0.211 & 0.207 & 0.950 & & 2.879 & 0.208 & 0.212 & 0.955 \\
				\\
				8  & Naive & -71.406 & 0.073 & 0.073 & 0.346 & & -77.833 & 0.070 & 0.071 & 0.250 & & -70.692 & 0.071 & 0.073 & 0.498 & & -75.456 & 0.071 & 0.071 & 0.434 \\
				& TRUE & 0.983 & 0.058 & 0.060 & 0.954 & & -0.714 & 0.060 & 0.060 & 0.950 & & 0.615 & 0.060 & 0.059 & 0.952 & & 0.378 & 0.059 & 0.059 & 0.949 \\
				& RCE & -0.431 & 0.237 & 0.234 & 0.944 & & -5.111 & 0.237 & 0.240 & 0.958 & & 0.476 & 0.232 & 0.234 & 0.950 & & -0.603 & 0.241 & 0.240 & 0.952 \\
				\\
				9  & Naive & -71.908 & 0.073 & 0.073 & 0.328 & & -77.891 & 0.070 & 0.071 & 0.245 & & -70.109 & 0.070 & 0.073 & 0.507 & & -75.328 & 0.071 & 0.071 & 0.438 \\
				& TRUE & 0.983 & 0.058 & 0.060 & 0.954 & & -0.714 & 0.060 & 0.060 & 0.950 & & 0.615 & 0.060 & 0.059 & 0.952 & & 0.378 & 0.059 & 0.059 & 0.949 \\
				& RCE & 0.428 & 0.285 & 0.285 & 0.948 & & -4.204 & 0.286 & 0.292 & 0.960 & & -0.500 & 0.280 & 0.284 & 0.950 & & -1.545 & 0.292 & 0.292 & 0.955 \\
				\hline 
			\end{tabular}
		\end{adjustbox}
	\end{center}
\end{table}

According to results in Tables 1 to 4, it is noted that naive estimation performs the worst as expected. Across all scenarios and sample sizes, the naive estimator exhibits substantial relative bias, typically ranging from -70\% to -80\% for the slope parameters associated with error-prone continuous covariates (i.e., $\widehat{\beta}_{11}$, $\widehat{\beta}_{21}$, $\widehat{\beta}_{12}$, $\widehat{\beta}_{22}$, $\widehat{\beta}_{13}$, $\widehat{\beta}_{23}$, $\widehat{\beta}_{14}$, $\widehat{\beta}_{24}$), indicating a systematic attenuation of true effect sizes when raw contaminated surrogates are directly incorporated into logistic regression without error adjustment. This severe attenuation towards zero is consistent with the well-known effect of classical measurement error in logistic regression \citep{Cao2023,Carroll2006,Cao2021}. Moreover, coverage probabilities for naive estimates are far below the nominal 95\% level, often falling below 0.50 and occasionally approaching zero or dropping to zero for 
$n=6000$. 
Such poor coverage is expected for naive analyses under severe measurement error, as bias dominates sampling variability. Although the bias patterns remain largely stable across different misclassification configurations of $V$ and $A$, which indicates the dominant source of bias in naive estimation may be due to the measurement error in continuous covariates $X_1$ and $X_2$, misclassification further distorts estimates in a non-negligible manner. While the TRUE estimator, which is obtained using unobserved true covariates $X_1$, $X_2$, $V$ and $A$, yields negligible relative bias since RB of most parameter estimates are within $\pm$ 3\% for $n=6000$. Across all settings, SEs are close to the SDs, and coverage probabilities match the nominal 0.95 level. Estimation results from TRUE further confirms the validity of the simulation setup and provides a reliable benchmark for evaluating the finite-sample performance of Naive and RCE.

By contrast, the proposed RCE substantially reduces the bias compared with naive estimation across all cases. For most slope parameters, the RB\% of RCE decreases from around -75\% to within $\pm$ 10\% for many cases and drops under $\pm$ 5\% when sample size increaes from $n=2000$ to $n=6000$. Regardless of sample size, misclassification rate and magnitude of true regression effects, coverage probabilities of all slope parameter estimates are close to 95\% with a small fluctuations. This indicates that variance estimation of RCE is reliable and also verifies that RCE has robust statistical inference independent of effect magnitude. However, RCE does not completely eliminate bias in several cases, especially those with higher misclassification rates (e.g., Case 3, Case 6 and Case 9 with $\pi_0^a=\pi_1^a=0.2$). RB\% values of certain parameter estimates remain moderately large (e.g., up to 24.787\% for $\widehat{\beta}_{14}$ at $n=2000$ for Case 4 in Table 2), indicating that the correction method is imperfect when one of misclassification error is large and sample size is $n=2000$. These results are consistent with findings in previous references \citep{Cao2021,Cao2023,Bartlett2018}. As noted by \citet{Carroll2006}, regression calibration yields approximately consistent estimators for nonlinear models and may retain non‑negligible bias when measurement error is substantial.

With regard of the effect of misclassification rates on RCE, it is noted that when the misclassification probabilities for $V$ and $A$ increase (e.g., misclassification for $A$ from Case 1 to Case 3, misclassification for $V$ from Case 1 to Case 4), the RB\% of RCE tends to increase in magnitude and the corresponding SEs become wider, reflecting the loss of information due to greater classification uncertainty. For instance, compared with Case 1 (with $\pi_0^a=\pi_1^a=0$), the RB\% of RCE for several parameters in Case 3 (with $\pi_0^a=\pi_1^a=0.2$) is doubled or tripled (e.g., $\widehat{\beta}_{13}$, $\widehat{\beta}_{23}$, $\widehat{\beta}_{14}$ in groups 3 and 4). This suggests that although RCE remains valid under moderate misclassification of one categorical variable, its performance drops significantly under high misclassification rate of the other categorical variable. Furthermore, when comparing results in Case 1 (or Case 2 or Case 3) with those in Case 4 (or Case 5 or Case 6), it can be seen that the maximum RB\% among slope parameter estimates will increase noticeably, especially when $n=2000$. This change is due to the increase in the maximum misclassification rate of $V$ from 0.1 to 0.23. Moreover, comparing results in Case 1 (or Case 2 or Case 3) with those in Case 7 (or Case 8 or Case 9), it can be seen that corresponding RB stays relatively constant when the misclassification rate of $A$ is fixed regardless of the misclassification rate of $V$. The phenomenon is due to the increasing proportion of true $V=1$ (i.e., $p_t^v$). 

As expected, the increase of sample size from $n=2000$ to $n=6000$ improves the performance of RCE across all cases resulting in RB\% decreasing markedly; SEs being reduced by roughly a factor of $\sqrt{2000/6000}\approx 0.577$; and CPs becoming more stable and closer to the nominal. 
For example, for Case 4 of Table 1, the RB\% for $\widehat{\beta}_{11}$ made a modest change; the SE decreased from 0.197 to 0.113; and the CP improved from 0.939 to 0.951 when the sample size increases from 2000 to 6000. This pattern confirms the asymptotic consistency of RCE, as the correction relies on the consistency of estimated nuisance parameters from the validation data. In addition, the patterns in Tables S1–S4 of the Web Appendix are similar to those in Tables 1–4. Furthermore, the comparison of effects of different parameter magnitudes (Tables S5-S8 versus Tables S1-S4 and Tables 1-4) shows that the relative bias of naive estimation remains its severity in proportional terms, whereas RCE continues to provide effective bias reduction when the true regression coefficients increase. Specifically, although the RB\% remains largely unchanged, the absolute bias grows with an increase in the magnitude of true regression coefficient. This indicates that RCE performs proportionally well regardless of the parameter magnitude, making it a reliable correction method across different effect sizes. 

Note that for Cases 1, 4 and 7, $A$ is not misclassified (i.e., $\pi_0^a=\pi_1^a=0$) and thus results from these cases are only affected by measurement error in covariates and one misclassified variable (i.e., $V$). Comparing naive estimation results from Cases 1-3 (or Cases 4-6 or Cases 7-9), it can be seen that the overall RB\% of naive estimation under Cases 2-3 (or Cases 5-6 or Cases 8-9) tend to be slightly larger than those under case 1 (or Case 4 or Case 7), which indicates misclassification on one more categorical variable will likely produce more bias of naive estimation. Furthermore, Cases 1, 4 and 7 are the same as three scenarios considered in \citet{Cao2023} since the variable $A$ is not classified with error. It can be seen that performances of RCE for these three cases are quite similar to those in Tables 1-2 of \citet{Cao2023}. That is, the maximum RB\% of RCE may be serious (can be as large as 24.787\%, see RB\% of $\widehat{\beta}_{14}$ at Case 4 in Table 2) due to small probabilities of $Pr(V=0|V_o=0)$ and $Pr(V=1|V_o=1)$, but increasing sample size can benefit RCE for reducing biases in estimation. Moreover, the overall RB\% from Case 7 are slightly smaller than those from Case 1, especially for $n=2000$, because $Pr(V=1|V_o=1)$ for Case 7 is larger. 

Overall, simulation results demonstrate that the proposed RCE substantially outperforms naive estimation in terms of  reduction of bias, stabilization of SE and achievement of the nominal CP across a wide range of measurement error and misclassification scenarios. The proposed method is robust to varying misclassification rates and sample sizes although its performance naturally deteriorates under conditions of high misclassification and small sample size. These findings support the practical applicability of the proposed RCE approach to the analysis of logistic regression model with both correlated measurement error in continuous covariates and misclassification in multiple categorical variables, which is commonly encountered in nutritional epidemiological studies such as the EPIC-InterAct Study \citep{InterAct2011}.

\section{Real Data Analysis}\label{sec:data_analysis}
\subsection{Data description}\label{sec:data_description}
We apply the proposed method to handle measurement error and misclassification involved in the EPIC-InterAct Study \citep{InterAct2011}, which is a large prospective case-cohort study nested within the European Prospective Investigation into Cancer and Nutrition (EPIC) Study \citep{Riboli2002}. 
In this study, participants were asked to report their long-term
dietary intake (e.g., vegetable, fruit and alcohol consumption), lifestyle characteristics, anthropometric measurements, and medical history in a food frequency questionnaire (FFQ). The detailed description of the data from the EPIC-InterAct Study were already given by related references \citep{InterAct2012,InterAct2013,InterAct2015}. Since most answers in the FFQ are subjective, 
it is very likely for those dietary intake assessed with a self-report instrument is prone to measurement error, especially considering their answers are based on recall past intakes over an extended period of time.

In the EPIC-InterAct Study, there were a total of 27,779 participants and
12,403 of them suffered from type 2 diabetes (T2D). The remaining participants were first randomly selected
stratified by center with a buffy coat from the 340,234 participants sampled from 8 of the 10 EPIC countries, then excluded if they finally developed incident T2D. We used the same procedure of \citet{Cao2021} for cleaning data in the main and validation studies. That is, we excluded
individuals if they had one or more missing variables, or suffered from stroke or myocardial infarction, or came from Denmark and Sweden. Finally, 15,257 (5785 males) and 1021 (408 males) participants remained in
the main study and the validation study, respectively.  Baseline
characteristics of interested variables such as age, 
protein, energy and physical activity in the main and validation studies
can be seen in Tables 3 and 4 of \citet{Cao2021}. The comparison of these two tables shows that characteristics in main and validation studies are similar. 

Note that our proposed correction method RCE is based on the normality assumption on true covariates and error variables (see the assumptions (A4) and (A6) in Section \ref{subsec:misclassification_measurement_error}). However, distributions of original measurements of most nutrients from the EPIC-InterAct Study deviate from normal distribution. But, after making the square-root transformation, normal Q--Q plots show that those nutrients of intake such as protein, fat and energy approximately follow normal distribution. See Figures 2.1-2.3 of \citet{Cao2019} in details. Thus, we made the square-root transformation on measurements of those continuous nutrients, such as protein, fat and energy, in the real data analysis. Besides, similar to \citet{Cao2021,Cao2022}, 
we applied standardization to all transformed continuous variables to make their mean and variance equal to 0 and 1, respectively.

\subsection{Estimation of misclassification rate and measurement error}\label{subsec:est_measurement_error}
To address both measurement error and misclassification problem in the EPIC-InterAct Study, 
\citet{Cao2023} investigated the association of T2D on fruit, protein and fat, which were measured with error; as well as a misclassified variable, physical activity (PA), based on a logistic regression model. Besides, \citet{Cao2025} applied Weibull regression to study the association of T2D on three error-prone covariates (fruit plus vegetable, protein, carbohydrate) and a misclassified categorical variable PA, along with some other error-free covariates. However, both references considered only one misclassified categorical variable, PA. This article aims to handle two misclassified variables together with several error-prone continuous covariates. So, we construct another misclassified categorical variable in addition to PA. It is noted that many studies \citep{Keefe2007,Baliunas2009,Thompson2013,Bell2017,Wood2018} reported a U--shaped or J--shaped trend where light-to-moderate drinkers had better health outcomes compared to non-drinkers and heavy drinkers. Besides, a J-- or U--shaped relationship between alcohol consumption and T2D was found in a recent study \citep{Dai2026}. This study further obtained the theoretical minimum risk exposure level (TMREL)
at 18g per day. Above the TMREL, the protective association weakens and disappears at 47g per day as the equivalence level of non-drinker. The consumption of alcohol beyond this point is associated with higher risk than no drinking. In other words, alcohol consumption in the middle group has different effect on the risk of T2D compared with other two groups if it is separated into three groups. Since the alcohol consumption is also self-reported and measured with error, it will be misclassified if we classify it into different groups based on chosen cut-off points. In this article, we first convert alcohol consumption into a three-level categorical variable by two cut-off points and then consider it as the second misclassified variable.

Let $A$ denote a three-level categorical variable for alcohol consumption with cut-off points $c_1$ and $c_2$, $A_o$ be its corresponding observed value. Based on observed measurement of alcohol consumption, we have
\begin{eqnarray*}
	A_o=\begin{cases}
		1\qquad {\rm if~alcohol~consumption}<c_1, \\
		2\qquad {\rm if}~c_1\leq {\rm alcohol~consumption} \leq c_2,\\
		3\qquad {\rm if}~c_2<{\rm alcohol~consumption}.
	\end{cases}
\end{eqnarray*}
To estimate corresponding misclassification rates, we adopt the same notation introduced in Section \ref{subsec:misclassification_measurement_error}, that is, $P^a_{obs,j}=Pr(A_o=j)$, $P^a_{T,j}=Pr(A=j)$, and $\phi_{jj'}=Pr(A_o=j'|A=j)$ are observed proportion, true proportion and misclassification probabilities, respectively, where $j,j'=1,2,3$. Besides, let $P_{jj'}=Pr({\rm observing}~j~{\rm in ~the~first~visit~and~observing}~j'~{\rm in~the~second~visit})$ and $n_{jj'}={\rm number~of~individuals~with}~j~{\rm in~the~first~determination~and}~j'~{\rm in~the~second~determination}$. Then, we have the following joint probabilities,
\begin{eqnarray*}
	P_{11}&=&P^a_{T,1}(1-\phi_{12}-\phi_{13})^2+P^a_{T,2}\phi_{21}^2+P^a_{T,3}\phi_{31}^2,\\
	P_{12}&=&P^a_{T,1}(1-\phi_{12}-\phi_{13})\phi_{12}+P^a_{T,2}\phi_{21}(1-\phi_{21}-\phi_{23})+P^a_{T,3}\phi_{31}\phi_{32},\\
	P_{13}&=&P^a_{T,1}(1-\phi_{12}-\phi_{13})\phi_{13}+P^a_{T,2}\phi_{21}\phi_{23}+P^a_{T,3}\phi_{31}(1-\phi_{31}-\phi_{32}),\\
	P_{22}&=&P^a_{T,1}\phi_{12}^2+P^a_{T,2}(1-\phi_{21}-\phi_{23})^2+P^a_{T,3}\phi_{32}^2,\\
	P_{23}&=&P^a_{T,1}\phi_{12}\phi_{13}+P^a_{T,2}(1-\phi_{21}-\phi_{23})\phi_{23}+P^a_{T,3}\phi_{32}(1-\phi_{31}-\phi_{32}),\\
	P_{33}&=&P^a_{T,1}\phi_{13}^2+P^a_{T,2}\phi_{23}^2+P^a_{T,3}(1-\phi_{31}-\phi_{32})^2,\\
	P_{21}&=&P_{12},\quad P_{31}=P_{13},\quad P_{32}=P_{23}.
\end{eqnarray*} 
Therefore, the likelihood function $L$ is proportional to $P_{11}^{n_{11}}P_{12}^{n_{12}+n_{21}}P_{13}^{n_{13}+n_{31}}P_{22}^{n_{22}}P_{23}^{n_{23}+n_{32}}\bigr[1-(P_{11}+2P_{12}+2P_{13}+P_{22}+2P_{23})\bigr]^{n_{33}}$. Note that we have five equations of $\partial \log L/\partial P_{11}=0$, $\partial \log L/\partial P_{12}=0$, $\partial \log L/\partial P_{13}=0$, $\partial \log L/\partial P_{22}=0$ and $\partial \log L/\partial P_{23}=0$ in terms of 8 unknown parameters, that is, $P^a_{T,1}$, $P^a_{T,2}$, $\phi_{12}$, $\phi_{13}$, $\phi_{21}$, $\phi_{23}$, $\phi_{31}$ and $\phi_{32}$. To estimate these unknown parameters, we make the following three assumptions: (1) $\phi_{12}=\phi_{21}=\phi_1$; (2) $\phi_{23}=\phi_{32}=\phi_2$; (3) $\phi_{13}=\phi_{31}=0$ in this article. The first two assumptions indicate that we classify $A$ into its neighbour level incorrectly with equal probability. It is reasonable to make this assumption when we have no other prior information about misclassification. Thus, sensitivity analyses would be desirable to assess the impact of these assumptions on adjacent levels, we will discuss this point at the end of this section. Since the categorical variable $A$ is classified based on a continuous covariate, it is unlikely that the true measurement belongs to the first group (i.e., $A=1$) but the observed measurement is classified to the third group (i.e., $A_o=3$); or the true measurement belongs to the third group (i.e., $A=3$) but the observed measurement is classified to the first group (i.e., $A_o=1$) if the measurement error of alcohol consumption is not huge. Therefore, we assume $\phi_{13}=\phi_{31}=0$.

In the above estimation strategy, $P_{ij}$ is the joint probability of repeated measures on $A_o$ in two visits for each individual.  Since the validation study of the EPIC-InterAct Study contains a single measure from FFQ and a single measure from a 24-hour recall (24HR) on each error-prone covariate, we take a common assumption of additive error model that their corresponding random errors are independent \citep{Keogh2020,Keogh2013}. We first denote $Q_i$ and $H_i$ as the observed FFQ and 24HR values of alcohol consumption for the $i$th individual ($i=1,\dots, n$). Then, we get two observed values of $A_o$ for the $i$th individual based on observed values of $Q_i$ and $H_i$ by cut-off points $c_1$ and $c_2$. Under the three assumptions of the above paragraph, misclassification rates ($\phi_1$ and $\phi_2$) and true probabilities ($P_{T,1}^a$ and $P_{T,2}^a$) can be estimated by maximizing the likelihood function $L$.  For choosing the best set of cut-off points ($c_1$ and $c_2$), we search cut-off points around 18 and 47 as found by \citet{Dai2026} by incrementing the lower cut-off point from 10 to 20 by one, and the upper cut-off point from 45 to 60 by one, so that the log-likelihood function $\log L$ is maximized. Data driven selection of cut‑points may introduce overfitting; this choice is conditional on our validation sample.

In the EPIC-InterAct Study, women drink much less alcohol than men.  It is noted that 86.9\% of female consumes less than 18g of alcohol per day and only 1.4\% of female consumes more than 47g of alcohol per day. Classifying alcohol consumption of female into three groups is inappropriate.  Therefore, we only consider the analysis of male data in this article.   
For male data from the validation study, $\log L$ is maximized when $(c_1,c_2)=(15,55)$, which results in the estimation results as $\widehat{P}^a_{T,1}=0.383$, $\widehat{P}^a_{T,2}=0.572$, $\widehat{\phi}_1=0.131$ and $\widehat{\phi}_2=0.164$. It means that the estimated proportions for light, moderate and heavy drinkers are 38.3\%, 57.2\% and 4.5\%, respectively; the probability of classifying light drinkers to moderate drinkers (or moderate drinkers to light drinkers) is 0.131, and the probability of classifying moderate drinkers and heavy drinkers incorrectly is 0.164. 
As the quality of estimation depends on assumptions behind, we apply chi-square goodness of fit test to check whether the observed number in the group of  $A_o=j$ and $A_o=j'$ based on observed values of FFQ and  24HR (i.e., $n_{jj'}$), matches the expected number under assumptions on misclassification rates. With regard to the chosen cut-off points (15 and 55), the test for male data from the validation study shows that assumptions on misclassification rates are satisfied with $\chi^2=4.33$ and P-value=$0.36$.

For the misclassification rate of PA, we adopt estimation results of $\pi^v_1$, $\pi^v_0$ and $P_t^v$ obtained in Section 5.3 of \citet{Cao2023}, that is, $\hat{\pi}^v_0=0.14$, $\hat{\pi}^v_1=0.23$ and $\widehat{P}_t^v=0.13$ for male. It is noted that \citet{Cao2021} studied the association of T2D on four error-prone covariates (i.e, fruit plus vegetable, protein, fat and energy); and \citet{Cao2023} also investigated the association of T2D on three error-prone covariates (i.e., fruit, protein and fat) and one misclassified categorical variable PA.  RC estimation results of these two studies based on logistic regression model show that no significant effects of fruit plus vegetable and fruit on the risk of T2D. As some dietary intakes, e.g., egg and meat, contain protein and fat, intakes of protein and fat are essentially correlated. Thus, in addition to energy, we consider either protein or fat, say, protein, in a model to reduce the effect of multicollinearity. Finally, we apply a logistic regression model for
studying the association of T2D on two error-prone nutrients, i.e., protein and energy, two misclassified variables, i.e., PA and discretized alcohol, together with some other error-free covariates in this article.

To apply RCE to handle measurement errors in nutrients, we need to estimate $E({\bX}|{\bW})$ and ${\rm Var}({\bX}|{\bW})$ for error-prone covariates and define ${\bX}=(X_{protein},X_{energy})^{\top}$ and ${\bW}=(Q_{protein},Q_{energy})^{\top}$. Using standardized measurement of plasma vitamin C and standardized square root transformed FFQ and 24HR measurements of protein and energy in the validation study, all unknown parameters in the additive error model (\ref{add_error_m}) can be estimated. According to the detailed procedure of the moment method for estimating ${\Sigma}_{xx}$, ${\Sigma}_{xw}$ and ${\Sigma}_{uu}$ described in Section 2.2 of \citet{Cao2021b}, $\widehat{\Sigma}_{xx}$,
$\widehat{\Sigma}_{xw}$ and $\widehat{\Sigma}_{uu}$ of these two nutrients for male data are as follows:
\begin{eqnarray*}\label{egm}
	\widehat{\bSigma}_{xx} = 
	\begin{pmatrix}
		1.155  & 0.337 \\
		0.337  & 1.139 \\
	\end{pmatrix},~~
	\widehat{\bSigma}_{xw}  = 
	\begin{pmatrix}
		0.388 & 0.093 \\
		0.124 & 0.315 \\
	\end{pmatrix},~~
	\widehat{\bSigma}_{uu} =
	\begin{pmatrix}
		0.869 & 0.752 \\
		0.752 & 0.913 \\
	\end{pmatrix}.
\end{eqnarray*}
Covariance matrix $\widehat{\bSigma}_{uu}$ shows that correlation coefficient of error variables in FFQ measurements of protein and energy is as high as 0.844. This result is similar to those results of Section 5.2 of \citet{Cao2021}. 

\subsection{Main study considering error-prone covariates and misclassified categorical variables}
In the data analysis, the response variable is binary, that is, T2D $\in \{0,1\}$, which denotes whether a participant suffers from T2D or not. Continuous covariates of interest include protein, energy and age, where the first two covariates are measured with error. For the misclassified categorical variable PA, we define it as 1 when its status is active and as 0 otherwise. For the alcohol consumption (alc), we classify it into a three-level categorical variable using the cut-off points 15 and 55, which is treated as the second misclassified variable taking values of 1, 2, 3. For other error-free covariates, education level (edu) is denoted as 1 when a participant has received secondary or higher education; edu=0 otherwise. If a participant is a current smoker, then his/her smoking status (smoke) is defined as 1, otherwise it is 0. As 5,785 male participants came from 14 different centers, center is also included in the model as a categorical variable. Then, we consider the logistic regression model \eqref{logistic} of T2D on the above covariates and interactions of error-prone covariates with PA and alc, that is, 
\begin{eqnarray}\label{logistic_rd}
	{\rm log}\bigg({P\over 1-P}\bigg)
	=\sum_{l=1}^6(\beta_{0l}+\beta_l^{\top}\bX)I(G=l)+\balpha^{\top} {\bZ},
\end{eqnarray}
where $P=Pr(Y=1|{\bX},{\bZ}, V, A)$, $Y=$T2D, ${\bX}=(X_{protein},X_{energy})^{\top}$, ${\bZ}=(Z_{age},Z_{edu},Z_{smoke},Z_{center})^{\top}$, $G=3*V+A$ for $V={\rm PA}\in \{0,1\}$ and $A={\rm alc}\in \{1,2,3\}$. Note that the absolute Pearson
correlation between the two error-prone nutrients (i.e., 
protein and energy) is smaller than 0.22, and
the absolute point-biserial correlation coefficient \citep{Rizopoulos2006} between any one of error-prone nutrients and any one
of binary error-free covariates (i.e., education level, smoke, and
center) is smaller than 0.13. Furthermore, the absolute point-biserial correlation coefficient between continuous alcohol consumption and PA is only 0.01 and thus we can infer that the three-level categorical alcohol $A$ is nearly independent with PA, i.e., the two misclassification models described in Section \ref{subsec:misclassification_measurement_error} can be treated to be independent. Moreover, according to the empirical knowledge, intakes of these two true nutrients $\bX$ will not be affected by $\bZ$ if $\bZ$ is fixed. Hence, we can roughly infer the distribution of $\bX|(\bW,\bZ)$ does not depend on $\bZ$ for fixed values of $\bZ$ and $\bW$, i.e., the assumption (A7) holds for the above logistic regression model.
Based on results of 
$\widehat{P}^a_{T,i}=Pr(A=i)$, $\widehat{\phi}_{ij}$ ($i,j=1,2,3$), $\widehat{P}_t^v$, $\hat{\pi}^v_0$, $\hat{\pi}^v_1$, 
$\widehat{\bSigma}_{xx}$ and $\widehat{\bSigma}_{xw}$ in Section \ref{subsec:est_measurement_error}, we thus obtain estimation results of naive estimation and RCE for males in Table 5.

\begin{table}
	\footnotesize
	\caption{Estimated regression coefficient, standard error, and P-value of Logistic regression using different methods of estimation based on model \eqref{logistic_rd} for male.}	
	\begin{center}
		\begin{adjustbox}{width=1\textwidth}
			\begin{tabular}{lclrrrrrrr}
				\hline
				& & &   \multicolumn{3}{c}{Naive estimation}& &\multicolumn{3}{c}{RC estimation}  \\
				\cline{4-6}\cline{8-10} PA & Alcohol (g per day) & Parameter &Estimate & SE & P-value & & Estimate & SE & P-value \\
				\hline
				Inactive & $<15$ & $\beta_{01}$      &-2.253 & 0.224 & $<0.001$ &  
				& -2.222  & 0.230 &  $<0.001$ \\ 
				&   & $\beta_{protein}$ & 0.291 & 0.085 & 0.001 &  & 0.310 & 0.137 & 0.024 \\ 
				&   & $\beta_{energy}$  &-0.215 & 0.083 & 0.009 &  & 0.064 & 0.193 & 0.741 \\ 
				\\
				Inactive & [15,55]  & $\beta_{02}$      &-2.274 & 0.226 &  $<0.001$ &  & -2.267 & 0.231 &  $<0.001$ \\ 
				&   & $\beta_{protein}$ & 0.319 & 0.096 & 0.001 &  & 0.328 & 0.146 & 0.024 \\
				&  & $\beta_{energy}$  &-0.257 & 0.098 & 0.008 &  & -0.078 & 0.214 & 0.718 \\ 
				\\
				Inactive & $>$55 & $\beta_{03}$      &-2.071 & 0.251 &  $<0.001$ &  & -1.599 & 0.540 & 0.003 \\ 
				&  & $\beta_{protein}$ &-0.077 & 0.150 & 0.607 &  & -0.953 & 0.741 & 0.198 \\ 
				&   & $\beta_{energy}$  & 0.110 & 0.168 & 0.515 &  & 0.866 & 1.254 & 0.490 \\ 
				\\
				Active &  $<15$ & $\beta_{04}$      &-2.500 & 0.229 &  $<0.001$ &  & -2.822 & 0.347 &  $<0.001$ \\
				&  & $\beta_{protein}$ & 0.373 & 0.158 & 0.018 &  & -0.138 & 0.574 & 0.810 \\ 
				&  & $\beta_{energy}$  &-0.473 & 0.150 & 0.002 &  & -1.484 & 0.758 & 0.050 \\ 
				\\
				Active & [15,55] & $\beta_{05}$      &-2.516 & 0.235 &  $<0.001$ &  & -2.912 & 0.361 &  $<0.001$ \\
				&   & $\beta_{protein}$ & 0.409 & 0.184 & 0.026 &  & 0.243 & 0.602 & 0.687 \\ 
				&  & $\beta_{energy}$  &-0.430 & 0.184 & 0.020 &  & -0.698 & 0.865 & 0.419 \\
				\\
				Active & $>$55 & $\beta_{06}$      &-2.005 & 0.329 &  $<0.001$ &  &  0.585 & 2.345 & 0.803 \\ 
				&  & $\beta_{protein}$ & 0.317 & 0.287 & 0.269 &  & 1.250 & 3.198 & 0.696 \\
				& & $\beta_{energy}$  &-0.326 & 0.344 & 0.343 &  & -1.241 & 5.729 & 0.828 \\
				\\ 
				&   & $\alpha_{age}$    & 0.048 & 0.004 &  $<0.001$ &  & 0.048 & 0.004 &  $<0.001$ \\ 
				&   & $\alpha_{edu}$    &-0.492 & 0.063 &  $<0.001$ &  & -0.492 & 0.063 &  $<0.001$ \\ 
				&   & $\alpha_{smoke}$  & 0.157 & 0.060 & 0.008 &  & 0.157 & 0.060 & 0.008 \\ 
				\hline				
			\end{tabular}
		\end{adjustbox}
	\end{center}	
	{\raggedright  Note: SE is standard error.} 
\end{table}

Naive estimation shows that for light or moderate drinkers, protein has positive association with the risk of T2D no matter individuals take active physical activity (i.e., PA=1) or not. But the positive effect disappears if the individual is a heavy drinker. For energy, its estimate shows a negatively significant effect on the risk of T2D for non-heavy drinkers. Similarly, the effect of energy for heavy drinkers is also insignificant. After correcting measurement error and misclassification, estimates from RC indicate that the true protein intake has a slightly greater effect on increasing the risk of T2D than the observed protein intake, but this significant effect only works on the group of non-heavy drinkers doing inactive physical activity. For men doing activity physical activity (i.e., PA=1), protein from RC estimation is insignificant no matter whether an individual belongs to the group of light, moderate or heavy drinkers. However, estimate of energy from RC are all insignificant under six groups except for the fourth group (P-value=0.05), which is consistent with result of the InterAct Consortium \citep{InterAct2013}.

For error-free covariates,  estimates by naive estimation and RCE are the same, which are similar to those in \citet{Cao2023}. That is, age is
positively associated with an increased risk of T2D but education level has the opposite effect. Besides, smoking increases the risk of T2D. These findings are consistent with those of related references \citep{Cao2023,Cao2021,InterAct2012,InterAct2015}.

As we pointed out in Section \ref{subsec:logistic}, we can easily obtain the interaction effects between $\bX$, $V$ and $A$ based on model \eqref{logistic}. By using results in Table 5, we find that for naive estimation, the interaction effect between heavy drinkers (i.e. ${\rm alc}=3$) and protein is significantly negative on the risk of T2D (P-value=0.026). The interaction between heavy drinkers and energy is weakly positively associated with the risk of T2D (P-value=0.078). However, after correcting measurement error and misclassification, RC estimates show that the interaction effect between heavy drinker group and protein becomes marginally significant negative (P-value = 0.083), 
and the interaction effect between PA and energy is negative and marginally significant on the risk of T2D (P-value = 0.072). Note that for the group of men not doing active physical activity, protein for the heavy drinker group has marginally significant smaller effect on the risk of T2D than the light drinker group.  According to Table 5, protein has positive effect on the risk of T2D if an individual is a light drinker.  
As the interaction between the heavy drinker group and protein is negatively aligned and marginally significant, protein has no effect on the risk of T2D if an individual becomes a heavy drinker. 
The negative interaction between PA and energy implies that energy for the light drinker group has weakly significant negative effect on the risk of T2D if he does active physical activity. 
Beyond that, there are no other significant interactions between PA, alc and protein (or energy).

Results in Table 5 are based on the cut-off points 15 and 55, which were obtained via the data-driven procedure described in Section \ref{subsec:est_measurement_error}. To assess how different cut-off points affect coefficient estimation, we selected two additional cut-off points for sensitivity analyses. The first setting is \((c_1,c_2)=(15,50)\), where only the second cut-off point is changed to 50, corresponding to the 85th percentile of alcohol consumption. The second setting is \((c_1,c_2)=(17.6,58.1)\), where 17.6 (the first component) corresponds to approximately the 45th percentile of alcohol consumption, and 58.1 (the second component) corresponds to approximately the 90th percentile of alcohol consumption. The corresponding results are provided in Tables S9 and S10 of the Web Appendix. Estimates across Table 5, Table S9, and Table S10 are broadly similar. For instance, the RC-based estimate for protein among the first group (i.e., light drinkers without doing active physical activity) is positively significant in both Table 5 and Table S9, and marginally positively significant (P-value=0.087) in Table S10. The protein estimate for the second group (i.e., moderate drinkers without doing active physical activity) is positively significant in Table 5 and Table S10, and marginally positively significant (P-value=0.065) in Table S9. For the fifth group (i.e., moderate drinkers doing active physical activity), both protein and energy estimates from the naive estimator are significant in Table 5. By contrast, only the naive energy estimate remains significant in Table S9 and becomes marginally significant (P-value=0.098) in Table S10.
While the naive protein estimate is marginally positively significant (P-value=0.071) in Table S9 and becomes insignificant in Table S10. All remaining estimates of protein and energy retain the same sign and statistical significance across the three tables. These findings suggest that the RCE estimates are relatively robust when cut-off points do not deviate substantially from the data-driven selected values.

\section{Discussion}\label{discussion}  
In this paper, we have developed a regression calibration (RC) framework for logistic regression on multiple correlated additive measurement error continuous covariates and two misclassified categorical variables. This work extends the RC method of \citet{Cao2023} from a single misclassified binary variable to a more general setting involving one binary misclassified variable $V$, one three-level misclassified variable $A$, and retaining correlated measurement error among the continuous covariates $\bX$. This proposed approach combines $V$ and $A$ into a composite categorical variable $G$ with six levels, constructs the correction matrix 
$R^G$ through Kronecker products of the corresponding measurement error and misclassification correction matrices. The resulting estimator is easy to implement using a standard logistic regression software after the corrected design matrix is constructed. In principle, this framework is not limited to two misclassified categorical variables and can be extended to more misclassified categorical predictors by forming a product-level categorical variable.

Extensive simulation studies demonstrate clear advantages of the proposed RCE over the naive estimation, which uses error-contaminated surrogate covariates directly. Across all simulation scenarios, naive estimators exhibit severe attenuation of slope coefficients toward zero, with relative bias between -70\% and -80\%, accompanied by severely degraded 95\% coverage probability.  In contrast, the RC estimator has substantially smaller relative bias (i.e., within ±10\% for many slope parameters and even within ±5\% for some cases and at large sample size, say 6000); standard error closer to the Monte Carlo standard deviation; and coverage probability closer to the nominal level.  These results support the asymptotic validity and inferential reliability of the proposed correction method under moderate measurement error and misclassification. However, RCE is not bias-free. When measurement error and misclassification rate are substantial, the RCE cannot entirely eliminate bias. This is consistent with the fact that RC is an approximately consistent estimator for a nonlinear regression model and can retain non-negligible bias when measurement error is large or when nuisance parameters are estimated with limited validation data \citep{Carroll2006}. Simulation results therefore suggest that the proposed method should be applied with appropriate caution when both measurement error on error-prone covariates and misclassification on multiple categorical variables are severe.

The real data application to the EPIC-InterAct Study illustrates the practical value of our proposed method in nutritional epidemiology. In the male population, we jointly corrected for correlated measurement error in protein and energy, and for misclassification in physical activity and a three-level alcohol consumption variable. The naive analysis produced apparently significant but nonintuitive associations, such as a negative energy effect and positive protein effects in several groups. After RCE correction,  energy associations became insignificant or marginally significant across the six $V \times A$ groups, which is consistent with the InterAct Consortium findings \citep{InterAct2013}. While protein remained positively associated with T2D risk only among non-heavy drinkers with PA=0. Effects of age, education, and smoking were consistent with previous literature \citep{Cao2023,Cao2021}, and the interaction analysis showed that some naive interaction signals changed after correction. These findings highlight that ignoring multiple misclassification and correlated measurement error can lead to misleading conclusions, not only for main effects but also for interaction effects.

One advantage of the proposed RC framework is to offer an explicit and computationally efficient correction to accommodate correlated measurement error in several continuous covariates and multiple misclassified categorical variables. Besides, it also provides a unified framework for identifying interaction effects between error-prone covariates and misclassified categorical variables through the composite variable $G$. However, several limitations should be acknowledged. First, the proposed RC procedure treats parameters in the measurement error and misclassification models as known quantities obtained from a validation study or external source. However, their estimation uncertainty cannot be quantified. Although this is a common feature of many RC methods \citep{Keogh2020}, it may lead to underestimated standard errors of estimates, especially when the sample size of validation study is relatively small. Future work could incorporate bootstrap, multiple imputation, or Bayesian methods to propagate nuisance parameter uncertainty. Second, our derivation assumes independent misclassification mechanisms across different categorical covariates. In settings where misclassification error mechanism for categorical variables are correlated, the proposed estimator may be biased. Developing methods relaxing this independence assumption is an important avenue for future research. Third, for the three-level alcohol consumption variable, we assumed symmetric misclassification between adjacent levels and zero misclassification between non‑adjacent levels. Sensitivity analyses varying these misclassification constraints would benefit applied practice. Fourth, the choice of cut-off points for alcohol consumption is data-driven. Different cut-off points could affect the estimation of misclassification rates and regression coefficients. Finally, our proposed method is based on non-differential measurement error and non-differential misclassification assumptions. If these assumptions are not satisfied, the proposed method may be invalid.  

Future research can proceed in several directions. Methodologically, it would be valuable to relax the assumption of independent misclassification mechanisms and allow correlated misclassification between multiple categorical covariates. It would also be useful to develop RC or hybrid methods that account for nuisance parameter uncertainty, possibly through semiparametric efficiency theory or robust variance estimation. Another natural extension is to survival models, such as Cox or Weibull regression, with multiple misclassified categorical variables and correlated measurement error. Some progress has been made for a single misclassified variable \citep{Cao2025}, but more general settings remain open.  In addition, the proposed framework could be extended to high-dimensional covariates and non-normal measurement errors, which are our future research work.
\clearpage



\section*{Acknowledgements}
The EPIC-InterAct study received funding from the European Union (Integrated Project LSHM-CT-2006-037197 in the Framework Programme 6 of the European Community). We thank all participants and staff for their contribution to this study. We thank staff from the Technical, Field Epidemiology and Data teams of the MRC Epidemiology Unit in Cambridge, UK, for carrying out sample preparation, DNA provision and quality control, genotyping and data-handling work. We thank the EPIC-InterAct PI and management team for enabling us to use data from the study in this paper. 

\section*{Conflict of interest}
The authors declare no potential conflict of interests.

\section*{Supporting information}
R code used in this article for simulation studies can be found at \url{https://github.com/Zhiqiangcao/multiple_measurement_error_and_misclassification}.


\bibliography{refs.bib}

\begin{thebibliography}{}

\bibitem[\protect\citeauthoryear{Agogo, van~der Voet, van't Veer, Ferrari,
  Leeders, Muller, Sanchez-Cantalejo, Bamia, Braaten, Knuppel, Johansson, van
  Eeuwijk, and Boshuizen}{Agogo et~al.}{2014}]{Agogo2014}
Agogo, G., H.~van~der Voet, P.~van't Veer, P.~Ferrari, M.~Leeders, D.~Muller,
  E.~Sanchez-Cantalejo, C.~Bamia, T.~Braaten, S.~Knuppel, I.~Johansson, F.~van
  Eeuwijk, and H.~Boshuizen (2014).
\newblock Use of two-part regression calibration model to correct for
  measurement error in episodically consumed foods in a single-replicate study
  design: Epic case study.
\newblock {\em PLoS One\/}~{\em 9\/}(11), e113160.

\bibitem[\protect\citeauthoryear{Augustin}{Augustin}{1999}]{Augustin1999}
Augustin, T. (1999).
\newblock Correcting for measurement error in parametric duration models by
  quasi-likelihood.
\newblock {\em Sonderforschungsbereich\/}~{\em 386}, 157.

\bibitem[\protect\citeauthoryear{Baliunas, Taylor, Irving, Roerecke, Patra,
  Mohapatra, and Rehm}{Baliunas et~al.}{2009}]{Baliunas2009}
Baliunas, D., B.~Taylor, H.~Irving, M.~Roerecke, J.~Patra, S.~Mohapatra, and
  J.~Rehm (2009).
\newblock Alcohol as a risk factor for type 2 diabetes: a systematic review and
  meta-analysis.
\newblock {\em Diabetes Care\/}~{\em 32\/}(11), 2123--2132.

\bibitem[\protect\citeauthoryear{Bartlett and Keogh}{Bartlett and
  Keogh}{2018}]{Bartlett2018}
Bartlett, J. and R.~Keogh (2018).
\newblock Bayesian correction for covariate measurement error: A frequentist
  evaluation and comparison with regression calibration.
\newblock {\em Stat Methods Med Res\/}~{\em 27\/}(6), 1695--1708.

\bibitem[\protect\citeauthoryear{Bell, Daskalopoulou, Rapsomaniki, George,
  Britton, Bobak, Casas, Dale, Denaxas, Shah, and Hemingway}{Bell
  et~al.}{2017}]{Bell2017}
Bell, S., M.~Daskalopoulou, E.~Rapsomaniki, J.~George, A.~Britton, M.~Bobak,
  J.~Casas, C.~Dale, S.~Denaxas, A.~Shah, and H.~Hemingway (2017).
\newblock Association between clinically recorded alcohol consumption and
  initial presentation of 12 cardiovascular diseases: population based cohort
  study using linked health records.
\newblock {\em BMJ\/}~{\em 256}, j909.

\bibitem[\protect\citeauthoryear{Cao}{Cao}{2019}]{Cao2019}
Cao, Z. (2019).
\newblock {\em Statistical Methods for Dealing with Measurement Error in
  Covariates: EPIC-InterAct Study}.
\newblock Hong Kong, PQDT-Global: The Hong Kong University of Science and
  Technology.

\bibitem[\protect\citeauthoryear{Cao and Wong}{Cao and Wong}{2021a}]{Cao2021}
Cao, Z. and M.~Wong (2021a).
\newblock Approximate maximum likelihood estimation for logistic regression
  with covariate measurement error.
\newblock {\em Biometrical J\/}~{\em 63\/}(1), 27--45.

\bibitem[\protect\citeauthoryear{Cao and Wong}{Cao and Wong}{2021b}]{Cao2021b}
Cao, Z. and M.~Wong (2021b).
\newblock Moment estimation method of parameters in additive measurement error
  model.
\newblock {\em Comput Meth Prog Bio\/}~{\em 206}, 106090.

\bibitem[\protect\citeauthoryear{Cao and Wong}{Cao and Wong}{2022}]{Cao2022}
Cao, Z. and M.~Wong (2022).
\newblock Approximate profile likelihood estimation for cox regression with
  covariate measurement error.
\newblock {\em Stat Med\/}~{\em 41\/}(5), 910--931.

\bibitem[\protect\citeauthoryear{Cao and Wong}{Cao and Wong}{2025}]{Cao2025}
Cao, Z. and M.~Wong (2025).
\newblock Weibull regression with both measurement error and misclassification
  in covariates.
\newblock {\em Biometrical J\/}~{\em 65\/}(7), e70083.

\bibitem[\protect\citeauthoryear{Cao, Wong, and Cheng}{Cao
  et~al.}{2023}]{Cao2023}
Cao, Z., M.~Wong, and G.~Cheng (2023).
\newblock Logistic regression with correlated measurement error and
  misclassification in covariates.
\newblock {\em Stat Methods Med Res\/}~{\em 32\/}(4), 789--805.

\bibitem[\protect\citeauthoryear{Carlsson, Hammar, and Grill}{Carlsson
  et~al.}{2005}]{Carlsson2005}
Carlsson, S., N.~Hammar, and V.~Grill (2005).
\newblock Alcohol consumption and type 2 diabetes meta-analysis of
  epidemiological studies indicates a u-shaped relationship.
\newblock {\em Diabetologia\/}~{\em 48\/}(6), 1051--1054.

\bibitem[\protect\citeauthoryear{Carroll, Ruppert, Stefanski, and
  Crainiceanu}{Carroll et~al.}{2006}]{Carroll2006}
Carroll, R., D.~Ruppert, L.~Stefanski, and C.~Crainiceanu (2006).
\newblock {\em Measurement Error in Nonlinear Models: A Modern Perspective. 2nd
  ed.}
\newblock New York, NY: Chapman and Hall/CRC.

\bibitem[\protect\citeauthoryear{Consortium}{Consortium}{2011}]{InterAct2011}
Consortium, I. (2011).
\newblock Design and cohort description of the interact project: an examination
  of the interaction of genetic and lifestyle factors on the incidence of type
  2 diabetes in the epic study.
\newblock {\em Diabetologia\/}~{\em 54\/}(9), 2272--2282.

\bibitem[\protect\citeauthoryear{Consortium}{Consortium}{2012}]{InterAct2012}
Consortium, I. (2012).
\newblock Physical activity reduces the risk of incident type 2 diabetes in
  general and in abdominally lean and obese men and women: the epic-interact
  study.
\newblock {\em Diabetologia\/}~{\em 55\/}(7), 944--952.

\bibitem[\protect\citeauthoryear{Consortium}{Consortium}{2013}]{InterAct2013}
Consortium, I. (2013).
\newblock The association between dietary energy density and type 2 diabetes in
  europe: Results from the epic- interact study.
\newblock {\em PLoS One\/}~{\em 8\/}(5), e59947.

\bibitem[\protect\citeauthoryear{Consortium}{Consortium}{2015}]{InterAct2015}
Consortium, I. (2015).
\newblock Dietary fibre and incidence of type 2 diabetes in eight european
  countries: the epic-interact study and a meta-analysis of prospective
  studies.
\newblock {\em Diabetologia\/}~{\em 58\/}(7), 1394--1408.

\bibitem[\protect\citeauthoryear{Cook and Stefanski}{Cook and
  Stefanski}{1994}]{Cook1994}
Cook, J. and L.~Stefanski (1994).
\newblock Simulation-extrapolation estimation in parametric measurement error
  models.
\newblock {\em J Am Stat Asso\/}~{\em 89\/}(428), 1314–1328.

\bibitem[\protect\citeauthoryear{Dai, Nicholson, Lawlor, Carr, Steinmeta, Chen,
  Mclaughlin, Hay, et~al.}{Dai et~al.}{2026}]{Dai2026}
Dai, X., S.~Nicholson, H.~Lawlor, S.~Carr, J.~Steinmeta, N.~Chen,
  S.~Mclaughlin, O.~K. Hay, S.I., et~al. (2026).
\newblock Health effects associated with alcohol consumption: a burden of proof
  study.
\newblock {\em Nat Health\/}~{\em 1\/}(9), 979--999.

\bibitem[\protect\citeauthoryear{Falley, Stamey, and Beaujean}{Falley
  et~al.}{2018}]{Falley2018}
Falley, B., J.~Stamey, and A.~Beaujean (2018).
\newblock Bayesian estimation of logistic regression with misclassified
  covariates and response.
\newblock {\em J Appl Stat\/}~{\em 45\/}(10), 1756--1769.

\bibitem[\protect\citeauthoryear{Fraser and Stram}{Fraser and
  Stram}{2001}]{Fraser2001}
Fraser, G. and D.~Stram (2001).
\newblock Regression calibration in studies with correlated variables measured
  with error.
\newblock {\em Am J Epidemiol\/}~{\em 154\/}(9), 836--844.

\bibitem[\protect\citeauthoryear{Harding, Wareham, Bingham, Khaw, Luben, Welch,
  and Forouhi}{Harding et~al.}{2008}]{Harding2008}
Harding, A., N.~Wareham, S.~Bingham, K.~Khaw, R.~Luben, A.~Welch, and
  N.~Forouhi (2008).
\newblock Plasma vitamin c level, fruit and vegetable consumption, and the risk
  of new-onset type 2 diabetes mellitus: the european prospective investigation
  of cancer–norfolk prospective study.
\newblock {\em Arch Intern Med\/}~{\em 168\/}(14), 1493--1499.

\bibitem[\protect\citeauthoryear{Hu, Tang, Yang, Dai, Xiao, and Zhao}{Hu
  et~al.}{2022}]{Hu2022}
Hu, Y., D.~Tang, F.~Yang, S.~Dai, X.~Xiao, and X.~Zhao (2022).
\newblock The impacts of measurement errors on a dietary pattern analyses:a
  simulation study based on dietary data from the china multi-ethnic cohort
  (cmec) study.
\newblock {\em Am J Clin Nutr\/}~{\em 116\/}(2), 523--530.

\bibitem[\protect\citeauthoryear{Keogh, Shaw, Gustafson, Carroll, Deffner,
  Dodd, Kuchenhoff, Tooze, Wallace, Kipnis, and Freedman}{Keogh
  et~al.}{2020}]{Keogh2020}
Keogh, R., P.~Shaw, P.~Gustafson, R.~Carroll, V.~Deffner, K.~Dodd,
  H.~Kuchenhoff, J.~Tooze, M.~Wallace, V.~Kipnis, and L.~Freedman (2020).
\newblock Stratos guidance document on measurement error and misclassification
  of variables in observational epidemiology: Part 1–basic theory and simple
  methods of adjustment.
\newblock {\em Stat Med\/}~{\em 39\/}(16), 2197--2231.

\bibitem[\protect\citeauthoryear{Keogh, White, and Rodwell}{Keogh
  et~al.}{2013}]{Keogh2013}
Keogh, R., I.~White, and S.~Rodwell (2013).
\newblock Using surrogate biomarkers to improve measurement error models in
  nutritional epidemiology.
\newblock {\em Stat Med\/}~{\em 32\/}(22), 3838--3861.

\bibitem[\protect\citeauthoryear{Kuchenhoff, Mwalili, and Lesaffre}{Kuchenhoff
  et~al.}{2006}]{Kuchenhoff2006}
Kuchenhoff, H., S.~Mwalili, and E.~Lesaffre (2006).
\newblock A general method for dealing with misclassification in regression:
  the misclassification simex.
\newblock {\em Biometrics\/}~{\em 62\/}(1), 85–96.

\bibitem[\protect\citeauthoryear{Li, Imamura, Sharp, Schulze, Zheng, Amiano,
  Ardanaz, Bergmann, Chirlque, Fagherazzi, Franks, Grioni, Ibsen, Jakszyn,
  Johansson, Katzke, Laouali, Mancini, Overvad, Palli, Panico, Redondo-Sanchez,
  Ricceri, Rolandsson, Srour, Tjonneland, Tong, van~der Schouw, Riboli,
  Langenberg, Forouhi, and Wareham}{Li et~al.}{2026}]{Li2026}
Li, S., F.~Imamura, S.~Sharp, M.~Schulze, J.-S. Zheng, P.~Amiano, E.~Ardanaz,
  M.~Bergmann, M.-D. Chirlque, G.~Fagherazzi, P.~Franks, S.~Grioni, D.~Ibsen,
  P.~Jakszyn, I.~Johansson, V.~Katzke, N.~Laouali, F.~Mancini, K.~Overvad,
  D.~Palli, S.~Panico, D.~Redondo-Sanchez, F.~Ricceri, O.~Rolandsson, B.~Srour,
  A.~Tjonneland, T.~Tong, Y.~van~der Schouw, E.~Riboli, C.~Langenberg,
  N.~Forouhi, and N.~Wareham (2026).
\newblock Interactions between genetic predisposition to obesity, insulin
  resistance and type 2 diabetes risk, and food or beverage intake for incident
  type 2 diabetes: European prospective investigation into cancer and nutrition
  (epic) interact case–cohort study.
\newblock {\em Am J Clin Nutr\/}~{\em 123\/}(3), 101198.

\bibitem[\protect\citeauthoryear{MacLehose, Olshan, Herring, Honein, Shaw, and
  Romitti}{MacLehose et~al.}{2018}]{MacLehose2018}
MacLehose, R., A.~Olshan, A.~Herring, M.~Honein, G.~Shaw, and P.~Romitti
  (2018).
\newblock Bayesian methods for correcting misclassification an example from
  birth defects epidemiology.
\newblock {\em Epidemiology\/}~{\em 20\/}(1), 27--35.

\bibitem[\protect\citeauthoryear{Nishinakagawa, Sakurai, Nemoto, Matsunaga,
  Takebayashi, and Fujiwara}{Nishinakagawa et~al.}{2023}]{Nishinakagawa2023}
Nishinakagawa, M., R.~Sakurai, Y.~Nemoto, H.~Matsunaga, T.~Takebayashi, and
  Y.~Fujiwara (2023).
\newblock Influence of education and subjective financial status on dietary
  habits among young, middle-aged, and older adults in japan: a cross-sectional
  study.
\newblock {\em BMC Public Health\/}~{\em 23}, 1230.

\bibitem[\protect\citeauthoryear{Novick and Stefanski}{Novick and
  Stefanski}{2002}]{Novick2002}
Novick, S. and L.~Stefanski (2002).
\newblock Corrected score estimation via complex variable simulation
  extrapolation.
\newblock {\em J Am Stat Asso\/}~{\em 97\/}(458), 472–481.

\bibitem[\protect\citeauthoryear{O'Keefe, Bybee, and Lavie}{O'Keefe
  et~al.}{2007}]{Keefe2007}
O'Keefe, J., K.~Bybee, and C.~Lavie (2007).
\newblock Alcohol and cardiovascular health: the razor-sharp double-edged
  sword.
\newblock {\em J Am Coll Cardiol\/}~{\em 50\/}(11), 1009--1014.

\bibitem[\protect\citeauthoryear{Riboli, Hunt, Slimani, Ferrari, Norat, Fahey,
  Charrondiere, Hemon, Casagrande, Vignat, Overvad, Tjonneland,
  Clavel-Chapelon, Thiebaut, Wahrendorf, Boeing, Trichopoulos, Trichopoulou,
  Vineis, Palli, Bueno-de Mesquita, Peeters, Lund, Engeset, Gonzalez,
  Barricarte, Berglund, Hallmans, Day, Key, Kaaks, and Saracci}{Riboli
  et~al.}{2002}]{Riboli2002}
Riboli, E., K.~Hunt, N.~Slimani, P.~Ferrari, T.~Norat, M.~Fahey,
  U.~Charrondiere, B.~Hemon, C.~Casagrande, J.~Vignat, K.~Overvad,
  A.~Tjonneland, F.~Clavel-Chapelon, A.~Thiebaut, J.~Wahrendorf, H.~Boeing,
  D.~Trichopoulos, A.~Trichopoulou, P.~Vineis, D.~Palli, H.~B. Bueno-de
  Mesquita, P.~Peeters, E.~Lund, D.~Engeset, C.~Gonzalez, A.~Barricarte,
  G.~Berglund, G.~Hallmans, N.~Day, T.~Key, R.~Kaaks, and R.~Saracci (2002).
\newblock European prospective investigation into cancer and nutrition (epic):
  study populations and data collection.
\newblock {\em Public Health Nutr\/}~{\em 5\/}(6B), 1113--1124.

\bibitem[\protect\citeauthoryear{Rizopoulos}{Rizopoulos}{2006}]{Rizopoulos2006}
Rizopoulos, D. (2006).
\newblock An r package for latent variable modelling and item response theory
  analyses.
\newblock {\em J Stat Softw\/}~{\em 17\/}(5), 1--25.

\bibitem[\protect\citeauthoryear{Rosner, Willett, and Spiegelman}{Rosner
  et~al.}{1989}]{Rosner1989}
Rosner, B., W.~Willett, and D.~Spiegelman (1989).
\newblock Correction of logistic regression relative risk estimates and
  confidence intervals for systematic within-person measurement error.
\newblock {\em Stat Med\/}~{\em 8\/}(9), 1051--1069.

\bibitem[\protect\citeauthoryear{Schneeweiss and Kukush}{Schneeweiss and
  Kukush}{2009}]{Schneeweiss2009}
Schneeweiss, H. and A.~Kukush (2009).
\newblock Comparing the efficiency of structural and functional methods in
  measurement error models.
\newblock {\em Theor Probab Math St\/}~{\em 80}, 119–129.

\bibitem[\protect\citeauthoryear{Sevilimedu, Yu, and Samawi}{Sevilimedu
  et~al.}{2023}]{Sevilimedu2023}
Sevilimedu, V., L.~Yu, and H.~Samawi (2023).
\newblock Misclassification simulation extrapolation method for a weibull
  accelerated failure time model.
\newblock {\em Stat Methods Med Res\/}~{\em 32\/}(8), 1478--1493.

\bibitem[\protect\citeauthoryear{Shao}{Shao}{1999}]{Shao1999}
Shao, J. (1999).
\newblock {\em Mathematical Statistics}.
\newblock New York, NY: Springer.

\bibitem[\protect\citeauthoryear{Shaw, Deffner, Keogh, Tooze, Dodd, Kuchenhoff,
  Kipnis, and Freedman}{Shaw et~al.}{2018}]{Shaw2018}
Shaw, P., V.~Deffner, R.~Keogh, J.~Tooze, K.~Dodd, H.~Kuchenhoff, V.~Kipnis,
  and L.~Freedman (2018).
\newblock Epidemiologic analyses with error-prone exposures: review of current
  practice and recommendations.
\newblock {\em Ann Epidemiol\/}~{\em 28\/}(11), 821--828.

\bibitem[\protect\citeauthoryear{Shaw, Gustafson, Carroll, Deffner, Dodd,
  Keogh, Kipnis, Tooze, Wallace, Kuchenhoff, and Freedman}{Shaw
  et~al.}{2020}]{Shaw2020}
Shaw, P., P.~Gustafson, R.~Carroll, V.~Deffner, K.~Dodd, R.~Keogh, V.~Kipnis,
  J.~Tooze, M.~Wallace, H.~Kuchenhoff, and L.~Freedman (2020).
\newblock Stratos guidance document on measurement error and misclassification
  of variables in observational epidemiology: Part 2--more complex methods of
  adjustment and advanced topics.
\newblock {\em Stat Med\/}~{\em 39\/}(16), 2232--2263.

\bibitem[\protect\citeauthoryear{Sobiecki, Imamura, Davis, Sharp, Koulman,
  Hodgson, Guevara, Schulze, Zheng, Agnoli, Bonet, M.Colorado-Yohar,
  Fagherazzi, Franks, E.Gundersen, Jannasch, Kaaks, Katzke, Molina-Montes,
  Nilsson, Palli, Panico, Papier, Rolandsson, Sacerdote, Tjonneland, Tong,
  van~der Schouw, Danesh, Butterworth, Riboli, Murphy, Wareham, and
  Forouhi}{Sobiecki et~al.}{2023}]{Sobiecki2023}
Sobiecki, J., F.~Imamura, C.~Davis, S.~Sharp, A.~Koulman, J.~Hodgson,
  M.~Guevara, M.~Schulze, J.-S. Zheng, C.~Agnoli, C.~Bonet,
  S.~M.Colorado-Yohar, G.~Fagherazzi, P.~Franks, T.~E.Gundersen, F.~Jannasch,
  R.~Kaaks, V.~Katzke, E.~Molina-Montes, P.~Nilsson, D.~Palli, S.~Panico,
  K.~Papier, O.~Rolandsson, C.~Sacerdote, A.~Tjonneland, T.~Tong, Y.~van~der
  Schouw, J.~Danesh, A.~Butterworth, E.~Riboli, K.~Murphy, N.~Wareham, and
  N.~Forouhi (2023).
\newblock A nutritional biomarker score of the mediterranean diet and incident
  type 2 diabetes: Integrated analysis of data from the medley randomised
  controlled trial and the epic-interact case-cohort study.
\newblock {\em PLos Med\/}~{\em 20\/}(4), e1004221.

\bibitem[\protect\citeauthoryear{Spiegelman, Rosner, and Logan}{Spiegelman
  et~al.}{2000}]{Spiegelman2000}
Spiegelman, D., B.~Rosner, and R.~Logan (2000).
\newblock Estimation and inference for logistic regression with covariate
  misclassification and measurement error in main study/validation study
  designs.
\newblock {\em J Am Stat Asso\/}~{\em 95\/}(449), 51–61.

\bibitem[\protect\citeauthoryear{Stefanski and Carroll}{Stefanski and
  Carroll}{1987}]{Stefanski1987}
Stefanski, L. and R.~Carroll (1987).
\newblock Conditional scores and optimal scores for generalized linear
  measurement error models.
\newblock {\em Biometrika\/}~{\em 74\/}(4), 703–716.

\bibitem[\protect\citeauthoryear{Takahashi, Okada, Hashimoto, Kurogi, Murata,
  Ito, and Fukui}{Takahashi et~al.}{2024}]{Takahashi2024}
Takahashi, F., H.~Okada, Y.~Hashimoto, K.~Kurogi, H.~Murata, M.~Ito, and
  M.~Fukui (2024).
\newblock Association between alcohol consumption and incidence of type 2
  diabetes in middle-aged japanese from panasonic cohort study.
\newblock {\em cientific Reports\/}~{\em 14}, 20315.

\bibitem[\protect\citeauthoryear{Thompson}{Thompson}{2013}]{Thompson2013}
Thompson, P. (2013).
\newblock J-curve revisited: cardiovascular benefits of moderate alcohol use
  cannot be dismissed.
\newblock {\em Med J Aust\/}~{\em 198\/}(8), 419--422.

\bibitem[\protect\citeauthoryear{Wood, Kaptoge, Butterworth, Willeit,
  Warnakula, Bolton, Paige, Paul, Sweeting, et~al.}{Wood
  et~al.}{2018}]{Wood2018}
Wood, A., S.~Kaptoge, A.~Butterworth, P.~Willeit, S.~Warnakula, T.~Bolton,
  E.~Paige, D.~Paul, M.~Sweeting, et~al. (2018).
\newblock Risk thresholds for alcohol consumption: combined analysis of
  individual-participant data for 599 912 current drinkers in 83 prospective
  studies.
\newblock {\em Lancet\/}~{\em 391\/}(10129), 1513--1523.

\bibitem[\protect\citeauthoryear{Yi, Ma, Spiegelman, and Carroll}{Yi
  et~al.}{2015}]{Yi2015}
Yi, G., Y.~Ma, D.~Spiegelman, and R.~Carroll (2015).
\newblock Functional and structural methods with mixed measurement error and
  misclassification in covariates.
\newblock {\em J Am Stat Asso\/}~{\em 110\/}(510), 681–696.

\bibitem[\protect\citeauthoryear{Young, Conover, and Funk}{Young
  et~al.}{2018}]{Young2018}
Young, J., M.~Conover, and M.~Funk (2018).
\newblock Measurement error and misclassification in electronic medical
  records: Methods to mitigate bias.
\newblock {\em Curr Epidemiol Rep\/}~{\em 5\/}(1), 343--356.

\bibitem[\protect\citeauthoryear{Zucker, Gorfine, Li, Tadesse, and
  Spiegelman}{Zucker et~al.}{2013}]{Zucker2013}
Zucker, D., M.~Gorfine, Y.~Li, M.~Tadesse, and D.~Spiegelman (2013).
\newblock A regularization corrected score method for nonlinear regression
  models with covariate error.
\newblock {\em Biometrics\/}~{\em 69\/}(1), 80–90.

\bibitem[\protect\citeauthoryear{Zucker and Spiegelman}{Zucker and
  Spiegelman}{2008}]{Zuck2008}
Zucker, D. and D.~Spiegelman (2008).
\newblock Corrected score estimation in the proportional hazards model with
  misclassified discrete covariates.
\newblock {\em Stat Med\/}~{\em 27\/}(11), 1911--1933.

\end{thebibliography}
\clearpage

\appendix

\renewcommand{\thesection}{Wed Appendix A\arabic{section}}
\renewcommand{\thesubsection}{Web Appendix A\arabic{section}.\arabic{subsection}}
\renewcommand{\thesubsubsection}{Web Appendix A\arabic{section}.\arabic{subsection}.\arabic{subsubsection}}
\renewcommand{\theequation} {A.\arabic{equation}}
\renewcommand{\thetable}{S\arabic{table}}
\renewcommand{\thefigure}{S\arabic{figure}}

\renewcommand{\thetable}{S\arabic{table}}
\setcounter{table}{0}

\renewcommand{\thefigure}{S\arabic{figure}}
\setcounter{figure}{0}

\section{Additional Simulation results}
In this section, we report simulation results for $\btheta=(0, 0.25, 0.25, -0.1, 0.2, 0.2, 0.1, 0.3, 0.3, 0, 0.24, 0.24)^{\top}$ and $\btheta=(0, 0.5, 0.5, -0.1, 0.42, 0.42, 0.1, 0.4, 0.4, 0, 0.33, 0.33)^{\top}$ under different methods, sample sizes and cases of misclassification rates in Tables S1-S8. The 9 cases are Case 1: $P^v_t=0.15$, $\pi^v_0=\pi^v_1=0.1$ and $P^a_{t}=0.5$,  $\pi^a_0=\pi^a_1=0$;
Case 2: $P^v_t=0.15$, $\pi^v_0=\pi^v_1=0.1$ and $P^a_{t}=0.5$,  $\pi^a_0=\pi^a_1=0.1$;
Case 3:  $P^v_t=0.15$, $\pi^v_0=\pi^v_1=0.1$ and $P^a_{t}=0.5$,  $\pi^a_0=\pi^a_1=0.2$;
Case 4: $P^v_t=0.13$, $\pi^v_0=0.14$, $\pi^v_1=0.23$ and $P^a_{t}=0.5$,  $\pi^a_0=\pi^a_1=0$;
Case 5: $P^v_t=0.13$, $\pi^v_0=0.14$, $\pi^v_1=0.23$ and $P^a_{t}=0.5$,  $\pi^a_0=\pi^a_1=0.1$;
Case 6: $P^v_t=0.13$, $\pi^v_0=0.14$, $\pi^v_1=0.23$ and $P^a_{t}=0.5$,  $\pi^a_0=\pi^a_1=0.2$;
Case 7: $P^v_t=0.4$, $\pi^v_0=\pi^v_1=0.2$ and $P^a_{t}=0.5$,  $\pi^a_0=\pi^a_1=0$;
Case 8: $P^v_t=0.4$, $\pi^v_0=\pi^v_1=0.2$ and $P^a_{t}=0.5$,  $\pi^a_0=\pi^a_1=0.1$;
Case 9: $P^v_t=0.4$, $\pi^v_0=\pi^v_1=0.2$ and $P^a_{t}=0.5$,  $\pi^a_0=\pi^a_1=0.2$.

\begin{table}[h]
	\caption{\footnotesize Simulation results for regression coefficients of $X_1$ and $X_2$ at $G=1$ and $G=2$, i.e., $\beta_{11}$, $\beta_{21}$, $\beta_{12}$ and $\beta_{22}$ when $\btheta=(0, 0.25, 0.25, -0.1, 0.2, 0.2, 0.1, 0.3, 0.3, 0, 0.24, 0.24)^{\top}$ under $n=2000$ and different cases. RB (\%) = relative bias (percent), SD = Monte-Carlo standard 
		deviation, SE = Monte-Carlo standard error, CP = empirical 95\% coverage probability.}
	\label{simu:tabs1}
	\begin{center}
		\begin{adjustbox}{width=1\textwidth}
			\begin{tabular}{llrrrrrrrrrrrrrrrrrrr}
				\hline
				&& \multicolumn{9}{c}{$G=1$} & &\multicolumn{9}{c}{$G=2$} \\
				\cline{3-11}\cline{13-21}
				& &\multicolumn{4}{c}{$\hat{\beta}_{11}$}& 
				&\multicolumn{4}{c}{ $\hat{\beta}_{21}$} & &\multicolumn{4}{c}{ $\hat{\beta}_{12}$}
				& &\multicolumn{4}{c}{ $\hat{\beta}_{22}$}
				\\
				\cline{3-6}\cline{8-11}\cline{13-16}\cline{18-21} Case & Method & RB(\%) & SD & SE & CP & &RB(\%) & SD & SE & CP & & RB(\%) & SD& SE & CP & &RB(\%) & SD & SE & CP\\
				\hline
				1  & Naive & -73.706 & 0.092 & 0.095 & 0.494 & & -76.899 & 0.094 & 0.092 & 0.449 & & -73.873 & 0.096 & 0.095 & 0.647 & & -77.744 & 0.094 & 0.092 & 0.604 \\
				& TRUE & -0.215 & 0.070 & 0.071 & 0.957 & & 0.603 & 0.070 & 0.071 & 0.955 & & 1.071 & 0.070 & 0.070 & 0.952 & & 0.413 & 0.071 & 0.070 & 0.944 \\
				& RCE & -1.571 & 0.177 & 0.183 & 0.954 & & -1.349 & 0.191 & 0.188 & 0.950 & & -3.362 & 0.184 & 0.183 & 0.953 & & -3.895 & 0.190 & 0.188 & 0.947 \\
				\\
				2  & Naive & -73.555 & 0.091 & 0.095 & 0.496 & & -77.015 & 0.088 & 0.092 & 0.454 & & -73.475 & 0.095 & 0.095 & 0.660 & & -76.299 & 0.096 & 0.092 & 0.617 \\
				& TRUE & 1.442 & 0.071 & 0.071 & 0.953 & & 0.313 & 0.072 & 0.071 & 0.950 & & -0.839 & 0.071 & 0.070 & 0.948 & & 0.001 & 0.073 & 0.070 & 0.943 \\
				& RCE & 1.026 & 0.206 & 0.208 & 0.950 & & 0.806 & 0.210 & 0.213 & 0.956 & & -2.938 & 0.208 & 0.207 & 0.947 & & -2.273 & 0.220 & 0.213 & 0.943 \\
				\\
				3  & Naive & -74.020 & 0.091 & 0.095 & 0.499 & & -77.521 & 0.088 & 0.092 & 0.438 & & -72.894 & 0.096 & 0.095 & 0.654 & & -75.635 & 0.095 & 0.092 & 0.628 \\
				& TRUE & 1.442 & 0.071 & 0.071 & 0.953 & & 0.313 & 0.072 & 0.071 & 0.950 & & -0.839 & 0.071 & 0.070 & 0.948 & & 0.001 & 0.073 & 0.070 & 0.943 \\
				& RCE & 1.904 & 0.252 & 0.252 & 0.951 & & 1.385 & 0.254 & 0.258 & 0.952 & & -3.971 & 0.259 & 0.252 & 0.945 & & -2.887 & 0.269 & 0.258 & 0.939 \\
				\\
				4  & Naive & -74.035 & 0.095 & 0.095 & 0.503 & & -76.929 & 0.093 & 0.092 & 0.448 & & -74.956 & 0.094 & 0.095 & 0.653 & & -75.662 & 0.092 & 0.092 & 0.631 \\
				& TRUE & 0.429 & 0.070 & 0.070 & 0.948 & & 0.979 & 0.070 & 0.070 & 0.951 & & 0.368 & 0.071 & 0.070 & 0.953 & & 0.307 & 0.069 & 0.070 & 0.951 \\
				& RCE & -2.914 & 0.204 & 0.197 & 0.937 & & -2.369 & 0.208 & 0.202 & 0.944 & & -4.173 & 0.197 & 0.196 & 0.946 & & -1.569 & 0.204 & 0.201 & 0.955 \\
				\\
				5  & Naive & -74.125 & 0.096 & 0.095 & 0.497 & & -77.663 & 0.093 & 0.092 & 0.433 & & -72.253 & 0.098 & 0.095 & 0.648 & & -75.699 & 0.093 & 0.092 & 0.615 \\
				& TRUE & 0.900 & 0.069 & 0.070 & 0.962 & & 0.041 & 0.071 & 0.070 & 0.947 & & -0.414 & 0.071 & 0.070 & 0.949 & & 0.270 & 0.069 & 0.070 & 0.954 \\
				& RCE & -2.682 & 0.226 & 0.223 & 0.942 & & -2.984 & 0.230 & 0.228 & 0.946 & & 0.973 & 0.224 & 0.222 & 0.951 & & 1.026 & 0.226 & 0.228 & 0.952 \\
				\\
				6  & Naive & -75.272 & 0.096 & 0.095 & 0.480 & & -77.805 & 0.093 & 0.092 & 0.446 & & -70.925 & 0.098 & 0.095 & 0.665 & & -75.528 & 0.093 & 0.092 & 0.619 \\
				& TRUE & 0.900 & 0.069 & 0.070 & 0.962 & & 0.041 & 0.071 & 0.070 & 0.947 & & -0.414 & 0.071 & 0.070 & 0.949 & & 0.270 & 0.069 & 0.070 & 0.954 \\
				& RCE & -5.143 & 0.275 & 0.270 & 0.950 & & -4.029 & 0.277 & 0.277 & 0.951 & & 3.719 & 0.275 & 0.270 & 0.942 & & 2.126 & 0.274 & 0.277 & 0.953 \\
				\\
				7  & Naive & -73.533 & 0.115 & 0.112 & 0.628 & & -76.065 & 0.109 & 0.109 & 0.581 & & -72.190 & 0.112 & 0.112 & 0.737 & & -76.791 & 0.109 & 0.109 & 0.705 \\
				& TRUE & 0.992 & 0.082 & 0.085 & 0.958 & & 0.267 & 0.085 & 0.085 & 0.952 & & -0.020 & 0.086 & 0.084 & 0.941 & & 1.342 & 0.082 & 0.084 & 0.958 \\
				& RCE & -2.554 & 0.266 & 0.267 & 0.954 & & -1.548 & 0.268 & 0.274 & 0.955 & & -0.635 & 0.270 & 0.267 & 0.947 & & -1.690 & 0.277 & 0.274 & 0.952 \\
				\\
				8  & Naive & -73.516 & 0.110 & 0.112 & 0.622 & & -76.241 & 0.109 & 0.109 & 0.584 & & -71.552 & 0.116 & 0.112 & 0.751 & & -76.644 & 0.110 & 0.109 & 0.702 \\
				& TRUE & 1.874 & 0.086 & 0.085 & 0.950 & & 1.028 & 0.086 & 0.085 & 0.954 & & 0.867 & 0.085 & 0.084 & 0.948 & & 1.527 & 0.085 & 0.084 & 0.946 \\
				& RCE & -0.765 & 0.303 & 0.303 & 0.950 & & 0.580 & 0.317 & 0.310 & 0.948 & & 0.214 & 0.310 & 0.302 & 0.938 & & -2.508 & 0.314 & 0.310 & 0.953 \\
				\\
				9  & Naive & -74.363 & 0.110 & 0.112 & 0.617 & & -76.720 & 0.109 & 0.109 & 0.585 & & -70.586 & 0.116 & 0.112 & 0.743 & & -76.073 & 0.111 & 0.109 & 0.705 \\
				& TRUE & 1.874 & 0.086 & 0.085 & 0.950 & & 1.028 & 0.086 & 0.085 & 0.954 & & 0.867 & 0.085 & 0.084 & 0.948 & & 1.527 & 0.085 & 0.084 & 0.946 \\
				& RCE & -2.159 & 0.367 & 0.367 & 0.948 & & 0.489 & 0.384 & 0.377 & 0.950 & & 1.595 & 0.373 & 0.367 & 0.945 & & -2.611 & 0.384 & 0.376 & 0.949 \\
				\hline 
			\end{tabular}
		\end{adjustbox}
	\end{center}
\end{table}

\begin{table}[h]
	\caption{\footnotesize Simulation results for regression coefficients of $X_1$ and $X_2$ at $G=3$ and $G=4$, i.e., $\beta_{13}$, $\beta_{23}$, $\beta_{14}$ and $\beta_{24}$ when  $\btheta=(0, 0.25, 0.25, -0.1, 0.2, 0.2, 0.1, 0.3, 0.3, 0, 0.24, 0.24)^{\top}$ under $n=2000$ and different cases. RB (\%) = relative bias (percent), SD = Monte-Carlo standard 
		deviation, SE = Monte-Carlo standard error, CP = empirical 95\% coverage probability.}
	\label{simu:tabs2}
	\begin{center}
		\begin{adjustbox}{width=1\textwidth}
			\begin{tabular}{llrrrrrrrrrrrrrrrrrrr}
				\hline
				&& \multicolumn{9}{c}{$G=3$} & &\multicolumn{9}{c}{$G=4$} \\
				\cline{3-11}\cline{13-21}
				& &\multicolumn{4}{c}{$\hat{\beta}_{13}$}& 
				&\multicolumn{4}{c}{ $\hat{\beta}_{23}$} & &\multicolumn{4}{c}{ $\hat{\beta}_{14}$}
				& &\multicolumn{4}{c}{ $\hat{\beta}_{24}$}
				\\
				\cline{3-6}\cline{8-11}\cline{13-16}\cline{18-21} Case & Method & RB(\%) & SD & SE & CP & &RB(\%) & SD & SE & CP & & RB(\%) & SD& SE & CP & &RB(\%) & SD & SE & CP\\
				\hline
				1  & Naive & -74.712 & 0.186 & 0.181 & 0.757 & & -79.404 & 0.180 & 0.177 & 0.727 & & -75.048 & 0.183 & 0.181 & 0.830 & & -78.408 & 0.177 & 0.176 & 0.813 \\
				& TRUE & 4.744 & 0.181 & 0.176 & 0.949 & & 4.532 & 0.183 & 0.176 & 0.949 & & 6.273 & 0.181 & 0.174 & 0.946 & & 3.091 & 0.175 & 0.173 & 0.960 \\
				& RCE & -0.846 & 0.588 & 0.571 & 0.945 & & -3.297 & 0.603 & 0.586 & 0.946 & & 1.290 & 0.588 & 0.571 & 0.954 & & 1.293 & 0.591 & 0.585 & 0.955 \\
				\\
				2  & Naive & -76.453 & 0.186 & 0.181 & 0.746 & & -77.858 & 0.179 & 0.177 & 0.728 & & -75.482 & 0.183 & 0.181 & 0.819 & & -78.663 & 0.181 & 0.176 & 0.820 \\
				& TRUE & 1.780 & 0.180 & 0.175 & 0.955 & & 2.438 & 0.180 & 0.176 & 0.956 & & 4.706 & 0.183 & 0.174 & 0.943 & & 5.118 & 0.181 & 0.174 & 0.944 \\
				& RCE & -1.600 & 0.660 & 0.646 & 0.950 & & 1.397 & 0.669 & 0.663 & 0.956 & & -5.676 & 0.637 & 0.646 & 0.957 & & -6.514 & 0.666 & 0.662 & 0.951 \\
				\\
				3  & Naive & -77.502 & 0.189 & 0.181 & 0.737 & & -78.893 & 0.180 & 0.177 & 0.729 & & -74.118 & 0.181 & 0.181 & 0.830 & & -77.672 & 0.179 & 0.176 & 0.814 \\
				& TRUE & 1.780 & 0.180 & 0.175 & 0.955 & & 2.438 & 0.180 & 0.176 & 0.956 & & 4.706 & 0.183 & 0.174 & 0.943 & & 5.118 & 0.181 & 0.174 & 0.944 \\
				& RCE & -7.231 & 0.804 & 0.785 & 0.951 & & -3.699 & 0.811 & 0.805 & 0.950 & & 0.773 & 0.772 & 0.784 & 0.952 & & -1.310 & 0.810 & 0.804 & 0.950 \\
				\\
				4  & Naive & -75.713 & 0.180 & 0.180 & 0.750 & & -79.675 & 0.177 & 0.176 & 0.716 & & -73.948 & 0.186 & 0.181 & 0.832 & & -76.709 & 0.178 & 0.176 & 0.816 \\
				& TRUE & 4.624 & 0.198 & 0.189 & 0.947 & & 1.554 & 0.197 & 0.190 & 0.945 & & 1.831 & 0.190 & 0.187 & 0.946 & & 10.398 & 0.190 & 0.188 & 0.957 \\
				& RCE & 0.437 & 0.830 & 0.820 & 0.954 & & -1.983 & 0.846 & 0.840 & 0.955 & & 21.271 & 0.847 & 0.823 & 0.948 & & 19.858 & 0.855 & 0.842 & 0.951 \\
				\\
				5  & Naive & -76.268 & 0.185 & 0.180 & 0.755 & & -79.127 & 0.182 & 0.176 & 0.714 & & -74.513 & 0.182 & 0.181 & 0.834 & & -76.906 & 0.178 & 0.176 & 0.815 \\
				& TRUE & 2.669 & 0.192 & 0.189 & 0.951 & & 4.952 & 0.197 & 0.190 & 0.950 & & 4.612 & 0.190 & 0.187 & 0.951 & & 3.962 & 0.196 & 0.187 & 0.951 \\
				& RCE & 2.373 & 0.954 & 0.929 & 0.949 & & 2.912 & 0.987 & 0.952 & 0.947 & & 8.306 & 0.952 & 0.931 & 0.948 & & 10.299 & 0.990 & 0.953 & 0.943 \\
				\\
				6  & Naive & -76.471 & 0.184 & 0.180 & 0.750 & & -79.188 & 0.182 & 0.176 & 0.718 & & -74.022 & 0.180 & 0.180 & 0.839 & & -76.943 & 0.177 & 0.176 & 0.826 \\
				& TRUE & 2.669 & 0.192 & 0.189 & 0.951 & & 4.952 & 0.197 & 0.190 & 0.950 & & 4.612 & 0.190 & 0.187 & 0.951 & & 3.962 & 0.196 & 0.187 & 0.951 \\
				& RCE & 7.904 & 1.145 & 1.128 & 0.945 & & 7.381 & 1.194 & 1.156 & 0.943 & & 2.599 & 1.156 & 1.130 & 0.948 & & 5.099 & 1.204 & 1.157 & 0.942 \\
				\\
				7  & Naive & -74.890 & 0.133 & 0.127 & 0.571 & & -78.003 & 0.129 & 0.124 & 0.513 & & -73.900 & 0.129 & 0.127 & 0.703 & & -78.007 & 0.127 & 0.123 & 0.673 \\
				& TRUE & 1.190 & 0.107 & 0.106 & 0.956 & & 2.102 & 0.106 & 0.106 & 0.952 & & 2.224 & 0.103 & 0.104 & 0.951 & & 1.212 & 0.103 & 0.104 & 0.959 \\
				& RCE & -0.825 & 0.370 & 0.362 & 0.940 & & -1.013 & 0.381 & 0.371 & 0.949 & & 1.835 & 0.369 & 0.361 & 0.949 & & 1.122 & 0.379 & 0.370 & 0.948 \\
				\\
				8  & Naive & -74.728 & 0.130 & 0.127 & 0.582 & & -78.626 & 0.125 & 0.124 & 0.513 & & -75.894 & 0.124 & 0.127 & 0.693 & & -77.214 & 0.125 & 0.123 & 0.676 \\
				& TRUE & 1.963 & 0.105 & 0.106 & 0.956 & & 1.233 & 0.109 & 0.106 & 0.941 & & 0.888 & 0.104 & 0.104 & 0.952 & & 1.938 & 0.104 & 0.104 & 0.956 \\
				& RCE & 1.132 & 0.422 & 0.409 & 0.941 & & -0.539 & 0.427 & 0.420 & 0.952 & & -7.456 & 0.403 & 0.408 & 0.954 & & -3.803 & 0.428 & 0.419 & 0.946 \\
				\\
				9  & Naive & -75.124 & 0.128 & 0.127 & 0.557 & & -79.444 & 0.126 & 0.124 & 0.504 & & -75.253 & 0.126 & 0.127 & 0.697 & & -76.343 & 0.126 & 0.123 & 0.677 \\
				& TRUE & 1.963 & 0.105 & 0.106 & 0.956 & & 1.233 & 0.109 & 0.106 & 0.941 & & 0.888 & 0.104 & 0.104 & 0.952 & & 1.938 & 0.104 & 0.104 & 0.956 \\
				& RCE & 1.959 & 0.508 & 0.497 & 0.949 & & -1.664 & 0.524 & 0.510 & 0.944 & & -8.218 & 0.488 & 0.496 & 0.958 & & -2.546 & 0.517 & 0.509 & 0.949 \\
				\hline 
			\end{tabular}
		\end{adjustbox}
	\end{center}
\end{table}

\begin{table}[h]
	\caption{\footnotesize Simulation results for regression coefficients of $X_1$ and $X_2$ at $G=1$ and $G=2$, i.e., $\beta_{11}$, $\beta_{21}$, $\beta_{12}$ and $\beta_{22}$ when $\btheta=(0, 0.25, 0.25, -0.1, 0.2, 0.2, 0.1, 0.3, 0.3, 0, 0.24, 0.24)^{\top}$ under $n=6000$ and different cases. RB (\%) = relative bias (percent), SD = Monte-Carlo standard 
		deviation, SE = Monte-Carlo standard error, CP = empirical 95\% coverage probability.}
	\label{simu:tabs3}
	\begin{center}
		\begin{adjustbox}{width=1\textwidth}
			\begin{tabular}{llrrrrrrrrrrrrrrrrrrr}
				\hline
				&& \multicolumn{9}{c}{$G=1$} & &\multicolumn{9}{c}{$G=2$} \\
				\cline{3-11}\cline{13-21}
				& &\multicolumn{4}{c}{$\hat{\beta}_{11}$}& 
				&\multicolumn{4}{c}{ $\hat{\beta}_{21}$} & &\multicolumn{4}{c}{ $\hat{\beta}_{12}$}
				& &\multicolumn{4}{c}{ $\hat{\beta}_{22}$}
				\\
				\cline{3-6}\cline{8-11}\cline{13-16}\cline{18-21} Case & Method & RB(\%) & SD & SE & CP & &RB(\%) & SD & SE & CP & & RB(\%) & SD& SE & CP & &RB(\%) & SD & SE & CP\\
				\hline
				1  & Naive & -74.719 & 0.055 & 0.054 & 0.073 & & -77.539 & 0.053 & 0.053 & 0.044 & & -74.951 & 0.053 & 0.054 & 0.208 & & -75.750 & 0.053 & 0.053 & 0.182 \\
				& TRUE & -0.378 & 0.042 & 0.041 & 0.945 & & -0.127 & 0.043 & 0.041 & 0.940 & & -0.672 & 0.041 & 0.040 & 0.948 & & 0.360 & 0.041 & 0.041 & 0.949 \\
				& RCE & -5.016 & 0.103 & 0.105 & 0.953 & & -4.639 & 0.104 & 0.108 & 0.961 & & -2.899 & 0.103 & 0.105 & 0.958 & & -0.379 & 0.106 & 0.108 & 0.959 \\
				\\
				2  & Naive & -74.571 & 0.054 & 0.054 & 0.064 & & -77.536 & 0.053 & 0.053 & 0.044 & & -72.788 & 0.053 & 0.054 & 0.232 & & -77.103 & 0.053 & 0.053 & 0.174 \\
				& TRUE & 0.703 & 0.040 & 0.041 & 0.957 & & 0.461 & 0.040 & 0.041 & 0.953 & & 0.402 & 0.040 & 0.040 & 0.954 & & 1.172 & 0.041 & 0.041 & 0.949 \\
				& RCE & -2.773 & 0.118 & 0.119 & 0.947 & & -2.345 & 0.122 & 0.122 & 0.956 & & -2.094 & 0.118 & 0.119 & 0.951 & & -3.190 & 0.124 & 0.122 & 0.944 \\
				\\
				3  & Naive & -75.013 & 0.054 & 0.054 & 0.066 & & -78.008 & 0.053 & 0.053 & 0.046 & & -72.279 & 0.054 & 0.054 & 0.245 & & -76.505 & 0.053 & 0.053 & 0.180 \\
				& TRUE & 0.703 & 0.040 & 0.041 & 0.957 & & 0.461 & 0.040 & 0.041 & 0.953 & & 0.402 & 0.040 & 0.040 & 0.954 & & 1.172 & 0.041 & 0.041 & 0.949 \\
				& RCE & -2.507 & 0.145 & 0.145 & 0.950 & & -2.086 & 0.149 & 0.148 & 0.960 & & -2.522 & 0.145 & 0.145 & 0.943 & & -3.565 & 0.151 & 0.148 & 0.946 \\
				\\
				4  & Naive & -73.671 & 0.054 & 0.055 & 0.077 & & -77.871 & 0.052 & 0.053 & 0.040 & & -72.387 & 0.055 & 0.055 & 0.250 & & -77.115 & 0.054 & 0.053 & 0.170 \\
				& TRUE & 0.512 & 0.041 & 0.040 & 0.947 & & -0.175 & 0.040 & 0.040 & 0.950 & & 0.736 & 0.039 & 0.040 & 0.957 & & -0.234 & 0.041 & 0.040 & 0.947 \\
				& RCE & -3.448 & 0.113 & 0.113 & 0.952 & & -4.425 & 0.114 & 0.116 & 0.954 & & 1.209 & 0.114 & 0.113 & 0.947 & & -0.122 & 0.115 & 0.116 & 0.950 \\
				\\
				5  & Naive & -74.409 & 0.056 & 0.055 & 0.074 & & -77.231 & 0.055 & 0.053 & 0.052 & & -73.034 & 0.056 & 0.055 & 0.260 & & -77.111 & 0.055 & 0.053 & 0.174 \\
				& TRUE & 0.551 & 0.041 & 0.040 & 0.950 & & 0.408 & 0.040 & 0.040 & 0.946 & & 0.242 & 0.041 & 0.040 & 0.941 & & 0.534 & 0.041 & 0.040 & 0.939 \\
				& RCE & -1.971 & 0.129 & 0.128 & 0.946 & & -1.511 & 0.133 & 0.131 & 0.950 & & -3.393 & 0.127 & 0.128 & 0.959 & & -4.253 & 0.133 & 0.131 & 0.942 \\
				\\
				6  & Naive & -75.080 & 0.056 & 0.055 & 0.068 & & -77.759 & 0.054 & 0.053 & 0.046 & & -72.222 & 0.056 & 0.055 & 0.252 & & -76.456 & 0.054 & 0.053 & 0.190 \\
				& TRUE & 0.551 & 0.041 & 0.040 & 0.950 & & 0.408 & 0.040 & 0.040 & 0.946 & & 0.242 & 0.041 & 0.040 & 0.941 & & 0.534 & 0.041 & 0.040 & 0.939 \\
				& RCE & -2.711 & 0.157 & 0.155 & 0.950 & & -2.013 & 0.158 & 0.159 & 0.953 & & -2.559 & 0.155 & 0.155 & 0.952 & & -3.682 & 0.158 & 0.159 & 0.948 \\
				\\
				7  & Naive & -73.752 & 0.066 & 0.064 & 0.190 & & -76.360 & 0.063 & 0.063 & 0.144 & & -73.489 & 0.064 & 0.064 & 0.369 & & -76.866 & 0.063 & 0.063 & 0.306 \\
				& TRUE & -0.212 & 0.048 & 0.049 & 0.956 & & 0.262 & 0.048 & 0.049 & 0.952 & & -0.214 & 0.047 & 0.048 & 0.957 & & -0.153 & 0.049 & 0.048 & 0.941 \\
				& RCE & -3.352 & 0.157 & 0.153 & 0.947 & & -2.496 & 0.157 & 0.157 & 0.953 & & -4.431 & 0.153 & 0.153 & 0.954 & & -4.269 & 0.157 & 0.157 & 0.951 \\
				\\
				8  & Naive & -74.537 & 0.065 & 0.064 & 0.174 & & -77.037 & 0.062 & 0.063 & 0.138 & & -71.934 & 0.064 & 0.064 & 0.382 & & -75.971 & 0.062 & 0.063 & 0.328 \\
				& TRUE & 0.696 & 0.047 & 0.049 & 0.960 & & 0.847 & 0.048 & 0.049 & 0.951 & & -0.040 & 0.047 & 0.048 & 0.953 & & -0.147 & 0.049 & 0.048 & 0.944 \\
				& RCE & -4.934 & 0.175 & 0.174 & 0.943 & & -3.151 & 0.175 & 0.178 & 0.956 & & -0.428 & 0.176 & 0.173 & 0.945 & & -0.856 & 0.179 & 0.178 & 0.947 \\
				\\
				9  & Naive & -75.118 & 0.066 & 0.064 & 0.180 & & -77.377 & 0.063 & 0.063 & 0.134 & & -71.192 & 0.065 & 0.064 & 0.392 & & -75.534 & 0.063 & 0.063 & 0.328 \\
				& TRUE & 0.696 & 0.047 & 0.049 & 0.960 & & 0.847 & 0.048 & 0.049 & 0.951 & & -0.040 & 0.047 & 0.048 & 0.953 & & -0.147 & 0.049 & 0.048 & 0.944 \\
				& RCE & -5.683 & 0.214 & 0.211 & 0.946 & & -3.377 & 0.214 & 0.216 & 0.954 & & 0.569 & 0.214 & 0.211 & 0.945 & & -0.536 & 0.217 & 0.216 & 0.948 \\
				\hline 
			\end{tabular}
		\end{adjustbox}
	\end{center}
\end{table}

\begin{table}[h]
	\caption{\footnotesize Simulation results for regression coefficients of $X_1$ and $X_2$ at $G=3$ and $G=4$, i.e., $\beta_{13}$, $\beta_{23}$, $\beta_{14}$ and $\beta_{24}$ when  $\btheta=(0, 0.25, 0.25, -0.1, 0.2, 0.2, 0.1, 0.3, 0.3, 0, 0.24, 0.24)^{\top}$ under $n=6000$ and different cases. RB (\%) = relative bias (percent), SD = Monte-Carlo standard 
		deviation, SE = Monte-Carlo standard error, CP = empirical 95\% coverage probability.}
	\label{simu:tabs4}
	\begin{center}
		\begin{adjustbox}{width=1\textwidth}
			\begin{tabular}{llrrrrrrrrrrrrrrrrrrr}
				\hline
				&& \multicolumn{9}{c}{$G=3$} & &\multicolumn{9}{c}{$G=4$} \\
				\cline{3-11}\cline{13-21}
				& &\multicolumn{4}{c}{$\hat{\beta}_{13}$}& 
				&\multicolumn{4}{c}{ $\hat{\beta}_{23}$} & &\multicolumn{4}{c}{ $\hat{\beta}_{14}$}
				& &\multicolumn{4}{c}{ $\hat{\beta}_{24}$}
				\\
				\cline{3-6}\cline{8-11}\cline{13-16}\cline{18-21} Case & Method & RB(\%) & SD & SE & CP & &RB(\%) & SD & SE & CP & & RB(\%) & SD& SE & CP & &RB(\%) & SD & SE & CP\\
				\hline
				1  & Naive & -76.313 & 0.101 & 0.103 & 0.384 & & -78.370 & 0.099 & 0.101 & 0.352 & & -73.766 & 0.106 & 0.103 & 0.584 & & -79.609 & 0.103 & 0.100 & 0.506 \\
				& TRUE & 1.744 & 0.102 & 0.100 & 0.950 & & 2.443 & 0.098 & 0.100 & 0.956 & & 2.303 & 0.101 & 0.098 & 0.944 & & 2.099 & 0.100 & 0.098 & 0.950 \\
				& RCE & -2.696 & 0.323 & 0.326 & 0.950 & & -1.348 & 0.335 & 0.334 & 0.949 & & 3.034 & 0.327 & 0.324 & 0.954 & & -2.263 & 0.331 & 0.333 & 0.956 \\
				\\
				2  & Naive & -75.666 & 0.102 & 0.103 & 0.404 & & -79.083 & 0.099 & 0.101 & 0.344 & & -75.484 & 0.103 & 0.103 & 0.569 & & -76.395 & 0.101 & 0.100 & 0.542 \\
				& TRUE & 2.259 & 0.100 & 0.100 & 0.948 & & 0.987 & 0.100 & 0.100 & 0.948 & & 0.918 & 0.096 & 0.098 & 0.958 & & 1.332 & 0.096 & 0.098 & 0.957 \\
				& RCE & -0.188 & 0.356 & 0.368 & 0.963 & & -1.323 & 0.363 & 0.378 & 0.958 & & 0.303 & 0.367 & 0.367 & 0.959 & & 4.939 & 0.380 & 0.377 & 0.952 \\
				\\
				3  & Naive & -76.794 & 0.102 & 0.103 & 0.380 & & -79.028 & 0.101 & 0.101 & 0.344 & & -74.091 & 0.103 & 0.103 & 0.579 & & -76.446 & 0.102 & 0.100 & 0.539 \\
				& TRUE & 2.259 & 0.100 & 0.100 & 0.948 & & 0.987 & 0.100 & 0.100 & 0.948 & & 0.918 & 0.096 & 0.098 & 0.958 & & 1.332 & 0.096 & 0.098 & 0.957 \\
				& RCE & -2.532 & 0.432 & 0.447 & 0.959 & & -1.607 & 0.444 & 0.458 & 0.957 & & 3.263 & 0.443 & 0.446 & 0.950 & & 5.353 & 0.458 & 0.457 & 0.946 \\
				\\
				4  & Naive & -76.613 & 0.105 & 0.103 & 0.392 & & -78.869 & 0.098 & 0.100 & 0.342 & & -76.499 & 0.104 & 0.103 & 0.569 & & -79.097 & 0.101 & 0.100 & 0.525 \\
				& TRUE & 1.438 & 0.112 & 0.107 & 0.943 & & 2.228 & 0.109 & 0.107 & 0.946 & & 1.682 & 0.104 & 0.105 & 0.952 & & 0.829 & 0.107 & 0.105 & 0.949 \\
				& RCE & -1.051 & 0.483 & 0.468 & 0.944 & & 1.527 & 0.478 & 0.479 & 0.959 & & -5.945 & 0.462 & 0.467 & 0.953 & & -3.722 & 0.472 & 0.479 & 0.956 \\
				\\
				5  & Naive & -77.428 & 0.102 & 0.103 & 0.369 & & -79.094 & 0.100 & 0.100 & 0.340 & & -75.729 & 0.102 & 0.102 & 0.574 & & -78.466 & 0.099 & 0.100 & 0.532 \\
				& TRUE & 1.681 & 0.110 & 0.107 & 0.940 & & 2.080 & 0.109 & 0.107 & 0.940 & & 1.943 & 0.106 & 0.105 & 0.949 & & 2.691 & 0.102 & 0.105 & 0.958 \\
				& RCE & -3.981 & 0.524 & 0.529 & 0.951 & & -1.556 & 0.541 & 0.543 & 0.955 & & 0.383 & 0.526 & 0.529 & 0.953 & & 1.728 & 0.547 & 0.542 & 0.949 \\
				\\
				6  & Naive & -77.820 & 0.103 & 0.103 & 0.365 & & -79.186 & 0.101 & 0.100 & 0.345 & & -75.150 & 0.103 & 0.102 & 0.581 & & -78.417 & 0.099 & 0.100 & 0.531 \\
				& TRUE & 1.681 & 0.110 & 0.107 & 0.940 & & 2.080 & 0.109 & 0.107 & 0.940 & & 1.943 & 0.106 & 0.105 & 0.949 & & 2.691 & 0.102 & 0.105 & 0.958 \\
				& RCE & -1.173 & 0.650 & 0.643 & 0.944 & & 2.414 & 0.664 & 0.659 & 0.946 & & -2.787 & 0.650 & 0.642 & 0.956 & & -3.243 & 0.659 & 0.658 & 0.952 \\
				\\
				7  & Naive & -75.308 & 0.071 & 0.073 & 0.119 & & -78.286 & 0.070 & 0.071 & 0.087 & & -74.206 & 0.074 & 0.073 & 0.322 & & -78.031 & 0.069 & 0.071 & 0.237 \\
				& TRUE & -0.135 & 0.062 & 0.061 & 0.946 & & 0.625 & 0.060 & 0.061 & 0.953 & & 1.048 & 0.060 & 0.060 & 0.952 & & 0.995 & 0.060 & 0.060 & 0.949 \\
				& RCE & -2.554 & 0.206 & 0.207 & 0.955 & & -2.587 & 0.214 & 0.213 & 0.951 & & 1.958 & 0.210 & 0.207 & 0.949 & & 1.193 & 0.207 & 0.212 & 0.954 \\
				\\
				8  & Naive & -74.771 & 0.073 & 0.073 & 0.134 & & -80.273 & 0.071 & 0.071 & 0.074 & & -73.469 & 0.071 & 0.073 & 0.313 & & -78.214 & 0.071 & 0.071 & 0.248 \\
				& TRUE & 0.938 & 0.058 & 0.061 & 0.958 & & -0.243 & 0.060 & 0.061 & 0.955 & & 0.582 & 0.060 & 0.060 & 0.950 & & 0.340 & 0.060 & 0.060 & 0.947 \\
				& RCE & -2.252 & 0.238 & 0.235 & 0.951 & & -6.078 & 0.239 & 0.241 & 0.961 & & -0.076 & 0.234 & 0.234 & 0.946 & & -1.518 & 0.244 & 0.240 & 0.949 \\
				\\
				9  & Naive & -75.294 & 0.073 & 0.073 & 0.129 & & -80.362 & 0.070 & 0.071 & 0.082 & & -72.833 & 0.071 & 0.073 & 0.328 & & -78.047 & 0.071 & 0.071 & 0.256 \\
				& TRUE & 0.938 & 0.058 & 0.061 & 0.958 & & -0.243 & 0.060 & 0.061 & 0.955 & & 0.582 & 0.060 & 0.060 & 0.950 & & 0.340 & 0.060 & 0.060 & 0.947 \\
				& RCE & -1.452 & 0.288 & 0.285 & 0.945 & & -4.980 & 0.291 & 0.292 & 0.958 & & -1.032 & 0.286 & 0.285 & 0.941 & & -2.745 & 0.296 & 0.292 & 0.951 \\
				\hline 
			\end{tabular}
		\end{adjustbox}
	\end{center}
\end{table}

\begin{table}[h]
	\caption{\footnotesize Simulation results for regression coefficients of $X_1$ and $X_2$ at $G=1$ and $G=2$, i.e., $\beta_{11}$, $\beta_{21}$, $\beta_{12}$ and $\beta_{22}$ when ${\btheta}=(0, 0.5, 0.5, -0.1, 0.42, 0.42, 0.1, 0.4, 0.4, 0, 0.33, 0.33)^\top$ under $n=2000$ and different cases. RB (\%) = relative bias (percent), SD = Monte-Carlo standard 
		deviation, SE = Monte-Carlo standard error, CP = empirical 95\% coverage probability.}
	\label{simu:tabs5}
	\begin{center}
		\begin{adjustbox}{width=1\textwidth}
			\begin{tabular}{llrrrrrrrrrrrrrrrrrrr}
				\hline
				&& \multicolumn{9}{c}{$G=1$} & &\multicolumn{9}{c}{$G=2$} \\
				\cline{3-11}\cline{13-21}
				& &\multicolumn{4}{c}{$\hat{\beta}_{11}$}& 
				&\multicolumn{4}{c}{ $\hat{\beta}_{21}$} & &\multicolumn{4}{c}{ $\hat{\beta}_{12}$}
				& &\multicolumn{4}{c}{ $\hat{\beta}_{22}$}
				\\
				\cline{3-6}\cline{8-11}\cline{13-16}\cline{18-21} Case & Method & RB(\%) & SD & SE & CP & &RB(\%) & SD & SE & CP & & RB(\%) & SD& SE & CP & &RB(\%) & SD & SE & CP\\
				\hline
				1  & Naive & -74.794 & 0.095 & 0.095 & 0.029 & & -79.480 & 0.093 & 0.093 & 0.015 & & -75.612 & 0.093 & 0.095 & 0.088 & & -77.884 & 0.094 & 0.093 & 0.070 \\
				& TRUE & 1.510 & 0.076 & 0.077 & 0.953 & & 0.863 & 0.077 & 0.077 & 0.950 & & 0.197 & 0.076 & 0.075 & 0.942 & & 1.175 & 0.077 & 0.075 & 0.945 \\
				& RCE & -7.373 & 0.189 & 0.185 & 0.943 & & -8.895 & 0.194 & 0.190 & 0.936 & & -6.974 & 0.179 & 0.184 & 0.952 & & -6.155 & 0.189 & 0.189 & 0.952 \\
				\\
				2  & Naive & -76.836 & 0.096 & 0.095 & 0.025 & & -78.249 & 0.095 & 0.093 & 0.014 & & -74.161 & 0.094 & 0.095 & 0.094 & & -78.456 & 0.093 & 0.093 & 0.058 \\
				& TRUE & -0.021 & 0.078 & 0.077 & 0.948 & & 0.502 & 0.075 & 0.077 & 0.953 & & 0.924 & 0.075 & 0.075 & 0.951 & & 0.339 & 0.075 & 0.075 & 0.956 \\
				& RCE & -9.387 & 0.211 & 0.209 & 0.938 & & -7.520 & 0.220 & 0.214 & 0.932 & & -5.752 & 0.207 & 0.209 & 0.957 & & -7.019 & 0.217 & 0.214 & 0.947 \\
				\\
				3  & Naive & -76.858 & 0.096 & 0.095 & 0.020 & & -78.738 & 0.094 & 0.093 & 0.011 & & -74.205 & 0.094 & 0.095 & 0.090 & & -77.867 & 0.093 & 0.093 & 0.062 \\
				& TRUE & -0.021 & 0.078 & 0.077 & 0.948 & & 0.502 & 0.075 & 0.077 & 0.953 & & 0.924 & 0.075 & 0.075 & 0.951 & & 0.339 & 0.075 & 0.075 & 0.956 \\
				& RCE & -8.773 & 0.253 & 0.254 & 0.943 & & -7.219 & 0.265 & 0.260 & 0.939 & & -6.641 & 0.248 & 0.254 & 0.957 & & -7.474 & 0.263 & 0.260 & 0.944 \\
				\\
				4  & Naive & -75.300 & 0.094 & 0.095 & 0.028 & & -78.685 & 0.090 & 0.093 & 0.010 & & -76.706 & 0.095 & 0.095 & 0.082 & & -77.818 & 0.093 & 0.093 & 0.067 \\
				& TRUE & 1.132 & 0.077 & 0.076 & 0.948 & & 0.404 & 0.076 & 0.076 & 0.950 & & 0.069 & 0.073 & 0.074 & 0.955 & & 0.817 & 0.074 & 0.074 & 0.953 \\
				& RCE & -7.064 & 0.195 & 0.198 & 0.952 & & -7.360 & 0.196 & 0.203 & 0.956 & & -9.318 & 0.198 & 0.198 & 0.942 & & -7.424 & 0.205 & 0.203 & 0.949 \\
				\\
				5  & Naive & -76.278 & 0.097 & 0.095 & 0.026 & & -79.379 & 0.095 & 0.093 & 0.011 & & -73.745 & 0.095 & 0.095 & 0.096 & & -78.686 & 0.094 & 0.093 & 0.060 \\
				& TRUE & -0.050 & 0.076 & 0.076 & 0.952 & & 0.559 & 0.078 & 0.076 & 0.946 & & 0.934 & 0.077 & 0.074 & 0.948 & & -0.163 & 0.076 & 0.074 & 0.943 \\
				& RCE & -9.784 & 0.225 & 0.224 & 0.943 & & -9.675 & 0.233 & 0.230 & 0.938 & & -4.728 & 0.223 & 0.224 & 0.952 & & -6.299 & 0.229 & 0.230 & 0.949 \\
				\\
				6  & Naive & -76.622 & 0.097 & 0.095 & 0.025 & & -79.650 & 0.095 & 0.093 & 0.014 & & -73.427 & 0.096 & 0.095 & 0.098 & & -78.303 & 0.093 & 0.093 & 0.061 \\
				& TRUE & -0.050 & 0.076 & 0.076 & 0.952 & & 0.559 & 0.078 & 0.076 & 0.946 & & 0.934 & 0.077 & 0.074 & 0.948 & & -0.163 & 0.076 & 0.074 & 0.943 \\
				& RCE & -10.231 & 0.274 & 0.272 & 0.950 & & -9.717 & 0.278 & 0.279 & 0.955 & & -4.341 & 0.272 & 0.272 & 0.951 & & -6.232 & 0.274 & 0.279 & 0.956 \\
				\\
				7  & Naive & -76.135 & 0.111 & 0.113 & 0.082 & & -78.776 & 0.106 & 0.110 & 0.054 & & -75.620 & 0.113 & 0.112 & 0.204 & & -78.217 & 0.111 & 0.109 & 0.162 \\
				& TRUE & 1.948 & 0.092 & 0.092 & 0.949 & & 1.225 & 0.090 & 0.092 & 0.957 & & 0.581 & 0.088 & 0.089 & 0.956 & & 0.892 & 0.088 & 0.089 & 0.951 \\
				& RCE & -7.693 & 0.269 & 0.270 & 0.949 & & -6.980 & 0.273 & 0.277 & 0.947 & & -5.078 & 0.268 & 0.269 & 0.955 & & -4.206 & 0.276 & 0.276 & 0.959 \\
				\\
				8  & Naive & -76.921 & 0.114 & 0.113 & 0.078 & & -79.290 & 0.115 & 0.110 & 0.058 & & -75.974 & 0.114 & 0.112 & 0.182 & & -78.841 & 0.111 & 0.109 & 0.151 \\
				& TRUE & 0.388 & 0.091 & 0.092 & 0.949 & & 0.444 & 0.088 & 0.092 & 0.960 & & 0.647 & 0.089 & 0.089 & 0.955 & & 0.403 & 0.091 & 0.089 & 0.954 \\
				& RCE & -9.235 & 0.307 & 0.305 & 0.950 & & -8.683 & 0.324 & 0.313 & 0.938 & & -9.181 & 0.306 & 0.304 & 0.947 & & -8.881 & 0.313 & 0.312 & 0.951 \\
				\\
				9  & Naive & -76.999 & 0.115 & 0.113 & 0.075 & & -79.539 & 0.114 & 0.110 & 0.062 & & -75.958 & 0.115 & 0.112 & 0.194 & & -78.573 & 0.112 & 0.109 & 0.152 \\
				& TRUE & 0.388 & 0.091 & 0.092 & 0.949 & & 0.444 & 0.088 & 0.092 & 0.960 & & 0.647 & 0.089 & 0.089 & 0.955 & & 0.403 & 0.091 & 0.089 & 0.954 \\
				& RCE & -7.272 & 0.372 & 0.370 & 0.954 & & -7.289 & 0.389 & 0.380 & 0.941 & & -11.783 & 0.372 & 0.369 & 0.955 & & -10.809 & 0.380 & 0.379 & 0.954 \\
				\hline 
			\end{tabular}
		\end{adjustbox}
	\end{center}
\end{table}

\begin{table}[h]
	\caption{\footnotesize Simulation results for regression coefficients of $X_1$ and $X_2$ at $G=3$ and $G=4$, i.e., $\beta_{13}$, $\beta_{23}$, $\beta_{14}$ and $\beta_{24}$ when  ${\btheta}=(0, 0.5, 0.5, -0.1, 0.42, 0.42, 0.1, 0.4, 0.4, 0, 0.33, 0.33)^\top$ under $n=2000$ and different cases. RB (\%) = relative bias (percent), SD = Monte-Carlo standard 
		deviation, SE = Monte-Carlo standard error, CP = empirical 95\% coverage probability.}
	\label{simu:tabs6}
	\begin{center}
		\begin{adjustbox}{width=1\textwidth}
			\begin{tabular}{llrrrrrrrrrrrrrrrrrrr}
				\hline
				&& \multicolumn{9}{c}{$G=3$} & &\multicolumn{9}{c}{$G=4$} \\
				\cline{3-11}\cline{13-21}
				& &\multicolumn{4}{c}{$\hat{\beta}_{13}$}& 
				&\multicolumn{4}{c}{ $\hat{\beta}_{23}$} & &\multicolumn{4}{c}{ $\hat{\beta}_{14}$}
				& &\multicolumn{4}{c}{ $\hat{\beta}_{24}$}
				\\
				\cline{3-6}\cline{8-11}\cline{13-16}\cline{18-21} Case & Method & RB(\%) & SD & SE & CP & &RB(\%) & SD & SE & CP & & RB(\%) & SD& SE & CP & &RB(\%) & SD & SE & CP\\
				\hline
				1  & Naive & -74.082 & 0.179 & 0.182 & 0.625 & & -75.812 & 0.178 & 0.177 & 0.598 & & -72.223 & 0.186 & 0.182 & 0.741 & & -74.420 & 0.175 & 0.177 & 0.717 \\
				& TRUE & 2.352 & 0.185 & 0.182 & 0.945 & & 5.091 & 0.184 & 0.182 & 0.956 & & 2.894 & 0.179 & 0.177 & 0.952 & & 3.235 & 0.181 & 0.177 & 0.959 \\
				& RCE & -10.585 & 0.571 & 0.574 & 0.956 & & -6.868 & 0.597 & 0.590 & 0.953 & & -1.453 & 0.584 & 0.574 & 0.947 & & 0.044 & 0.582 & 0.588 & 0.956 \\
				\\
				2  & Naive & -72.589 & 0.185 & 0.182 & 0.636 & & -76.541 & 0.185 & 0.178 & 0.586 & & -69.153 & 0.182 & 0.181 & 0.753 & & -76.659 & 0.179 & 0.176 & 0.699 \\
				& TRUE & 1.037 & 0.180 & 0.181 & 0.953 & & 2.672 & 0.184 & 0.181 & 0.949 & & 5.094 & 0.178 & 0.177 & 0.956 & & 0.724 & 0.188 & 0.177 & 0.941 \\
				& RCE & -2.719 & 0.647 & 0.651 & 0.958 & & -4.411 & 0.676 & 0.668 & 0.950 & & 1.165 & 0.653 & 0.649 & 0.948 & & -4.111 & 0.673 & 0.665 & 0.949 \\
				\\
				3  & Naive & -72.822 & 0.188 & 0.182 & 0.633 & & -76.952 & 0.188 & 0.178 & 0.584 & & -69.014 & 0.183 & 0.181 & 0.748 & & -76.034 & 0.178 & 0.177 & 0.707 \\
				& TRUE & 1.037 & 0.180 & 0.181 & 0.953 & & 2.672 & 0.184 & 0.181 & 0.949 & & 5.094 & 0.178 & 0.177 & 0.956 & & 0.724 & 0.188 & 0.177 & 0.941 \\
				& RCE & -2.571 & 0.796 & 0.790 & 0.953 & & -3.853 & 0.834 & 0.811 & 0.943 & & 0.876 & 0.803 & 0.788 & 0.950 & & -4.545 & 0.818 & 0.808 & 0.949 \\
				\\
				4  & Naive & -72.026 & 0.186 & 0.181 & 0.627 & & -74.916 & 0.183 & 0.177 & 0.589 & & -71.690 & 0.185 & 0.181 & 0.728 & & -73.929 & 0.183 & 0.176 & 0.696 \\
				& TRUE & 2.203 & 0.203 & 0.196 & 0.950 & & 4.120 & 0.203 & 0.196 & 0.952 & & 4.838 & 0.193 & 0.191 & 0.961 & & 4.289 & 0.199 & 0.191 & 0.947 \\
				& RCE & -6.550 & 0.833 & 0.828 & 0.949 & & -4.619 & 0.859 & 0.849 & 0.950 & & -0.456 & 0.842 & 0.825 & 0.952 & & -0.402 & 0.880 & 0.845 & 0.944 \\
				\\
				5  & Naive & -71.512 & 0.182 & 0.181 & 0.630 & & -75.736 & 0.177 & 0.176 & 0.597 & & -67.595 & 0.180 & 0.181 & 0.764 & & -75.887 & 0.176 & 0.176 & 0.696 \\
				& TRUE & 3.091 & 0.205 & 0.196 & 0.947 & & 3.655 & 0.199 & 0.196 & 0.955 & & 4.889 & 0.197 & 0.191 & 0.949 & & 2.728 & 0.196 & 0.191 & 0.955 \\
				& RCE & 0.549 & 0.945 & 0.937 & 0.951 & & -0.117 & 0.965 & 0.961 & 0.952 & & 3.822 & 0.960 & 0.934 & 0.943 & & -3.691 & 0.980 & 0.957 & 0.949 \\
				\\
				6  & Naive & -71.142 & 0.181 & 0.181 & 0.635 & & -76.725 & 0.179 & 0.176 & 0.583 & & -67.936 & 0.181 & 0.181 & 0.767 & & -74.762 & 0.176 & 0.176 & 0.708 \\
				& TRUE & 3.091 & 0.205 & 0.196 & 0.947 & & 3.655 & 0.199 & 0.196 & 0.955 & & 4.889 & 0.197 & 0.191 & 0.949 & & 2.728 & 0.196 & 0.191 & 0.955 \\
				& RCE & 3.835 & 1.149 & 1.137 & 0.951 & & -0.807 & 1.187 & 1.166 & 0.953 & & 0.385 & 1.167 & 1.135 & 0.948 & & -2.925 & 1.187 & 1.162 & 0.954 \\
				\\
				7  & Naive & -72.508 & 0.126 & 0.127 & 0.370 & & -77.157 & 0.124 & 0.124 & 0.297 & & -71.956 & 0.128 & 0.127 & 0.528 & & -76.828 & 0.122 & 0.124 & 0.452 \\
				& TRUE & 1.824 & 0.111 & 0.109 & 0.947 & & 0.932 & 0.110 & 0.109 & 0.952 & & 1.470 & 0.104 & 0.106 & 0.957 & & 1.591 & 0.102 & 0.106 & 0.965 \\
				& RCE & -3.982 & 0.360 & 0.363 & 0.955 & & -5.848 & 0.376 & 0.372 & 0.948 & & -3.493 & 0.367 & 0.363 & 0.945 & & -5.646 & 0.369 & 0.371 & 0.956 \\
				\\
				8  & Naive & -73.848 & 0.127 & 0.127 & 0.361 & & -76.237 & 0.124 & 0.124 & 0.312 & & -72.122 & 0.127 & 0.127 & 0.539 & & -76.165 & 0.125 & 0.124 & 0.466 \\
				& TRUE & 1.160 & 0.110 & 0.109 & 0.950 & & 2.589 & 0.106 & 0.109 & 0.957 & & 1.952 & 0.106 & 0.106 & 0.954 & & 0.084 & 0.108 & 0.106 & 0.950 \\
				& RCE & -3.335 & 0.409 & 0.411 & 0.949 & & -2.088 & 0.420 & 0.421 & 0.953 & & -3.574 & 0.409 & 0.411 & 0.953 & & -4.642 & 0.420 & 0.421 & 0.953 \\
				\\
				9  & Naive & -74.716 & 0.127 & 0.127 & 0.342 & & -76.196 & 0.125 & 0.124 & 0.316 & & -71.207 & 0.127 & 0.127 & 0.543 & & -76.058 & 0.125 & 0.124 & 0.469 \\
				& TRUE & 1.160 & 0.110 & 0.109 & 0.950 & & 2.589 & 0.106 & 0.109 & 0.957 & & 1.952 & 0.106 & 0.106 & 0.954 & & 0.084 & 0.108 & 0.106 & 0.950 \\
				& RCE & -4.868 & 0.496 & 0.499 & 0.951 & & -1.287 & 0.514 & 0.512 & 0.946 & & -1.707 & 0.498 & 0.499 & 0.949 & & -5.230 & 0.515 & 0.511 & 0.949 \\
				\hline 
			\end{tabular}
		\end{adjustbox}
	\end{center}	
\end{table}

\begin{table}[h]
	\caption{\footnotesize Simulation results for regression coefficients of $X_1$ and $X_2$ at $G=1$ and $G=2$, i.e., $\beta_{11}$, $\beta_{21}$, $\beta_{12}$ and $\beta_{22}$ when ${\btheta}=(0, 0.5, 0.5, -0.1, 0.42, 0.42, 0.1, 0.4, 0.4, 0, 0.33, 0.33)^\top$ under $n=6000$ and different cases. RB (\%) = relative bias (percent), SD = Monte-Carlo standard 
		deviation, SE = Monte-Carlo standard error, CP = empirical 95\% coverage probability.}
	\label{simu:tabs7}
	\begin{center}
		\begin{adjustbox}{width=1\textwidth}
			\begin{tabular}{llrrrrrrrrrrrrrrrrrrr}
				\hline
				&& \multicolumn{9}{c}{$G=1$} & &\multicolumn{9}{c}{$G=2$} \\
				\cline{3-11}\cline{13-21}
				& &\multicolumn{4}{c}{$\hat{\beta}_{11}$}& 
				&\multicolumn{4}{c}{ $\hat{\beta}_{21}$} & &\multicolumn{4}{c}{ $\hat{\beta}_{12}$}
				& &\multicolumn{4}{c}{ $\hat{\beta}_{22}$}
				\\
				\cline{3-6}\cline{8-11}\cline{13-16}\cline{18-21} Case & Method & RB(\%) & SD & SE & CP & &RB(\%) & SD & SE & CP & & RB(\%) & SD& SE & CP & &RB(\%) & SD & SE & CP\\
				\hline
				1  & Naive & -75.943 & 0.056 & 0.055 & 0.000 & & -78.572 & 0.053 & 0.053 & 0.000 & & -75.454 & 0.055 & 0.055 & 0.000 & & -78.060 & 0.055 & 0.053 & 0.000 \\
				& TRUE & 0.226 & 0.044 & 0.044 & 0.952 & & 0.309 & 0.044 & 0.044 & 0.950 & & 0.083 & 0.042 & 0.043 & 0.955 & & 0.370 & 0.043 & 0.043 & 0.955 \\
				& RCE & -8.807 & 0.110 & 0.106 & 0.921 & & -8.355 & 0.111 & 0.109 & 0.929 & & -6.872 & 0.107 & 0.106 & 0.941 & & -6.371 & 0.112 & 0.109 & 0.940 \\
				\\
				2  & Naive & -76.001 & 0.053 & 0.055 & 0.000 & & -79.564 & 0.053 & 0.053 & 0.000 & & -74.645 & 0.055 & 0.055 & 0.000 & & -78.089 & 0.054 & 0.053 & 0.000 \\
				& TRUE & 0.262 & 0.044 & 0.044 & 0.962 & & -0.064 & 0.044 & 0.044 & 0.947 & & 0.039 & 0.043 & 0.043 & 0.951 & & -0.111 & 0.043 & 0.043 & 0.948 \\
				& RCE & -9.359 & 0.118 & 0.120 & 0.932 & & -9.913 & 0.123 & 0.123 & 0.924 & & -6.355 & 0.120 & 0.120 & 0.948 & & -6.485 & 0.123 & 0.123 & 0.943 \\
				\\
				3  & Naive & -76.373 & 0.053 & 0.055 & 0.000 & & -79.748 & 0.053 & 0.053 & 0.000 & & -74.222 & 0.055 & 0.055 & 0.000 & & -77.856 & 0.054 & 0.053 & 0.000 \\
				& TRUE & 0.262 & 0.044 & 0.044 & 0.962 & & -0.064 & 0.044 & 0.044 & 0.947 & & 0.039 & 0.043 & 0.043 & 0.951 & & -0.111 & 0.043 & 0.043 & 0.948 \\
				& RCE & -9.290 & 0.145 & 0.146 & 0.946 & & -9.685 & 0.151 & 0.150 & 0.935 & & -6.464 & 0.148 & 0.146 & 0.943 & & -6.747 & 0.151 & 0.150 & 0.942 \\
				\\
				4  & Naive & -75.892 & 0.055 & 0.055 & 0.000 & & -79.063 & 0.053 & 0.053 & 0.000 & & -75.516 & 0.054 & 0.055 & 0.000 & & -78.158 & 0.054 & 0.053 & 0.000 \\
				& TRUE & 0.138 & 0.044 & 0.044 & 0.947 & & 0.338 & 0.045 & 0.044 & 0.943 & & 0.092 & 0.042 & 0.042 & 0.957 & & 0.389 & 0.042 & 0.042 & 0.956 \\
				& RCE & -9.122 & 0.114 & 0.114 & 0.930 & & -9.274 & 0.117 & 0.117 & 0.930 & & -6.879 & 0.109 & 0.114 & 0.953 & & -6.451 & 0.114 & 0.117 & 0.947 \\
				\\
				5  & Naive & -76.221 & 0.055 & 0.055 & 0.000 & & -79.195 & 0.053 & 0.053 & 0.000 & & -75.073 & 0.054 & 0.055 & 0.000 & & -78.086 & 0.053 & 0.053 & 0.000 \\
				& TRUE & 0.150 & 0.044 & 0.044 & 0.954 & & 0.553 & 0.044 & 0.044 & 0.949 & & 0.135 & 0.042 & 0.042 & 0.958 & & 0.026 & 0.043 & 0.042 & 0.953 \\
				& RCE & -8.810 & 0.131 & 0.129 & 0.931 & & -8.806 & 0.133 & 0.132 & 0.935 & & -7.306 & 0.130 & 0.129 & 0.940 & & -7.146 & 0.134 & 0.132 & 0.940 \\
				\\
				6  & Naive & -76.525 & 0.055 & 0.055 & 0.000 & & -79.481 & 0.053 & 0.053 & 0.000 & & -74.725 & 0.054 & 0.055 & 0.000 & & -77.745 & 0.053 & 0.053 & 0.000 \\
				& TRUE & 0.150 & 0.044 & 0.044 & 0.954 & & 0.553 & 0.044 & 0.044 & 0.949 & & 0.135 & 0.042 & 0.042 & 0.958 & & 0.026 & 0.043 & 0.042 & 0.953 \\
				& RCE & -8.458 & 0.161 & 0.156 & 0.932 & & -8.538 & 0.162 & 0.161 & 0.944 & & -7.768 & 0.158 & 0.156 & 0.942 & & -7.499 & 0.161 & 0.160 & 0.942 \\
				\\
				7  & Naive & -76.406 & 0.065 & 0.065 & 0.000 & & -79.031 & 0.061 & 0.063 & 0.000 & & -75.976 & 0.066 & 0.065 & 0.003 & & -78.636 & 0.065 & 0.063 & 0.000 \\
				& TRUE & 0.374 & 0.053 & 0.053 & 0.953 & & 0.423 & 0.052 & 0.053 & 0.955 & & 0.354 & 0.050 & 0.051 & 0.951 & & 0.290 & 0.051 & 0.051 & 0.958 \\
				& RCE & -9.033 & 0.157 & 0.155 & 0.938 & & -8.554 & 0.157 & 0.159 & 0.952 & & -6.837 & 0.159 & 0.154 & 0.946 & & -6.175 & 0.164 & 0.158 & 0.943 \\
				\\
				8  & Naive & -76.188 & 0.065 & 0.065 & 0.000 & & -79.505 & 0.064 & 0.063 & 0.000 & & -75.416 & 0.064 & 0.065 & 0.004 & & -78.616 & 0.064 & 0.063 & 0.000 \\
				& TRUE & 0.437 & 0.053 & 0.053 & 0.950 & & 0.360 & 0.053 & 0.053 & 0.946 & & -0.113 & 0.052 & 0.051 & 0.948 & & -0.339 & 0.052 & 0.051 & 0.950 \\
				& RCE & -6.899 & 0.176 & 0.175 & 0.943 & & -7.512 & 0.181 & 0.179 & 0.943 & & -7.311 & 0.172 & 0.175 & 0.957 & & -7.234 & 0.179 & 0.179 & 0.950 \\
				\\
				9  & Naive & -76.567 & 0.066 & 0.065 & 0.000 & & -79.862 & 0.064 & 0.063 & 0.000 & & -74.963 & 0.065 & 0.065 & 0.004 & & -78.220 & 0.063 & 0.063 & 0.000 \\
				& TRUE & 0.437 & 0.053 & 0.053 & 0.950 & & 0.360 & 0.053 & 0.053 & 0.946 & & -0.113 & 0.052 & 0.051 & 0.948 & & -0.339 & 0.052 & 0.051 & 0.950 \\
				& RCE & -6.819 & 0.215 & 0.212 & 0.941 & & -7.693 & 0.219 & 0.218 & 0.949 & & -7.451 & 0.211 & 0.212 & 0.952 & & -7.115 & 0.217 & 0.218 & 0.953 \\
				\hline 
			\end{tabular}
		\end{adjustbox}
	\end{center}
\end{table}

\begin{table}[h]
	\caption{\footnotesize Simulation results for regression coefficients of $X_1$ and $X_2$ at $G=3$ and $G=4$, i.e., $\beta_{13}$, $\beta_{23}$, $\beta_{14}$ and $\beta_{24}$ when  ${\btheta}=(0, 0.5, 0.5, -0.1, 0.42, 0.42, 0.1, 0.4, 0.4, 0, 0.33, 0.33)^\top$ under $n=6000$ and different cases. RB (\%) = relative bias (percent), SD = Monte-Carlo standard 
		deviation, SE = Monte-Carlo standard error, CP = empirical 95\% coverage probability.}
	\label{simu:tabs8}
	\begin{center}
		\begin{adjustbox}{width=1\textwidth}
			\begin{tabular}{llrrrrrrrrrrrrrrrrrrr}
				\hline
				&& \multicolumn{9}{c}{$G=3$} & &\multicolumn{9}{c}{$G=4$} \\
				\cline{3-11}\cline{13-21}
				& &\multicolumn{4}{c}{$\hat{\beta}_{13}$}& 
				&\multicolumn{4}{c}{ $\hat{\beta}_{23}$} & &\multicolumn{4}{c}{ $\hat{\beta}_{14}$}
				& &\multicolumn{4}{c}{ $\hat{\beta}_{24}$}
				\\
				\cline{3-6}\cline{8-11}\cline{13-16}\cline{18-21} Case & Method & RB(\%) & SD & SE & CP & &RB(\%) & SD & SE & CP & & RB(\%) & SD& SE & CP & &RB(\%) & SD & SE & CP\\
				\hline
				1  & Naive & -72.969 & 0.104 & 0.103 & 0.185 & & -76.628 & 0.099 & 0.101 & 0.139 & & -72.103 & 0.102 & 0.103 & 0.368 & & -74.360 & 0.098 & 0.101 & 0.311 \\
				& TRUE & 0.525 & 0.101 & 0.102 & 0.961 & & 0.253 & 0.104 & 0.102 & 0.941 & & 2.247 & 0.099 & 0.100 & 0.957 & & 1.903 & 0.097 & 0.100 & 0.955 \\
				& RCE & -7.160 & 0.333 & 0.327 & 0.944 & & -7.848 & 0.334 & 0.335 & 0.947 & & -0.916 & 0.322 & 0.326 & 0.960 & & 0.765 & 0.326 & 0.334 & 0.955 \\
				\\
				2  & Naive & -72.923 & 0.104 & 0.103 & 0.196 & & -76.127 & 0.099 & 0.101 & 0.146 & & -70.993 & 0.105 & 0.103 & 0.375 & & -75.724 & 0.101 & 0.101 & 0.300 \\
				& TRUE & 1.315 & 0.102 & 0.103 & 0.953 & & 1.812 & 0.107 & 0.103 & 0.940 & & 1.214 & 0.104 & 0.100 & 0.948 & & 0.256 & 0.100 & 0.100 & 0.951 \\
				& RCE & -2.536 & 0.371 & 0.370 & 0.952 & & -1.475 & 0.373 & 0.379 & 0.951 & & -3.759 & 0.372 & 0.369 & 0.948 & & -5.559 & 0.376 & 0.378 & 0.951 \\
				\\
				3  & Naive & -73.409 & 0.104 & 0.103 & 0.191 & & -76.705 & 0.098 & 0.101 & 0.138 & & -70.345 & 0.104 & 0.103 & 0.381 & & -75.130 & 0.101 & 0.101 & 0.315 \\
				& TRUE & 1.315 & 0.102 & 0.103 & 0.953 & & 1.812 & 0.107 & 0.103 & 0.940 & & 1.214 & 0.104 & 0.100 & 0.948 & & 0.256 & 0.100 & 0.100 & 0.951 \\
				& RCE & -3.590 & 0.448 & 0.449 & 0.953 & & -2.630 & 0.450 & 0.460 & 0.954 & & -2.494 & 0.448 & 0.449 & 0.946 & & -4.454 & 0.459 & 0.460 & 0.952 \\
				\\
				4  & Naive & -72.531 & 0.106 & 0.103 & 0.205 & & -75.301 & 0.101 & 0.100 & 0.154 & & -71.259 & 0.100 & 0.103 & 0.371 & & -73.660 & 0.098 & 0.100 & 0.308 \\
				& TRUE & 1.240 & 0.112 & 0.110 & 0.945 & & 1.759 & 0.110 & 0.110 & 0.955 & & 1.745 & 0.112 & 0.108 & 0.944 & & 1.731 & 0.108 & 0.108 & 0.956 \\
				& RCE & -7.425 & 0.470 & 0.471 & 0.952 & & -5.592 & 0.475 & 0.482 & 0.954 & & -1.021 & 0.466 & 0.470 & 0.951 & & 1.071 & 0.483 & 0.481 & 0.942 \\
				\\
				5  & Naive & -72.745 & 0.103 & 0.103 & 0.196 & & -75.544 & 0.099 & 0.100 & 0.150 & & -70.497 & 0.103 & 0.103 & 0.368 & & -73.727 & 0.099 & 0.100 & 0.314 \\
				& TRUE & 0.837 & 0.110 & 0.110 & 0.950 & & 1.523 & 0.110 & 0.110 & 0.956 & & 1.744 & 0.110 & 0.108 & 0.952 & & 1.184 & 0.107 & 0.108 & 0.955 \\
				& RCE & -6.739 & 0.547 & 0.533 & 0.946 & & -5.201 & 0.548 & 0.546 & 0.946 & & -0.331 & 0.538 & 0.532 & 0.952 & & 0.511 & 0.546 & 0.545 & 0.960 \\
				\\
				6  & Naive & -73.363 & 0.103 & 0.103 & 0.182 & & -75.903 & 0.099 & 0.100 & 0.144 & & -69.731 & 0.104 & 0.103 & 0.384 & & -73.237 & 0.100 & 0.100 & 0.319 \\
				& TRUE & 0.837 & 0.110 & 0.110 & 0.950 & & 1.523 & 0.110 & 0.110 & 0.956 & & 1.744 & 0.110 & 0.108 & 0.952 & & 1.184 & 0.107 & 0.108 & 0.955 \\
				& RCE & -9.666 & 0.668 & 0.646 & 0.943 & & -7.268 & 0.675 & 0.662 & 0.944 & & 3.563 & 0.658 & 0.646 & 0.948 & & 3.430 & 0.666 & 0.662 & 0.951 \\
				\\
				7  & Naive & -72.914 & 0.074 & 0.073 & 0.024 & & -76.401 & 0.071 & 0.071 & 0.012 & & -72.284 & 0.075 & 0.073 & 0.101 & & -76.701 & 0.071 & 0.071 & 0.050 \\
				& TRUE & 1.202 & 0.061 & 0.062 & 0.957 & & 1.417 & 0.061 & 0.062 & 0.957 & & 0.362 & 0.061 & 0.061 & 0.948 & & 0.118 & 0.061 & 0.061 & 0.951 \\
				& RCE & -3.087 & 0.207 & 0.208 & 0.957 & & -3.349 & 0.211 & 0.214 & 0.959 & & -3.470 & 0.213 & 0.208 & 0.950 & & -4.967 & 0.213 & 0.213 & 0.949 \\
				\\
				8  & Naive & -74.562 & 0.073 & 0.073 & 0.022 & & -76.722 & 0.072 & 0.071 & 0.010 & & -71.975 & 0.073 & 0.073 & 0.107 & & -75.930 & 0.070 & 0.071 & 0.056 \\
				& TRUE & -0.137 & 0.063 & 0.062 & 0.948 & & 0.528 & 0.062 & 0.062 & 0.958 & & 0.799 & 0.061 & 0.061 & 0.950 & & 0.816 & 0.061 & 0.061 & 0.948 \\
				& RCE & -8.341 & 0.235 & 0.236 & 0.947 & & -6.354 & 0.243 & 0.242 & 0.949 & & -2.842 & 0.241 & 0.235 & 0.949 & & -3.670 & 0.245 & 0.241 & 0.955 \\
				\\
				9  & Naive & -75.163 & 0.073 & 0.073 & 0.021 & & -76.897 & 0.072 & 0.071 & 0.011 & & -71.288 & 0.074 & 0.073 & 0.108 & & -75.669 & 0.072 & 0.071 & 0.064 \\
				& TRUE & -0.137 & 0.063 & 0.062 & 0.948 & & 0.528 & 0.062 & 0.062 & 0.958 & & 0.799 & 0.061 & 0.061 & 0.950 & & 0.816 & 0.061 & 0.061 & 0.948 \\
				& RCE & -9.189 & 0.283 & 0.286 & 0.954 & & -5.934 & 0.292 & 0.294 & 0.956 & & -1.832 & 0.290 & 0.286 & 0.951 & & -4.056 & 0.297 & 0.293 & 0.946 \\
				\hline 
			\end{tabular}
		\end{adjustbox}
	\end{center}	
\end{table}

\section{Sensitivity analyses for real data analysis}
In this section, we consider two different cut-off points for real data analysis. The first one is $(c_1,c_2)=(15,50)$, where only the second cut-off point is changed to 50, which is equal to the 85\% percentile of alcohol consumption. The second one is  $(c_1,c_2)=(17.6,58.1)$, where both cut-off points are changed. Under this situation, cut-off points 17.6 and 58.1 fall around the 45th percentile and the 90th percentile of alcohol consumption, respectively. Corresponding results are presented in Tables S9 and S10, respectively.

\begin{table}
	\footnotesize
	\caption{Sensitivity analyses: Estimated regression coefficient, standard error, and P-value of Logistic regression using different methods of estimation based on model (9) of main text and new cut-off points $(15,50)$ for male.}	
	\begin{center}
		\begin{adjustbox}{width=1\textwidth}
			\begin{tabular}{lclrrrrrrr}
				\hline
				& & &   \multicolumn{3}{c}{Naive estimation}& &\multicolumn{3}{c}{RC estimation}  \\
				\cline{4-6}\cline{8-10} PA & Alcohol (g per day) & Parameter &Estimate & SE & P-value & & Estimate & SE & P-value \\
				\hline
				Inactive & $<15$ & $\beta_{01}$      &-2.259 & 0.224 & $<0.001$ &  
				& -2.231  & 0.230 &  $<0.001$ \\ 
				&   & $\beta_{protein}$ & 0.287 & 0.085 & 0.001 &  & 0.313 & 0.138 & 0.023 \\ 
				&   & $\beta_{energy}$  &-0.212 & 0.083 & 0.010 &  & -0.089 & 0.194 & 0.646 \\ 
				\\
				Inactive & [15,50]  & $\beta_{02}$      &-2.273 & 0.226 &  $<0.001$ &  & -2.262 & 0.233 &  $<0.001$ \\ 
				&   & $\beta_{protein}$ & 0.311 & 0.099 & 0.002 &  & 0.294 & 0.159 & 0.065 \\
				&  & $\beta_{energy}$  &-0.273 & 0.100 & 0.006 &  & -0.194 & 0.235 & 0.408 \\ 
				\\
				Inactive & $>$50 & $\beta_{03}$      &-2.180 & 0.245 &  $<0.001$ &  & -2.023 & 0.421 & $<0.001$ \\ 
				&  & $\beta_{protein}$ &-0.066 & 0.140 & 0.636 &  & -0.330 & 0.557 & 0.553 \\ 
				&   & $\beta_{energy}$  & 0.182 & 0.156 & 0.243 &  & 1.477 & 0.939 & 0.116 \\ 
				\\
				Active &  $<15$ & $\beta_{04}$      &-2.506 & 0.229 &  $<0.001$ &  & -2.828 & 0.348 &  $<0.001$ \\
				&  & $\beta_{protein}$ & 0.369 & 0.158 & 0.019 &  & -0.084 & 0.577 & 0.884 \\ 
				&  & $\beta_{energy}$  &-0.470 & 0.149 & 0.002 &  & -1.448 & 0.761 & 0.057 \\ 
				\\
				Active & [15,50] & $\beta_{05}$      &-2.514 & 0.237 &  $<0.001$ &  & -2.907 & 0.388 &  $<0.001$ \\
				&   & $\beta_{protein}$ & 0.349 & 0.193 & 0.071 &  & -0.010 & 0.676 & 0.988 \\ 
				&  & $\beta_{energy}$  &-0.410 & 0.192 & 0.032 &  & -0.835 & 0.957 & 0.383 \\
				\\
				Active & $>$50 & $\beta_{06}$      &-2.228 & 0.298 &  $<0.001$ &  &  -1.204 & 1.614 & 0.456 \\ 
				&  & $\beta_{protein}$ & 0.319 & 0.249 & 0.200 &  & 2.025 & 2.369 & 0.393 \\
				& & $\beta_{energy}$  &-0.210 & 0.296 & 0.478 &  & 0.428 & 4.131 & 0.917 \\
				\\ 
				&   & $\alpha_{age}$    & 0.048 & 0.004 &  $<0.001$ &  & 0.048 & 0.004 &  $<0.001$ \\ 
				&   & $\alpha_{edu}$    &-0.493 & 0.063 &  $<0.001$ &  & -0.493 & 0.063 &  $<0.001$ \\ 
				&   & $\alpha_{smoke}$  & 0.161 & 0.060 & 0.007 &  & 0.161 & 0.060 & 0.007 \\ 
				\hline				
			\end{tabular}
		\end{adjustbox}
	\end{center}	
	{\raggedright  Note: SE is standard error.} 
\end{table}

\begin{table}
	\footnotesize
	\caption{Sensitivity analyses: Estimated regression coefficient, standard error, and P-value of Logistic regression using different methods of estimation based on model (9) of main text and new cut-off points $(17.6,58.1)$ for male.}	
	\begin{center}
		\begin{adjustbox}{width=1\textwidth}
			\begin{tabular}{lclrrrrrrr}
				\hline
				& & &   \multicolumn{3}{c}{Naive estimation}& &\multicolumn{3}{c}{RC estimation}  \\
				\cline{4-6}\cline{8-10} PA & Alcohol (g per day) & Parameter &Estimate & SE & P-value & & Estimate & SE & P-value \\
				\hline
				Inactive & $<17.6$ & $\beta_{01}$   &-2.254 & 0.224 & $<0.001$ &  & -2.220  & 0.229 &  $<0.001$ \\ 
				&   & $\beta_{protein}$ & 0.264 & 0.081 & 0.001 &  & 0.227 & 0.133 & 0.087 \\ 
				&   & $\beta_{energy}$  &-0.210 & 0.079 & 0.008 &  & -0.006 & 0.188 & 0.974 \\ 
				\\
				Inactive & [17.6,58.1]  & $\beta_{02}$   &-2.285 & 0.227 &  $<0.001$ &  & -2.282 & 0.233 &  $<0.001$ \\ 
				&   & $\beta_{protein}$ & 0.340 & 0.101 & 0.001 &  & 0.406 & 0.156 & 0.009 \\
				&  & $\beta_{energy}$  &-0.252 & 0.103 & 0.014 &  & -0.008 & 0.230 & 0.971 \\ 
				\\
				Inactive & $>$58.1 & $\beta_{03}$  &-2.046 & 0.255 &  $<0.001$ &  & -1.435 & 0.631 & 0.023 \\ 
				&  & $\beta_{protein}$ &-0.056 & 0.158 & 0.723 &  & -1.232 & 0.885 & 0.164 \\ 
				&   & $\beta_{energy}$  & 0.095 & 0.180 & 0.596 &  & 0.891 & 1.514 & 0.556 \\ 
				\\
				Active &  $<17.6$ & $\beta_{04}$      &-2.501 & 0.228 &  $<0.001$ &  & -2.813 & 0.340 &  $<0.001$ \\
				&  & $\beta_{protein}$ & 0.408 & 0.153 & 0.008 &  & 0.125 & 0.557 & 0.822 \\ 
				&  & $\beta_{energy}$  &-0.494 & 0.145 & 0.001 &  & -1.447 & 0.737 & 0.050 \\ 
				\\
				Active & [17.6,58.1] & $\beta_{05}$      &-2.530 & 0.238 &  $<0.001$ &  & -2.947 & 0.384 &  $<0.001$ \\
				&   & $\beta_{protein}$ & 0.305 & 0.186 & 0.101 &  & -0.043 & 0.641 & 0.946 \\ 
				&  & $\beta_{energy}$  &-0.314 & 0.190 & 0.098 &  & -0.394 & 0.951 & 0.679 \\
				\\
				Active & $>$58.1 & $\beta_{06}$      &-1.858 & 0.342 &  $<0.001$ &  &  2.277 & 2.788 & 0.414 \\ 
				&  & $\beta_{protein}$ & 0.496 & 0.311 & 0.110 &  & 2.494 & 3.832 & 0.515 \\
				& & $\beta_{energy}$  &-0.571 & 0.367 & 0.120 &  & -5.174 & 6.764 & 0.444 \\
				\\ 
				&   & $\alpha_{age}$    & 0.048 & 0.004 &  $<0.001$ &  & 0.048 & 0.004 &  $<0.001$ \\ 
				&   & $\alpha_{edu}$    &-0.495 & 0.063 &  $<0.001$ &  & -0.495 & 0.063 &  $<0.001$ \\ 
				&   & $\alpha_{smoke}$  & 0.161 & 0.060 & 0.007 &  & 0.161 & 0.060 & 0.007 \\ 
				\hline				
			\end{tabular}
		\end{adjustbox}
	\end{center}	
	{\raggedright  Note: SE is standard error.} 
\end{table}

\bibliographystyle{chicago}
\end{document}